\documentclass[aps,prb,twocolumn,10pt]{revtex4-1}
\usepackage{ifpdf}
\usepackage{amsmath}
\usepackage{subcaption}
\usepackage{graphicx}
\usepackage{epstopdf}
\usepackage{mathtools}
\usepackage{color}
\begin{document}

\title{Frequency Dependent Functional Renormalization Group for Interacting Fermionic Systems}
\author{Nahom K. Yirga}
\author{David K. Campbell}
\affiliation{Department of Physics, Boston University, Boston, Massachusetts 02215, USA}
\date{\today}

\begin{abstract}
We derive an expansion of the functional renormalization (fRG) equations that treats the frequency and momentum dependencies of the vertices in a systematic manner. The scheme extends the channel-decomposed fRG equations to the frequency domain and reformulates them as a series of linear integral equations in the particle-particle, particle-hole and particle-hole exchange channels. We show that the linearity of the equations offers numerous computational advantages and leads to converged, stable solutions for a variety of Hamiltonians. As the expansion is in the coupling between channels, the truncations that are necessary to making the scheme computationally viable still lead to equations that treat contributions from all channels equally. As a first benchmark we apply the two-loop fRG equations to the single impurity Anderson model. We consider the sources of error within the fRG, the computational cost associated with each, and how the choice of regulator affects the flow of the fRG. We then use the optimal truncation scheme to study the Extended Hubbard Hamiltonian in one and two dimensions. We find that in many cases of interest the fRG flow converges to a stable vertex and self-energy from which we can extract the various correlation functions and susceptibilities of interest.
\end{abstract}

\maketitle
\section{Introduction and Background}
\label{Intro}
The functional renormalization group (fRG) represents a set of exact functional equations\cite{wetterich1993exact} which capture the evolution of a system from a microscopic Hamiltonian to a fully interacting macroscopic system. Although the framework is exact, it leads to an infinite hierarchy of equations, the solutions of which are beyond the scope of current and likely future computational resources. Truncated formulations of the fRG equations\cite{wetterich1993exact,shankar1994renormalization,zanchi2000weakly} have been instrumental in analyzing the stability of Fermi liquids\cite{shankar1994renormalization} and have been applied to numerous interacting fermion systems such as impurity Hamiltonians\cite{karrasch2008finite,hedden2004functional}, different families of Hubbard models, multi-band superconductors, and a variety of other systems of interest. Beyond helping to establish a complete phase diagram for the weak coupling regime\cite{halboth2000renormalization,honerkamp2001temperature,tam2006functional,eberlein2014superconductivity,kampf2003competing}, fRG methods also show us sub-dominant phases that can be enhanced when the system is perturbed\cite{thomale2009functional} and can in some cases describe strongly-coupled physics\cite{schutz2005collective,tsai2005renormalization} .

Most of the results above are obtained via the static one-particle irreducible(1PI) fRG equations at the one-loop level without self-energy feedback and involve a truncation of the fRG equations at the two-particle level. The static 1PI fRG equations were adopted out of necessity, as even the one and two dimensional Hamiltonians lead to very high dimensional fRG equations often far above the available computational resources. As the most singular contributions in the systems of interest occur at $\omega=0$, the static fRG captures much of the flow at a much lower computational cost. A further approximation employed in many fRG calculations is a patching of the Brillouin zone (BZ) based on relevance of operators in the non-interacting theory which reduces the equations but may miss the effects of nominally irrelevant operators on the effective Hamiltonian\cite{tam2006functional,menard2011renormalization}. Initial efforts to venture beyond these computational limits accounted for the frequency dependence of the vertices via a frequency channel truncation Ansatz \cite{karrasch2008finite}, an approximation that misses much of the non-separable frequency dependence of the two-particle vertex \cite{vilardi2017non}. Recent works have succeeded in a full accounting of frequency content of the vertex at the multiloop level\cite{tagliavini2019multiloop} with quantitative results on par with determinant Monte Carlo\cite{hille2020quantitative}. Access to the frequency content of the vertex along with the Schwinger-Dyson self energy has allowed for a study of the Hubbard model as a function of temperature and interaction strength at various loop orders\cite{hille2020pseudogap}.

The frequency dependence of the fRG equations is essential for going beyond one-loop approximations \cite{katanin2009two}, and recent successes demonstrate the quantitative power of the fRG given the full vertex. Such methods still require some tuning as the scaling of effort is quite large putting lower temperatures and multiband models out of reach. An alternative to these computationally intensive equations involves a generalized Hubbard-Stratonovich transformation to decouple the interaction between fermions via auxiliary boson fields. This can lead to a compact set of fRG equations that describe the various phases of interest \cite{friederich2011functional}. However, such methods can become very complicated for low-dimensional systems with multiple order parameters and are difficult to extend to a generalized model or its multi-band variants, as the complexity of the Fermionic interaction introduces ambiguity in the decoupling of the Fermionic fields.

The computational cost of the static fRG equations can be dramatically reduced if one parameterizes the equations via Bosonic propagators\cite{husemann2009efficient,lichtenstein2017high,PhysRevB.85.035414}. This parameterization, which will be the basis of our present work, leads to a decoupled computationally efficient set of fRG equations. These equations allow us to account for contributions from all momentum modes in a systematic manner. However, the current implementation of the fRG equations offers no pathway for a systematic decoupling of contributions from finite frequency modes. The next level of approximation often tends to be beyond the reach of computational resources which limits the predictive power of the fRG.

In this paper we derive a computational scheme for the efficient calculation of the finite frequency dependence of the fRG equations and address the common approximations implemented in most fRG calculations. The decoupled fRG equations \cite{lichtenstein2017high}, along with the frequency expansions we develop here, lead to well-known linear integral equations similar to the Bethe-Salpeter equation. The general Bethe-Salpeter equation is a linear integral equation which in many cases of interest reduces to an algebraic equation that is computationally trivial to solve. This simplification will be the foundation of most of our calculations. Of the approximations that are commonly implemented in applications of the fRG equations, the most impactful and necessary is the truncation of the infinite hierarchy of the fRG equations. We look at the effect of the frequency decoupling on a truncation at the two-loop level of the fRG in a solvable model, the Single Impurity Anderson Model (SIAM). Further, since truncations can also lead to results that are dependent upon the function chosen to regulate the flow, we study the sensitivity of our results to our choice of regulator. We use the results of these studies on the SIAM as reference to expand the vertices of Hamiltonians with spatial dependence.

The remainder of the paper is organized as follows. For the sake of completeness we begin in Sec.\ref{Method} with a short review of the fRG equations and present the reformulated version of the fRG equations. In Sec.\ref{SIAM} we use the single impurity Anderson model as a case study to systematically explore the impact the choice of regulator can have on converged fRG flows and to look at the systematics of the frequency dependence of the vertex. In Sec.\ref{EHModel} we apply our method, along with a momentum decoupling expansion, to calculate the correlation functions and the phase diagram of the one-dimensional extended Hubbard model. In Sec. \ref{2DModels} we apply the fRG to the two-dimensional extended Hubbard model, where we study the role of nearest-neighbor hopping ($t^\prime$) and nearest-neighbor density interactions ($V$) on the phase diagram and calculate the relevant susceptibilities of the system. Our conclusions and future challenges are given in Sec. \ref{Conclusion}.

\section{The functional Renormalization Group (fRG) Method} \label{Method}
The general interacting Fermi systems we wish to study require identifying the ordering tendencies of the many-body ground state of the system. For cases where the Hamiltonian can be separated into a free and an interacting piece ($\mathcal{H} = \mathcal{H}_0+\mathcal{H}_I$) one can use the fRG to construct the vertices and through the vertices the correlators that describe the full system. The action for the Hamiltonian $\mathcal{H}$ is given by
\begin{align}
S[\bar{\psi},\psi]=-\sum_{\alpha,\beta}\bar{\psi}_\alpha(\mathcal{G}_0^{-1})_{\alpha,\beta}\psi_\beta + V_I[\bar{\psi},\psi]
\end{align}
where $\mathcal{G}_0$ is the propagator of the free Hamiltonian, $\psi$,$\bar{\psi}$ are anticommuting Grassmann fields and $V_I$ is the vertex derived from interacting Hamiltonian. Extensive reviews of the fRG methods for fermions and the derivations of the fRG equations can be found in previous works \cite{metzner2012functional,kopietz2010introduction}. For completeness, a brief derivation of the fRG equations is given in Appendix.\ref{dfRGequations}.

Formally, the fRG equations are derived from a scale-dependent effective action constructed from the action $S$ that suppresses all contributions below a given scale. Using this scale-dependent effective action, one can derive a hierarchy of equations for the one-particle irreducible functions that describe the system at each scale $\Lambda$. Though the fRG equations for the 1PI vertex functions form an infinite hierarchy, the two functions of practical interest are the self-energy ($\Sigma_{\alpha\beta}$) and the two-particle vertex ($\Gamma_{\alpha\beta\gamma\delta}$). The self-energy is necessary to describe the renormalization of general single particle interaction effects such as the renormalization of the quasi-particle weight or shifts in the Fermi surface, while the two-particle vertex captures the various response functions of the system. The fRG equations for the self-energy and the two-particle vertex are given by
\begin{widetext}
\begin{equation}
\frac{d}{d\Lambda}\Sigma_{p}=\sum_{k}\partial_\Lambda G_{k}^\Lambda\Gamma^{(4)\Lambda}_{pkkp}
\label{selfEnergy}
\end{equation}
\begin{align}
\Gamma_{p_1p_2p_3p_4}=\Gamma_{p_1p_2p_3p_4}^{(0)} + \Phi_{p_1p_2p_3p_4}^{ph} + \Phi_{p_1p_2p_3p_4}^{phe} + \Phi_{p_1p_2p_3p_4}^{pp}\nonumber
\end{align}
\begin{align}
\frac{d}{d\Lambda}\Gamma_{p_1p_2p_3p_4}=&\sum_{k_1k_2}\left(\partial_{\Lambda,S}(G_{k_1}^\Lambda G_{k_2}^\Lambda)(\Gamma_{p_1k_1k_2p_4}^{(4)\Lambda}\Gamma_{k_2p_2p_3k_1}^{(4)\Lambda}-\Gamma_{p_2k_1k_2p_4}^{(4)\Lambda}\Gamma_{k_2p_1p_3k_1}^{(4)\Lambda})-\frac{1}{2}\partial_{\Lambda,S}(G_{k_1}^\Lambda G_{k_2}^\Lambda)\Gamma_{p_1p_2 k_2k_1}^{(4)\Lambda}\Gamma_{k_1k_2p_3p_4}^{(4)\Lambda}\right) + \nonumber\\
&\sum_{k}\partial_\Lambda G_{k}^{\Lambda}\Gamma_{p_1p_2kkp_3p_4}^{(6)\Lambda}
\label{fRGequations}
\end{align}
\end{widetext}
where the general indices $p_i$ label momentum, spin, band index, etc..., $G_{k}^{\Lambda}$ is the propagator and $\Gamma^{(6)\Lambda}$ is the three-particle vertex. The contributions to the particle-hole($\Phi^{ph}$), particle-hole exchange($\Phi^{phe}$) and particle-particle($\Phi^{pp}$) functions are given in the flow equation for the vertex and correspond to the first three terms, respectively. The final term in the vertex flow equation is driven by the three-particle vertex , which we need to approximate in order to close the system of equations. The standard computationally inexpensive approach is to discard the three-particle contributions. This is motivated by the result that the three-particle interactions are initially absent in most models of interest and consequently are generated only by the flow. This truncation limits the applicability of the fRG to the weak coupling regime and leads to divergent flows even at weak coupling. Improvements upon this approximation have been made by including contributions from non-overlapping two-loop diagrams in the three-particle vertex via Katanin's correction($\partial_{\Lambda,S}\rightarrow \frac{d}{d\Lambda}$)\cite{katanin2004quasiparticle} and overlapping two-loop diagrams via Eberlin's scheme \cite{eberlein2014Fermionic}. To fully close the fRG equations one needs to employ the multi-loop fRG\cite{kugler2018multiloop}, which systematically captures the full contribution of the three-particle vertex. At the one-loop level of approximation, the computational cost of the fRG is equal to solving an integral equation for the two-particle vertex, a three variable function, which leads to a power law scaling of the fRG. Specifically, evaluating the two-particle vertex amounts to solving $N_k^4N_w^4N_b^3$ fRG equations where $N_k$ represents the number of sites, $N_w$ the discretization of frequency axis and $N_b$ the number of bands. Furthermore, the computational cost depends linearly on the number of loops, which puts even simple one-band models beyond the full one-loop fRG. Methods such as the Truncated Unity-fRG (TUfRG)\cite{lichtenstein2017high} decouple the momentum dependence of the fRG equations in the three channels via an efficient decomposition of the vertex. For the right choice of basis functions, the decoupling scales only linearly with the momentum modes of the system. The computational gains are tremendous, which emphasizes the need for a similar decoupling of the frequency modes.
\begin{figure*}
\centering
\includegraphics[scale=0.63]{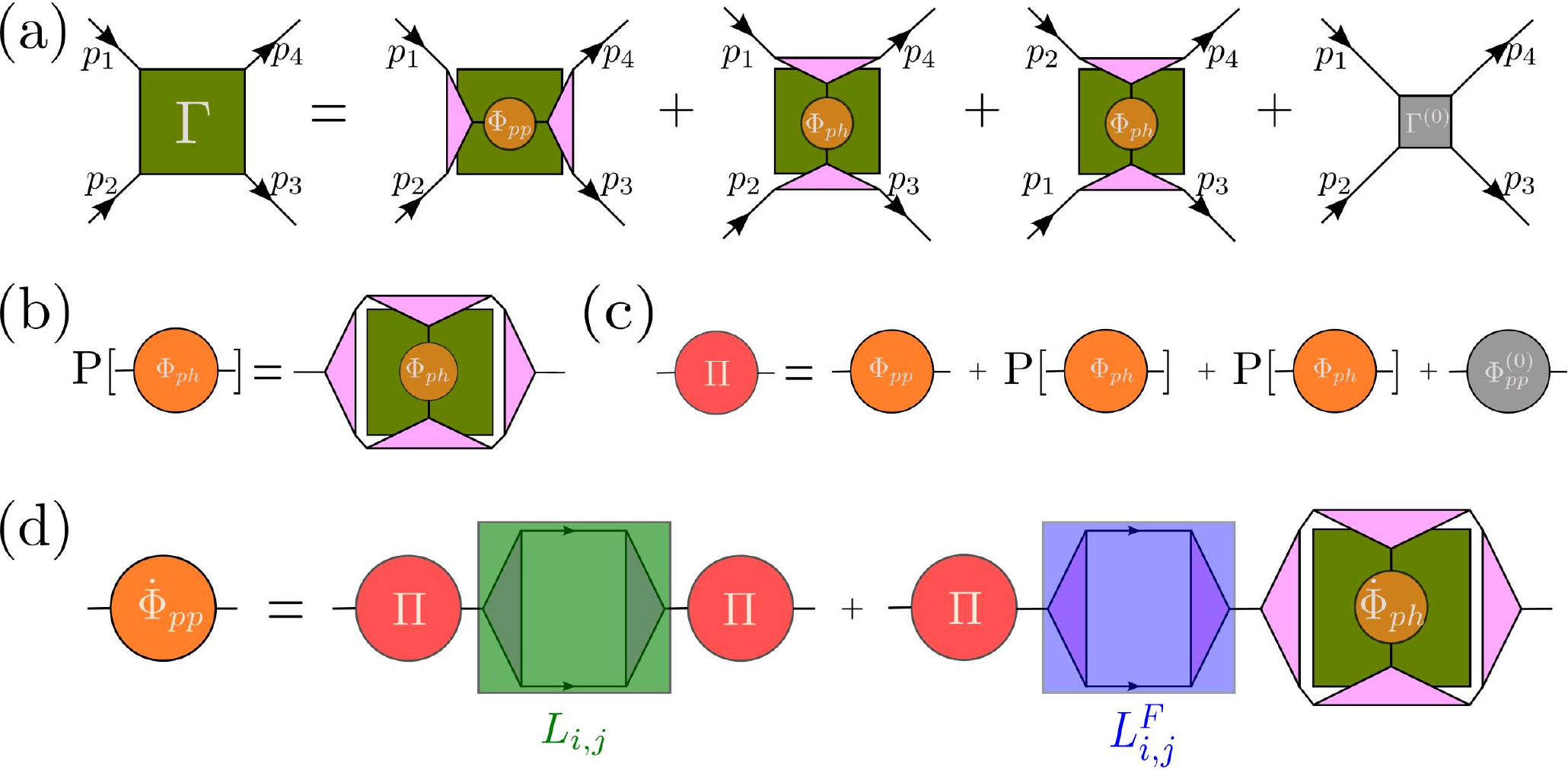}
\caption{A schematic representation of the channel decomposed fRG. (a) The decomposed contributions from the three channels to the two-particle vertex. (b) A projection from the ph to the pp channel. (c) The expansion of the vertex in the pp channel. (d) The two-loop fRG equations of the vertex in the pp channel.}
\label{fRGdiag}
\end{figure*}

The scale-dependent effective action that corresponds to the vertices defined above is constructed by modifying the bare propagator of our effective action via a regulator that suppresses all infrared modes below the current scale ($\Lambda$). Within the fRG, we have a choice of where to impose the regulator. Most numerical work in the fRG is performed via pure momentum or frequency regulators. For interacting electron systems, momentum regulators have a distinct disadvantage, as the Fermi surface, towards which the flow evolves, is derived from the flow itself, leading to a complicated self-consistent equation. To avoid this problem, we will employ a pure frequency cutoff for the propagators of all the systems we study in this paper. The regulated propagator is strongly suppressed below the current scale and, as it drives the flow of the 1PI-vertex functions, the 1PI vertices also exhibit structures of width $\Lambda$. To decouple the fRG equations one needs to efficiently expand the 1PI vertices to some order in the auxiliary variables.  Such an expansion is only feasible after a proper rescaling at all scales by some function $f(\Lambda)$. The function $f$ has a simple structure. At the start of the flow, it needs to be proportional to the scale $\Lambda$, and as the flow converges at the scale of interest it needs to approach one. For systems at finite temperature, the rescaling of the frequency axis corresponds to scaling the temperature ($\beta_\Lambda$) of the system (see Appendix.\ref{scaledVert}). The rescaling of the 1PI vertices is a direct consequence of the fRG methods, as the renormalization process generates a continuous set of actions ($\mathcal{S}_\Lambda$) each with vertices at some scale, $\Lambda$, that are only useful in describing the dynamics of the system at higher scales. The exact form of the function is model dependent. Thus in our numerical work we calculate $f(\Lambda)$ directly from the flow.

For translation-invariant systems, the contributions of the fRG flow to each channel are driven by exchange propagators, which are characterized via a singular Bosonic frequency\cite{karrasch2008finite} and two auxiliary frequencies that label the incoming and outgoing objects. We shall parameterize the two-particle vertex via these singular frequencies. The singular frequencies for the exchange propagators are given by
\begin{equation}
s_{pp}=p_1 +p_2, \quad s_{ph}=p_3-p_2, \quad s_{phe}=p_3-p_1
\end{equation}
for particle-particle, particle-hole and particle-hole exchange channels respectively. The parametrization ensures that the two-particle vertex conserves energy and momentum. The one-loop fRG equations can then be decoupled by expanding each of the channels around the corresponding Bosonic frequency \cite{husemann2009efficient}. We begin by defining the three expansions of the two-particle vertex. We have
\begin{align}
\Pi_{m,n}(s_{pp}) =\int\int ds_xds_y\Gamma_{s_{pp},s_{ph},s_{phe}}f_m(s_x)f_n(s_y)\nonumber\\
\Delta_{m,n}(s_{ph})=\int\int ds_xds_y\Gamma_{s_{pp},s_{ph},s_{phe}}f_m(s_x)f_n(s_y)\nonumber\\
X_{m,n}(s_{phe})=\int\int ds_xds_y\Gamma_{s_{pp},s_{ph},s_{phe}}f_m(s_x)f_n(s_y)
\label{threeExpansions}
\end{align}
where $f_n$ are our orthogonal set of basis functions and for each channel $s_x,s_y$ label the auxiliary frequencies. As the expansion of the vertex is over the full frequency domain, we parameterize the auxiliary labels with Bosonic frequencies which correspond to the sum and difference of the two other Bosonic frequencies (For the pp-channel: $s_x=s_{ph}-s_{phe}$, $s_y=s_{ph}+s_{phe}$). Details on the choice for the auxiliary variable can be found in Appendix.\ref{auxVarChoice}. This choice of parameterization comes with the added advantage of transforming all the symmetries of the vertex to simple sign changes\cite{karrasch2008finite}. Formally each of the three functions contains the full two-particle vertex, but a full expansion is redundant. Thus, we expect each expansion to be relevant near the corresponding singular frequency. The orthogonality of the basis functions allows us to easily invert the expression to find the two-particle vertex in terms of the three functions. To decouple the fRG equations, we formally expand the two-particle vertex in the contribution from each channel via the corresponding expansion. For example, for the particle-particle channel expanding $\Gamma^{(4)\Lambda}$ in terms of $\Pi_{m,n}$ leads to
\begin{align}
\Phi_{\alpha\beta\gamma\delta}^{\Lambda,PP}=\sum_{m,n,i,j}f_m(s_x)f_j(s_y)\Pi_{m,n}(s_{pp})\Pi_{i,j}(s_{pp})\times \nonumber\\ \sum_w f_n\left(s_{pp}-2w\right)f_i\left(2w-s_{pp}\right)\partial_\Lambda\left(G_w^\Lambda G_{s_{pp}-w}^\Lambda\right)
\end{align}
where the other singular frequencies occur only in the form factors ($f_m(s_x),f_j(s_y)$). Simply expanding the contribution from the particle-particle channel ($\Phi^{(4)\Lambda, PP}$) around $s_{PP}$ yields the flow equation of the particle-particle contributions. The natural limit of the expression is the single channel truncation of the fRG where we discard the contributions from the other channels. In this particular limit, the vertex is parametrized  by $s_{pp}$ ($\Gamma^{(4)\Lambda}=\Pi_{0,0}(s_{pp}$) only, and the equation above reduces to the Bethe-Salpeter equation in the particle-particle channel (see Appendix.\ref{asymP}).

The expansion described above is only as good as our choice of basis functions. As the Matsubara frequencies are continuous at zero temperature our expansion needs to efficiently capture the entire frequency domain. The need to rescale our vertices combined with the instantaneous nature of the interactions we are interested in restricts our available basis choices. A direct approach to handle the scale dependence of our vertex is to interpret the intermediate vertices at scale $\Lambda$, which only describe fluctuations at higher energy scales ($\omega >\Lambda$), as those of a system at a higher temperature. We can then patch the time interval $[0,\beta_\Lambda]$ and approximate the vertex by averaging over these patches. The patching scheme is akin to the dynamical cluster approximation (DCA)\cite{jarrell2008dynamical} except with averaging implemented over the imaginary time axis. We have
\begin{align}
\Pi_{\tau_m,\tau_n}(\omega_{pp})=\int_{\tau_1\in\Delta_m(f_\Lambda)}\int_{\tau_2\in\Delta_n(f_\Lambda)}\Pi_{\tau_1,\tau_2}(\omega_{pp})
\end{align}
with $\Pi_{\tau_1,\tau_2}(\omega_{pp})$ defined as
\begin{align}
\Pi_{\tau_1,\tau_2}(\omega_{pp})=\frac{1}{\beta^2}\sum_{\omega_{pp_x},\omega_{pp_y}}\Gamma_{\omega_{pp},\omega_{ph},\omega_{phe}}^{\Lambda}e^{i\omega_{pp_x}\tau_1-i\omega_{pp_y}\tau_2}
\end{align}
with $\tau_{1/2}$ corresponding to the fourier transform of our auxiliary channels. The patches are scale dependent with the cluster size scaling with $f_\Lambda$ ($A\sim\Delta\tau_m\Delta\tau_n/(f_\Lambda^2)$).

\begin{figure}
\centering
\includegraphics[scale=0.6]{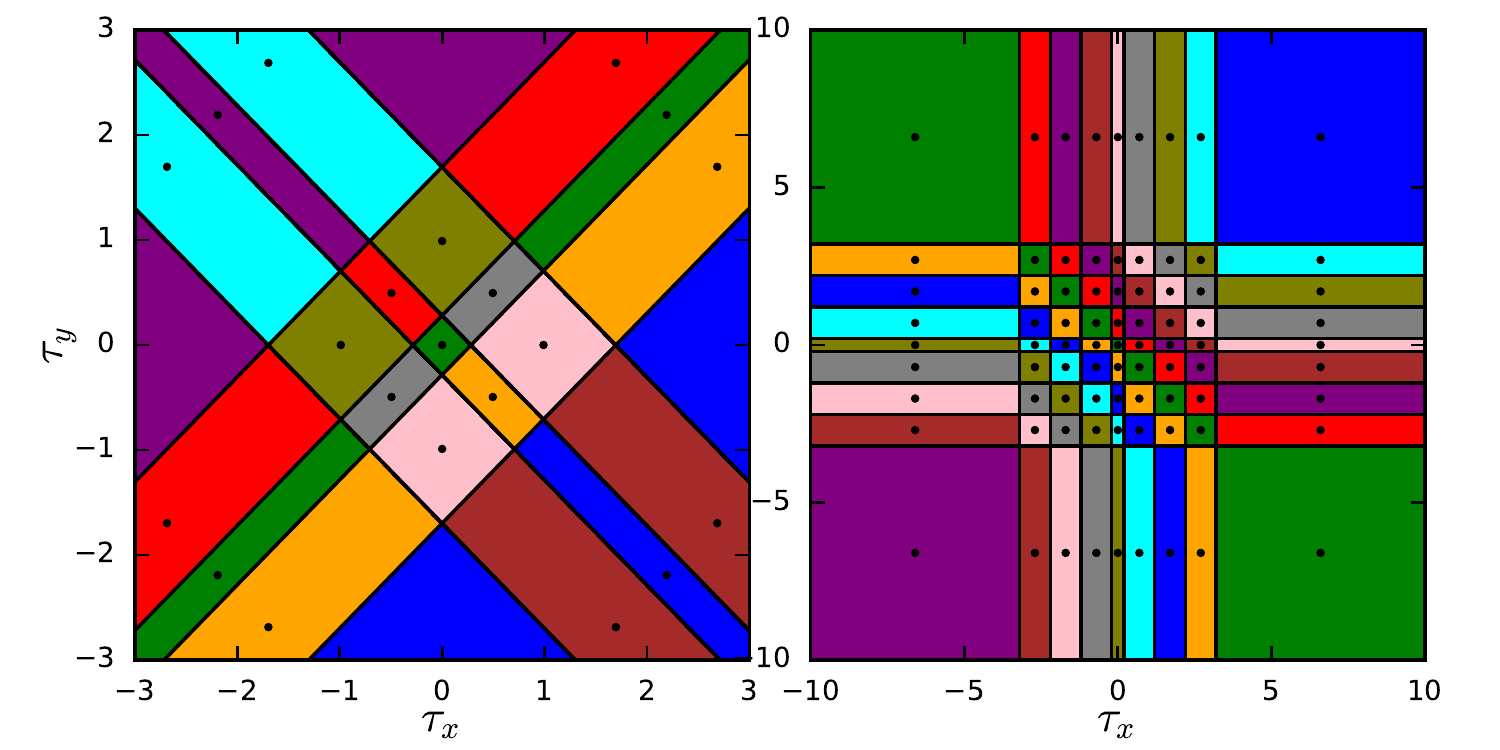}
\caption{Discretization of the imaginary time axis for $N_\omega=4$ at $\beta$=10 (left) and $N_\omega=8$ at $\beta$=20 (right) with $\Delta\tau=1.0$ and $\Delta\tau_{min}=0.1$.}
\label{Patches}
\end{figure}

The scheme requires further modification to account for instantaneous interactions with support at all frequencies. Our modification involves the insertion of a small patch ($\Delta \tau_{min}=1/\Lambda_0$) at all scales which allows support up to $\Lambda_0$. The rest of our patches are focused around small times ($\Delta\tau<<\beta_\Lambda/N_\omega$) to better resolve the full set of fluctuations at low frequencies. For low temperatures ($N_\omega\Delta\tau <\beta_\Lambda$) an additional patch is required to fully cover the interval. At high temperatures with a large basis set the patching scheme is identical to an equal partition of the interval($[0,\beta_\Lambda]$) but for low temperature calculations or higher dimensional models this patching scheme is superior as computational resources limit the maximum size of the available basis set ($N_\omega$). The scheme for two patch geometries is shown in Fig.\ref{Patches}. The scale dependent temperature $\beta_\Lambda$ is given as $f_\Lambda\beta$ with $f_\Lambda$ ideally calculated from the flow. We note that simple forms of $f_\Lambda$ $(1+\Lambda^n)^{1/n}$ work pretty well especially for small basis sets where the frequency modes are coarsely averaged over. As we integrate the flow $\beta_\Lambda$ quickly approaches the temperature of our system. Calculations at zero temperature can be simplified as the rescaling of the frequency modes along the flow can be recast as a derivative. For the self-energy we have
\begin{align}
\partial_l\Sigma(\omega_\Lambda)=-\omega\partial_\omega\Sigma(\omega_\Lambda)-\Lambda\sum_{s}\partial_\Lambda G_{s}^\Lambda\Gamma^{(4)\Lambda}_{\omega ss\omega}
\end{align}
where we have parameterized the cutoff ($\Lambda=\Lambda_0 \exp(-l)$) and suppressed momentum dependence of the vertices. A similar derivative term appears at the two-particle level which complicates the flow of our expansion as in addition to the rescaling of the singular frequency we have to rescale the auxiliary channels. For the particle-particle channel we have
\begin{align}
\partial_l\Phi_{m,n}^{pp}(\omega_{pp,\Lambda})&=-\omega\partial_\omega\Phi_{m,n}^{pp}(\omega) + F_\Lambda(\Gamma,\mathcal{G},\Gamma)\nonumber\\
&-\Lambda\sum_{a,b,s_x,s_y}\Phi_{a,b}^{pp}(\omega)\partial_\Lambda\left(f_a(s_x)f_b(s_y)\right)
\label{scaleVertexFlow}
\end{align}
where the functions $F_\Lambda$ represent unrescaled contributions to the flow with identical alterations for particle-hole and exchange channels. In the numerical implementation of the cluster expansion a direct calculation of the derivative can be noisy and sensitive to interpolation schemes used in conjunction with the flow. Rather than calculating the noisy numerical derivative at points of interest, we found estimating it indirectly from the interpolated vertices at the point in the flow to be the cleaner approach. Expressions for $F_\Lambda$ along with an expansion of the vertex calculated directly from full fRG is given in Appendix.\ref{exVert}.

A simpler approach to handle the instantaneous interactions with finite support at all frequencies is to map the Matsubara axis to a finite domain and use the scaled Legendre polynomials. At each scale $\Lambda$ we map the frequency axis to a finite domain via a scale dependent transformation. The dependence of the map on $f(\Lambda)$ enables an efficient expansion of the 1PI vertices at the given scale. After rescaling, we expand the singular frequency dependence along the imaginary time interval $[0,\beta_\Lambda]$ in terms of Legendre polynomials and expand the auxiliary frequencies via scaled Legendre polynomials\cite{shen2009some}. The Legendre polynomials have been shown\cite{boehnke2011orthogonal} to offer a compact representation of the frequency content of the vertex functions. The (scaled) algebraic transformation used to map the frequency axis to $[-1,1]$ is given by
\begin{equation}
z=\frac{\omega}{\sqrt{\omega^2+f(\Lambda)^2}}
\end{equation}
where $\Lambda$ is the current scale of the flow. The scale derivative of the vertex also leads to an additional term, akin to the $F_\Lambda$ term in the cluster expansion given in Eq.\ref{scaleVertexFlow}, as the expansion itself is now scale dependent. We subsume this additional term into the scale derivative. To further reduce the computational cost we use translation symmetry, energy conservation, and the hermiticity of the Hamiltonian to reduce the relevant basis set need for the two-particle vertex (see Appendix.\ref{symmV}).

As we will see below, both expansions appear viable for a variety of models, with the scaled cluster expansion showing a faster convergence despite issues of noise at the single particle level. Noise is a common issue with cluster averaging though in the decoupled fRG flow it appears restricted with observables at the two-particle level showing little sensitivity. Another advantage of the cluster expansion is sparsity, with the spectral weight of the three expansions of our vertex contained in a small subset of modes. This feature of the expansion is ideal, as we can qualitatively treat complex models by retaining a small but relevant (sorted via spectral weight) basis set. Finally, we note that we have not fully explored patch geometries over which we average our models. For zero temperature calculations a logarithmic patch works well but at finite temperatures, especially in cases where $\beta\rightarrow 0, $ the simpler linear patches shown in Fig.\ref{Patches} work remarkably well. Patch geometry and noise in the scaled cluster expansion show minimal impact with the various models we consider in this work. Hence we leave a full study of the optimal choice for future work.

Extension of the fRG equations to Hamiltonians with spatial dependence is direct, as the rescaling of the vertex is isolated to the frequency domain due to our choice of a pure frequency regulator. We decouple the momentum modes using a set of plane waves which have been shown to retain many of the computational benefits of the decomposition\cite{eckhardt2018truncated}. Further computational gains can be achieved by exploiting the symmetries of the lattice of interest\cite{platt2013functional}. Utilization of lattice symmetries to reduce the three expansions of vertex is carried out for the two dimensional square lattice in Appendix.\ref{symmV}. In order to avoid the errors from truncating the momentum expansion, we patch the BZ and average over the momenta modes. This gives us a full coverage of the BZ and allows us to reach large system sizes ($N\approx1024$). Limitations to system size are primarily due to the retention of large set of matsubara frequencies ($N_f\geq 40$). Even larger system sizes are accessible at higher temperatures due to the much smaller sets needed for a stable flow.

For a general single band Hamiltonian that obeys SU(2) spin symmetry we can use the expansions of the vertex ($\Gamma_{\uparrow\downarrow\downarrow\uparrow}$) to decompose all channels of the fRG equations. Hence we have
\begin{align}
\dot{\boldsymbol{\Phi}}^{pp}(s_{pp})=&\boldsymbol{\Pi}(s_{pp})\textbf{L}_{\Lambda}^{pp}(s_{pp},\Lambda)\boldsymbol{\Pi}(s_{pp})\\
\dot{\boldsymbol{\Phi}}^{ph}(s_{ph})=&\left(\boldsymbol{\Delta}(s_{ph})-\textbf{X}(s_{ph})\right)\textbf{L}_{\Lambda}^{ph}(s_{ph})\boldsymbol{\Delta}(s_{ph}) +\nonumber\\
&\boldsymbol{\Delta}(s_{ph})\textbf{L}^{ph}(s_{ph},\Lambda)(\boldsymbol{\Delta}(s_{ph})-\textbf{X}(s_{ph}))\\
\dot{\boldsymbol{\Phi}}^{phe}(s_{phe})=&\textbf{X}(s_{phe})\textbf{L}^{ph}(s_{phe},\Lambda)\textbf{X}(s_{phe})
\label{channelDerv}
\end{align}
where the bold letters denote matrices whose components correspond to the form factors of the frequency and momentum modes with the dots denoting a scale derivative. These equations are decoupled in the sense that once we calculate the particle-particle and particle-hole exchange propagators($L^{pp},L^{ph}$) we can easily use the matrix equations above to determine the contributions to the vertex at scale $\Lambda$.  The equations for the components of the exchange propagator are given by
\begin{align}
L_{m,n}^{\Lambda,ph/pp}(s_x)=&\sum_af_m\left(s_x\pm 2a\right)f_n\left(\pm s_x\pm 2a\right)\times\nonumber\\
&\partial_\Lambda\left(\mathcal{G}_a^\Lambda\mathcal{G}_{s_x\pm a}^\Lambda\right)
\label{exchangeP}
\end{align}
where a runs over frequency and momentum. After we find the contributions from the three channels, we can construct part of the vertex necessary for driving the flow of the self-energy (Eq.\ref{selfEnergy}). To close the fRG equations we need to address the projections of the three expansions of the vertex ($\Pi,\Delta,X$) into the various channels along the fRG flow. The contributions to the three expansions from their respective channels add directly, whereas one needs to project contributions across channels ($\dot{\Phi}^{PP}\rightarrow\Delta,X$). Projection of the frequency modes requires expanding the piece of the vertex we need to project in terms of our basis functions. We have
\begin{equation}
\Phi_{m,n}^{x}(i)=\int D\alpha\Phi_{m,n}^{x}(\alpha)f_i(\alpha)
\end{equation}
where $D\alpha$ contains the Jacobian for the scale transform as well our momentum and frequency normalizations. The contributions from $\Phi^x$ to the other channels can then be easily calculated via our expansion given in Eq.\ref{threeExpansions}. For example, a projection from $\Pi$ to $\Delta$ is given by
\begin{align}
\Delta_{a,b}(s_{ph})=\sum_{m,n,i}\int\int ds_xds_yf_a(s_x)f_b(s_y)\times\nonumber\\
f_m\left(s_{ph}-s_{phe}\right)f_n\left(s_{ph}+s_{phe}\right)f_i(s_{pp})\Phi_{m,n}^{pp}(i)
\label{projectV}
\end{align}
where $s_x$ and $s_y$ are the auxiliary variables for the ph channel. The integrals can be calculated separately for the momentum and frequency basis functions before initializing the flow. The various projections between the three channels are described in detail in Appendix.\ref{projectionD}.

The final term we need to account for within this channel-truncated scheme is the contribution to the flow from the three-particle vertex. An efficient and stable approach to accounting for these contributions is to decompose them into the three channels and express them in terms of the one- and two-particle vertices. The contributions can be divided into diagrams with overlapping and non-overlapping loops. Contributions from diagrams with non-overlapping loops can be accounted for via Katanin's correction, which requires inserting the self-energy flow ($\partial_\Lambda\Sigma$) into the exchange propagator. For the scheme above this insertion leads to a recalculation of $L_{m,n}^{\Lambda,ph/pp}$ with the full derivative of the propagator($\mathcal{G}$). Similarly the contributions from overlapping loops can be included via Eberlin's scheme, which recasts these terms as derivatives in the three channels. To implement this scheme, we need to calculate the full exchange propagators($L(\partial_\Lambda(\mathcal{G}\mathcal{G}))\rightarrow L_F(\mathcal{G}\mathcal{G})$) on top of the on scale exchange propagators calculated above. We also need to calculate the projections of the contributions in each channel ($\dot{\boldsymbol{\Phi}}^{x}$) to the other channels. The two-loop contributions to the particle-particle channel are given by
\begin{align}
\dot{\boldsymbol{\Phi}}_{2L}^{pp}(s_{pp})=&\dot{\boldsymbol{\Pi}}(s_{pp})\textbf{L}_{F,\Lambda}^{pp}(s_{pp},\Lambda)\boldsymbol{\Pi}(s_{pp})+\nonumber\\&\boldsymbol{\Pi}(s_{pp})\textbf{L}_{F,\Lambda}^{pp}(s_{pp},\Lambda)\dot{\boldsymbol{\Pi}}(s_{pp})\\
\dot{\boldsymbol{\Pi}}(s_{pp})=&P_{pp}(\dot{\boldsymbol{\Phi}}^{ph})(s_{pp})+P_{pp}(\dot{\boldsymbol{\Phi}}^{phe})(s_{pp})
\end{align}
where $P_{pp}()$ represents a projection to the particle-particle channel. The projection of the one-loop channel derivatives along with the full exchange propagators fully capture all contributions from the overlapping loops.

Up to this point we have not addressed the regulator in the scale-dependent fRG equations. Contributions at a particular scale are determined by a regulator ($R_\Lambda(s)$) which suppresses all contributions below the current
scale. For the full fRG equations,  the choice of a regulator is irrelevant, but as soon as one truncates the hierarchy of equations, the errors from the truncation can be amplified by the regulator. The dependence of the truncated fRG equations on a regulator has been previously studied and an optimized regulator that will reduce the errors from the truncation as well as lead to a faster convergence has been proposed\cite{litim2001optimized}. Throughout this work, at the two-loop level of approximation described above, we will evaluate the dependence of the solution to the fRG equations on the regulator of choice. To this end, we will consider four regulators: namely,  the optimized Litim cutoff, a smooth additive cutoff, a sharp and soft multiplicative cutoff. The form of the cutoff functions implemented over the frequency axis is given by
\begin{align}
R_\Lambda^{A_l}(\omega)=\quad&i\Theta(\Lambda-|\omega|)(\Lambda sgn(\omega)-\omega)\\
R_\Lambda^{A_s}(\omega)=\quad&i\left(sgn(\omega)(\Lambda^4+\omega^4)^{\frac{1}{4}}-\omega\right)\\
R_\Lambda^{M_s}(\omega)=\quad&\frac{\omega^2}{\omega^2+\Lambda^2}\\
R_\Lambda^{M_{sh}}(\omega)=\quad&\Theta(|\omega|-\Lambda)
\end{align}
where $\Theta$ is the Heaviside step function. Our results indicate that (as expected) at the one-loop level the optimized Litim regulator performs better but at the two-loop level there appears to be no clear favorite. We note that the multiplicative cutoff schemes show faster convergence particularly within the cluster decomposition so they might be the ideal choice for computationally demanding studies of one and two dimensional Hamiltonians. We note that for the Katanin and Eberlin truncation schemes, the computational burden of the multiplicative step function is slightly higher than those for the other regulators, as the single-scale propagator is a delta function and thus requires an evaluation of the exchange propagators at scale $\Lambda$.

To numerically solve the functional differential equations given above, we need to discretize both the frequency and the momentum dependencies of the vertices. Finite-size systems naturally have discrete momentum modes thus, we calculate the flow of the three expansions of the vertex on a subset of these modes (up to the full set depending on the size of the system). For large system size we can either cluster our modes or utilize symmetries, particularly lattice symmetries in the case of two dimensional systems, to reduce the number of modes that need to be retained. Retaining the full vertex is expensive, but for large systems the memory cost of retaining the full set of momentum modes is always more expensive than the computational cost of filling the vertex that comes with a reduced set. For small systems the choice is complicated by the dimensionality of the system, as interpolative schemes show small artifacts for one-dimensional models but can, depending on the choice of interpolator, lead to unphysical results in sparsely sampled two-dimensional systems. For the frequency axis a natural discretization is possible by doing our calculations at finite temperatures. At finite temperatures the self-energy is defined at Fermionic Matsubara frequencies, while the expansions in the three channels of the vertex depends on the Bosonic Matsubara frequencies. As the number of frequencies is still infinite for our implementation, we select a manageable logarithmic set of Bosonic frequencies. In instances when we need to evaluate the vertex (cross projections, for instance), we use linear or cubic interpolation to estimate the functions at the points of interest. We find that as long as our set of points is not too sparse, the linear and cubic interpolation methods perform equally well. For a very small set of points, the linear interpolator is incredibly cheap while cubic interpolator leads to a faster convergence.

A numerical burden that comes with choosing to work at finite temperatures is the sum over Fermionic Matsubara frequencies that occurs both in calculation of the self-energy, Eq.\ref{selfEnergy}, and the exchange propagators, Eq.\ref{exchangeP}. Within the Matsubara expansion of the Fermi function, these sums converge so slowly that they completely negate the advantages of working with the discrete frequencies. Truncated Matsubara sums can be performed much more effectively via a continued fraction expansion of the Fermi function pioneered by Ozaki\cite{ozaki2007continued}. Hence, throughout this work,  the sums over Fermionic variables are performed over the poles of the continued fraction expansion. For models with phase transitions, performing our calculations at finite temperatures comes with the added advantage of avoiding divergences, as we can always set the temperature of our fRG computation above the critical temperature.

\section{The Single Impurity Anderson Model: A model in zero spatial dimensions}\label{SIAM}
\begin{figure}
\centering
\includegraphics[scale=0.57]{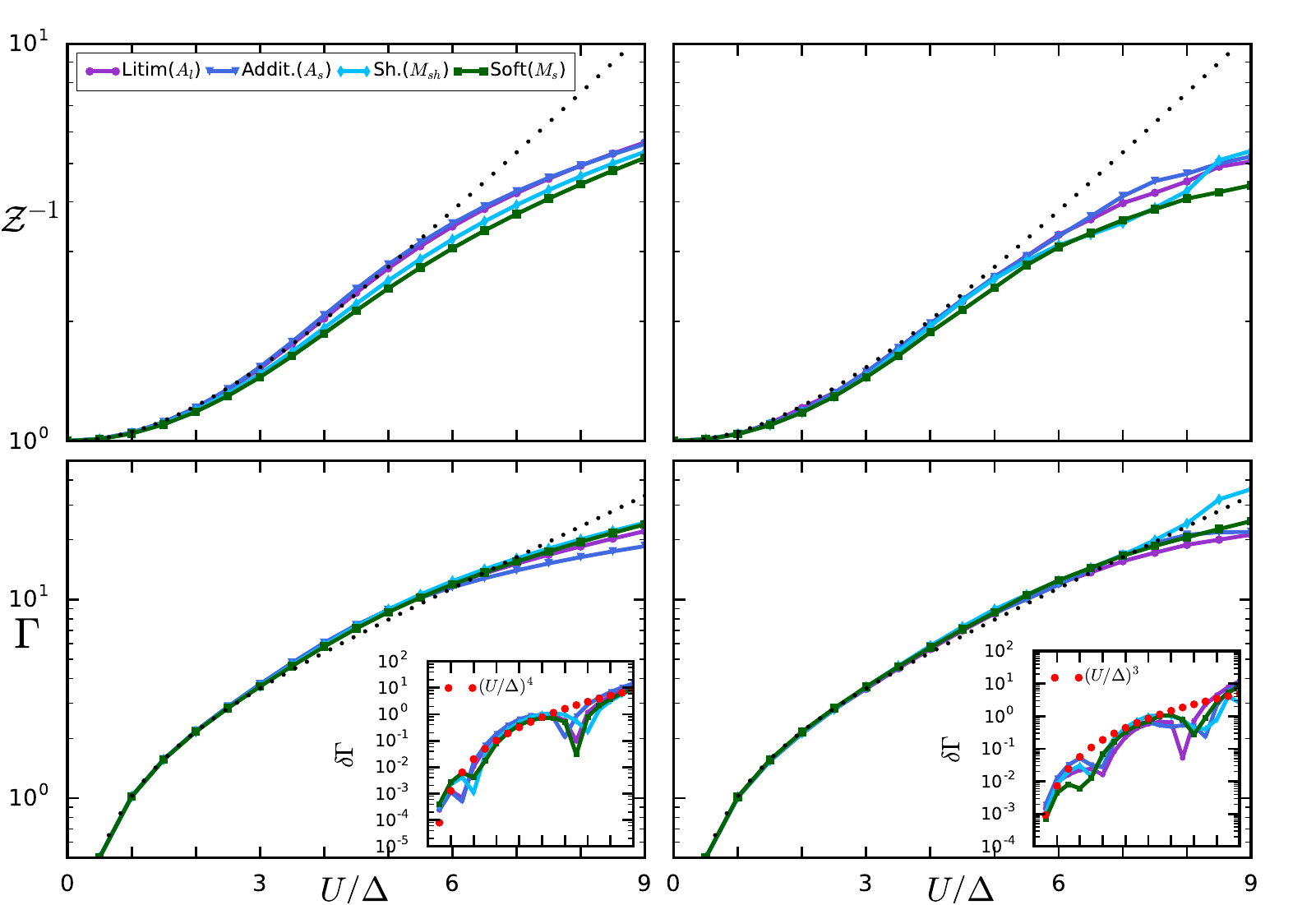}
\caption{Quasi-particle weight (Top) and vertex (Bottom) of the SIAM as a function of the local Hubbard interaction via the Legendre expansion (Left) and Cluster expansion (Right) along with the zero temperature Bethe Ansatz results for the Litim, Additive, Soft and Sharp regulators at the two-loop level. The fRG equation were solved for $\beta=80$, we retained $N_f=30$ Matsubara frequencies with $N_\omega=10$ even Legendre functions and $N_\omega=20$ patches for each expansion. The error at the level of the vertex is shown (Inset).}
\label{susceptB}
\end{figure}
In this section, we apply the decoupled fRG equations to compute single and two-particle properties of the Single Impurity Anderson Model (SIAM). The SIAM is a minimal model that captures the physics of a local site interacting with a quasi-continuous bath of free electrons. The Hamiltonian for the SIAM is given by
\begin{align}
\mathcal{H}=&\sum_\sigma \mu c_\sigma^\dagger c_\sigma +Un_\uparrow n_\downarrow + \sum_{k,\sigma}\xi_k a_{k\sigma}^\dagger a_{k\sigma} \nonumber\\
&+\sum_{k,\sigma}\left(V_ka_{k\sigma}^\dagger c_\sigma + h.c\right)
\end{align}
where $U$ is the local Hubbard repulsion, $\xi_k$ is the dispersion of the free electrons and $V_k$ is the coupling between local site and the free electrons. The operators $a$ and $c$ create and annihilate Fermions in the bath and at the local site. The SIAM was originally proposed to describe magnetic impurities embedded in a metallic host. It is closely related to the Kondo model which describes an impurity spin coupled to a reservoir of non-interacting electrons. As we discuss below, the model has been solved by several methods. The solution reveals that as one lowers the temperature, the SIAM forms a local moment at the impurity site. This local moment can be clearly seen as the Coulomb repulsion grows $U/\Delta\rightarrow\infty$ where charge fluctuations are completely frozen out, and the system is characterized by the local spin. If the temperature is further decreased, the impurity spin at any coupling will always be screened by the conduction electrons. Thus, for large Hubbard interactions at low temperatures, one can always expect an exponentially small quasi-particle weight.

The Hamiltonian is quadratic in the conduction elections so we can integrate them out, leading to an action with a dynamic term, the hybridization ($\Delta(\omega)$), for the local Hamiltonian. As the details of the band play a minor role in the physics of the SIAM, we consider only the hybridization of an infinite band, for which $\Delta(i\omega)=-i\pi|V|^2sgn(\omega)$. The SIAM has been investigated thoroughly via the Numerical Renormalization Group (NRG) pioneered by Wilson for the Kondo model\cite{krishna1980renormalization}. Besides the NRG and various other computational methods such a Monte-Carlo and exact diagonalization (ED) methods, the Bethe Ansatz has also been used to obtain closed-form expressions for various thermodynamic properties. Previous fRG studies of the SIAM have produced quantitative results up to intermediate coupling, and captured static thermodynamic variables at strong coupling\cite{karrasch2008finite,bartosch2009functional}.
\begin{figure}
\centering
\includegraphics[scale=0.57]{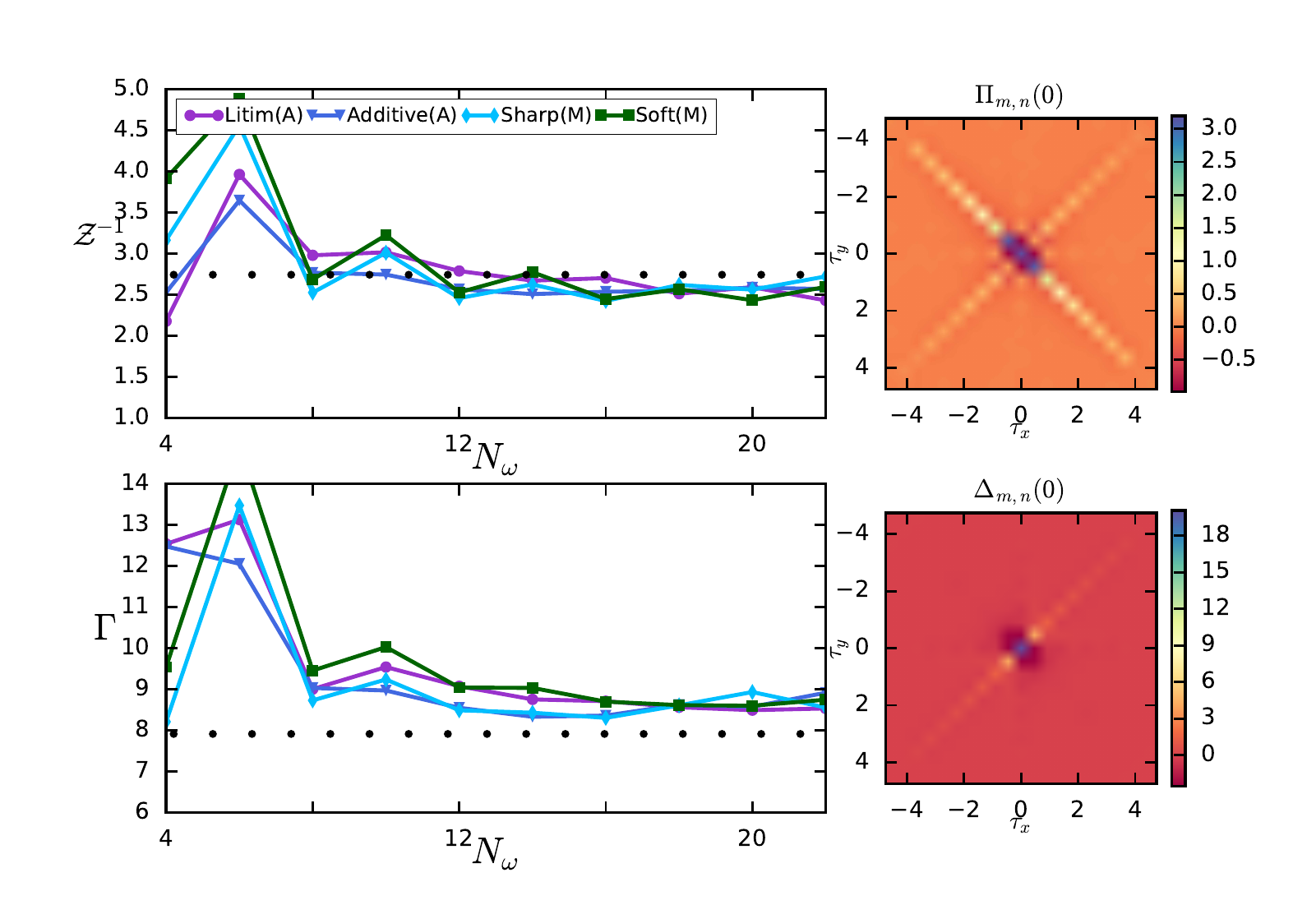}
\caption{Convergence of the quasi-particle weight and vertex as function of the number of clusters for the decoupled two-loop fRG equations at $U/\Delta$=5. (Right): The zero frequency particle-particle and particle-hole vertices as a function of $\tau$ obtained via the Litim regulator.}
\label{uNWsinc}
\end{figure}

In terms of computational demands the fRG equations for the SIAM are relatively modest, as the model is in zero spatial dimensions and thus only has a frequency dependence. We apply the two-loop fRG with the four regulators to calculate the self-energy and the vertex at various temperatures. Calculations at one-loop along with comparisons with the full fRG are detailed in Appendix.\ref{exVert}. Since our primary purpose in performing the SIAM calculations are to benchmark the performance of the decoupled fRG, we will compare to the Bethe Ansatz results for the spin and charge susceptibilities to look at the accuracy of the method at the single particle and two-particle levels. The Ward identities for the spin and charge susceptibilities are given by
\begin{align}
\chi_c+\chi_s = &\quad2\rho(0)\left(1-\partial_{i\omega}\Sigma(\omega)\biggl|_{\omega=0}\right)\\
\chi_c-\chi_s = &\quad2\rho(0)^2\Gamma^{(4)}(0,0,0)
\end{align}
where $\rho$ is the density of states. As discussed above, we parameterize the frequency dependence of the vertex via a logarithmic set of $N_f$ frequencies. For the self-energy, we retain the full set of frequencies obtained via the continued fraction expansion of the Fermi function.
\begin{figure}
\centering
\includegraphics[scale=0.57]{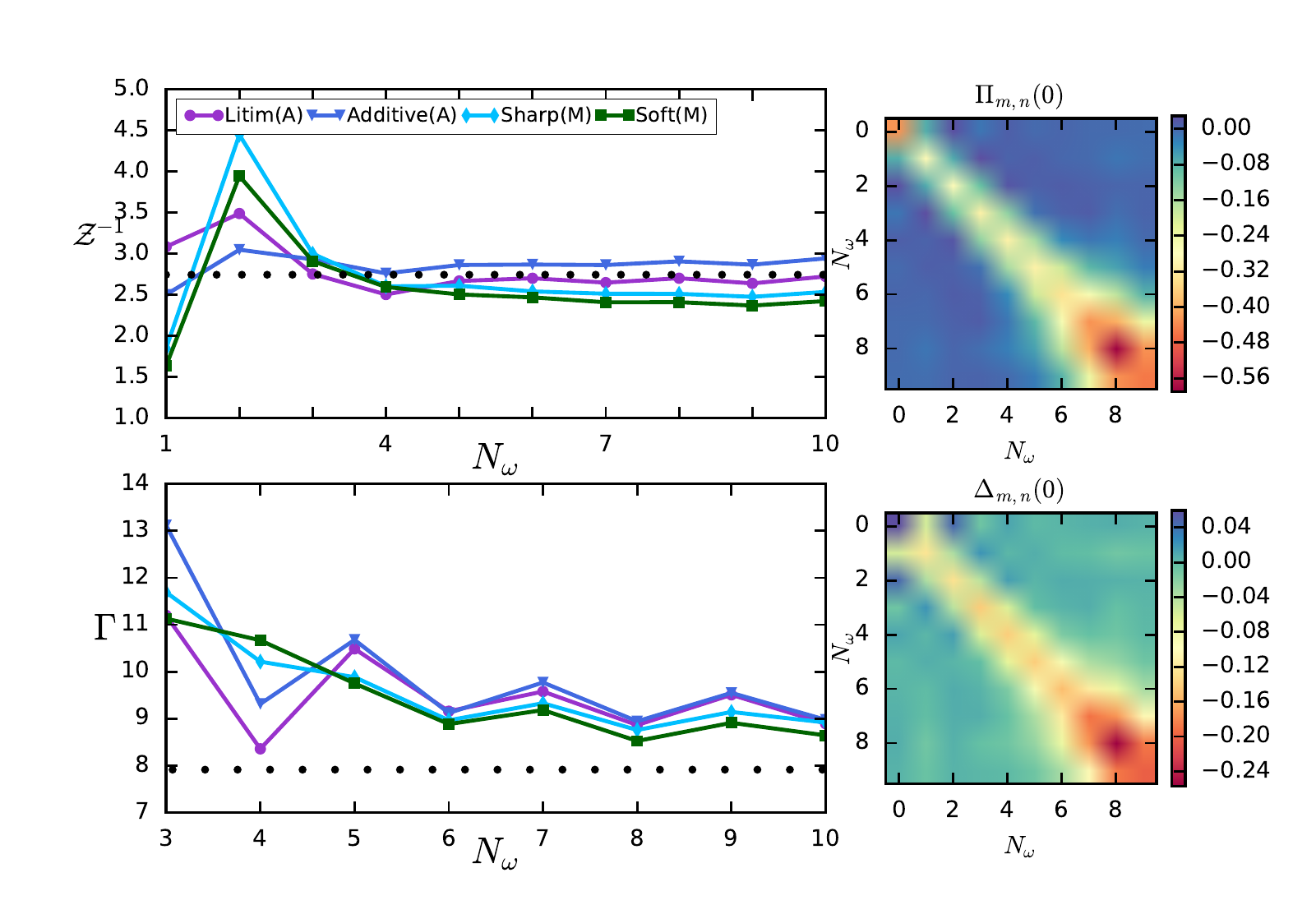}
\caption{Convergence of the quasi-particle weight and vertex as function of Legendere basis functions for the decoupled two-loop fRG equations via the four regulators at $U/\Delta$=5. (Right): The zero frequency particle-particle and particle-hole vertices obtained via the Litim regulator.}
\label{uNWleg}
\end{figure}

The computations relevant to the SIAM can be parametrized by the number of Matsubara frequencies ($N_f$), the number of basis functions ($N_w$) and the temperature of the model ($\beta$). For every scale, we start our fRG calculations by constructing our three vertices (Eq.\ref{threeExpansions}). This involves projection between channels, which scales linearly with $N_f$ (Legendere$\sim \mathcal{O}(N_fN_w^4N_{wm})$, Cluster$\sim\mathcal{O}(N_fN_w^3N_{wm})$). The final term $N_{wm}$ corresponds to the number of basis functions used for a full expansion of the singular channel. The projection of the vertices can be recast as a line integral which is the primary reason for the difference in scaling between our two basis sets. The Legendre expansion requires the full set of basis functions ($N_w^2$) to capture the line integral while for the cluster expansion we only need the patches in the path of the line ($\sim N_\omega$).  The next step in the fRG is the evaluation of the self-energy (Eq.\ref{selfEnergy}) which requires an expansion of the vertex in the three channels. The evaluation of the vertex scales linearly with the number of frequencies, so for the self-energy we have the scaling $\mathcal{O}(N_{fm}N_{fm}N_w^2)$. We note that the frequencies for the self-energy are the continued fraction Fermionic Matusbara frequencies. We then calculate the exchange propagators for the particle-particle and particle-hole channels ($\mathcal{O}(N_fN_w^2N_{fm})$). For the one-loop fRG the final step is to evaluate the derivatives in the three channels (Eq.\ref{channelDerv}). The evaluation requires $N_f$ matrix multiplications with $N_w$ basis functions leads to the scaling $\mathcal{O}(N_fN_w^3)$. A truncation at the two-loop level requires that we further calculate the full exchange propagators (which scale identically with the at-scale exchange propagators, Eq.\ref{exchangeP}) and project the three channel derivatives into the other channels. The memory costs of these fRG calculations for the SIAM are negligible, but we should note that most of the memory consumption occurs in the projection step. Thus, as we scale up for higher-dimensional models, we need to fully expand our vertices in a manageable basis set before projection.
\begin{figure}
\centering
\includegraphics[scale=0.58]{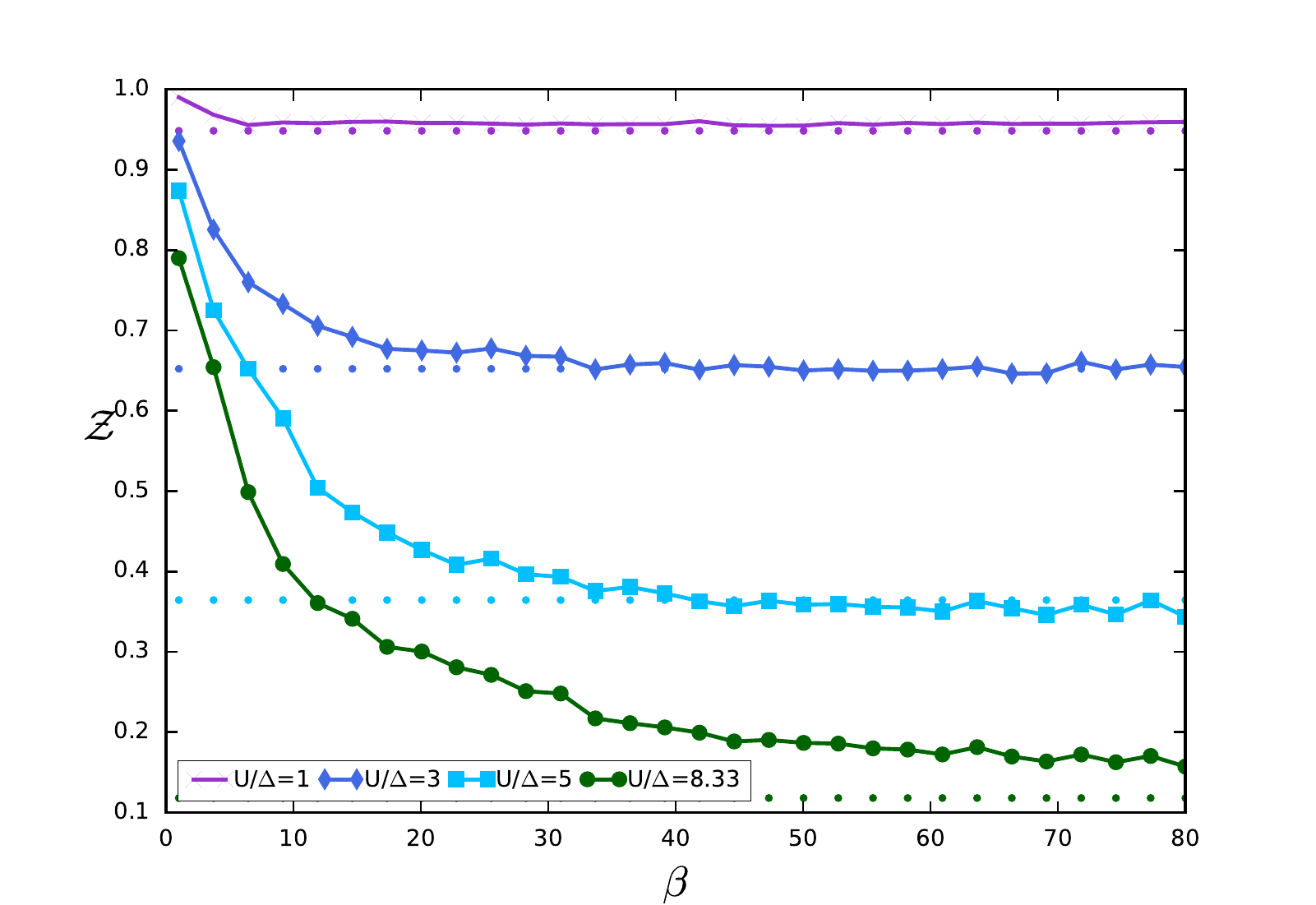}
\caption{The quasi-particle weight as a function of temperature for weak to strong coupling. Calculations where done via the cluster expansion with the Litim regulator for $N_f=30$,$N_\omega=20$. The zero temperature Bethe Ansatz results (dotted lines) are also shown.}
\label{tdepQWsinc}
\end{figure}

In addition to the discretization parameters discussed above for the cluster expansion we have a time step $\Delta\tau=\beta/N$ which for the SIAM we set in accordance with the hybridization to be $\Delta\tau \geq 0.25\Delta$ for all temperatures considered. This is the simplest choice given our noninteracting Hamiltonian but further refinements are possible as the physics of the SIAM primarily revolves around the interplay between $U$ and $\Delta$ ($\Delta\tau=f(U,\Delta)$). We follow this prescription for the general models studied in this work with the choice of $\Delta\tau$ dependent only on the noninteracting Hamiltonians. Further enhancements are specific to the model with the SIAM having a purely imaginary propagator and a real vertex, which within the Legendre expansion restricts us to an even set of basis functions. We used the first even $N_\omega$ rational Legendre polynomials as the basis set for all SIAM calculations. After choosing $\beta$, $N_f$ and $N_\omega$, we solved the decoupled fRG equations at each scale $\Lambda$ via an adaptive step Runge-Kutta method. We terminate the fRG flow when the scale reaches the lowest energy scale in the problem ($\Lambda=\omega_{min}$). The decoupled fRG equations return a convergent self-energy and vertex for all values of the coupling.
\begin{figure}
\centering
\includegraphics[scale=0.58]{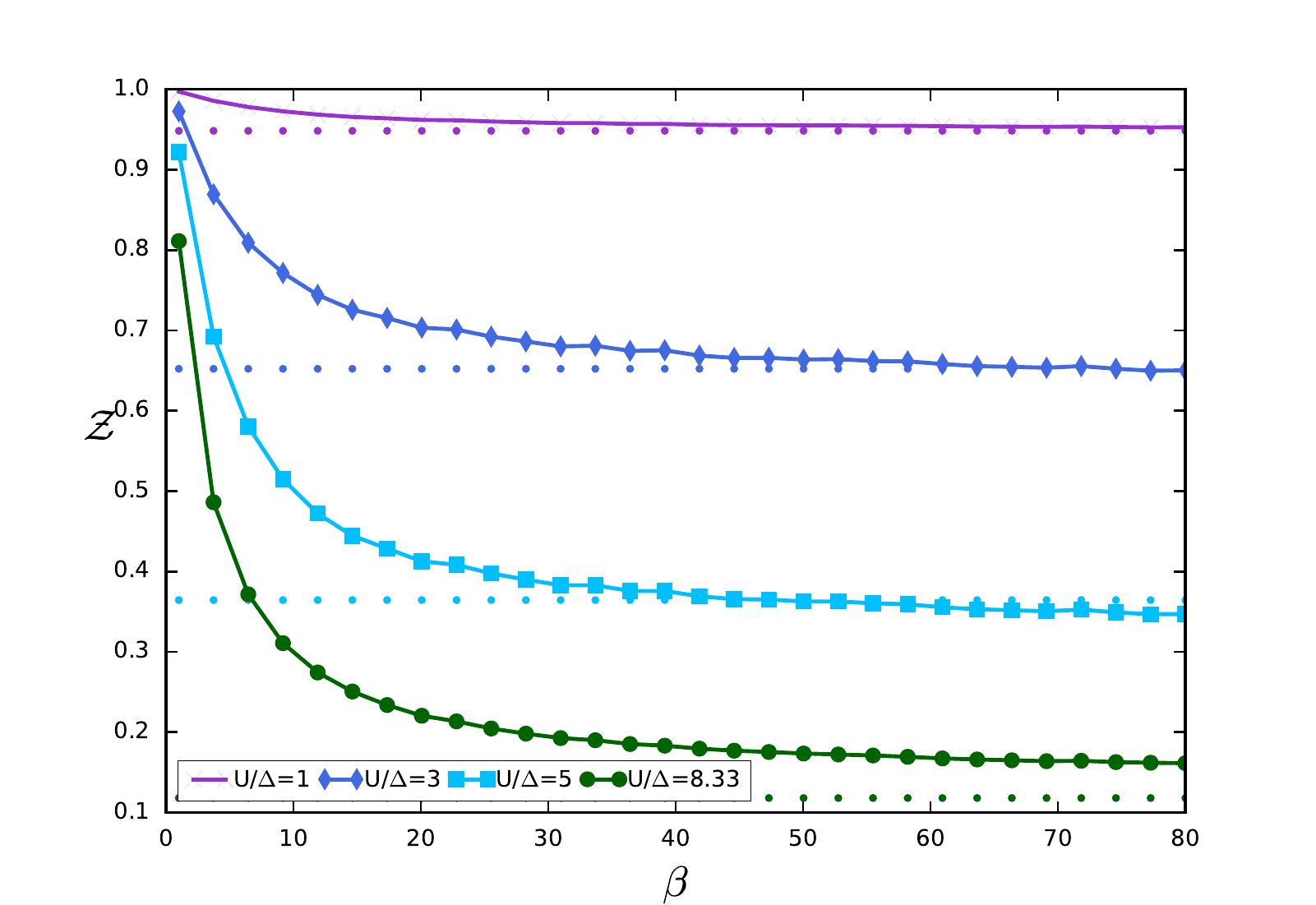}
\caption{The quasi-particle weight as a function of temperature for weak to strong coupling. Calculations where done via the Legendre expansion with the Litim regulator for $N_f=30$,$N_\omega=10$. The zero temperature Bethe Ansatz results (dotted lines) are also shown.}
\label{tdepQWleg}
\end{figure}

Our results from the decoupled fRG flow are presented below. In Fig.\ref{susceptB} we show the results of our calculations for the quasi-particle weight and the coupling at the two-loop level via the Cluster and Legendre expansions at $N_f=30$. The results at the one-loop level from the decoupled fRG are consistent with previous fRG studies\cite{karrasch2008finite}. At the two-loop level we see consistency across all regulators considered for the flow with results for the vertex showing agreement with Bethe Ansatz results up to moderate coupling for both expansions. The error at the vertex level scales as $\mathcal{O}((U/\Delta)^3)$ for the cluster expansion while it shows $\mathcal{O}((U/\Delta)^4)$ for the Legendre basis which is in line with the expected error of a two-loop calculation. The exact source of the discrepancy is unclear, although the restriction of the Legendre expansion to even functions may play a role in providing a faster cover. Despite the differences in scaling, both errors are small up to moderate coupling and are of the same order at strong coupling. Besides errors due to the truncated hierarchy at two-loop level the discretization of the imaginary time axis in the cluster expansion introduces noise which can be seen in the error calculation for the vertex. There are many filters available to tackle this noise but its impact at the two-loop level seems minimal. As we move to higher loop orders proper filtering may be crucial in ensuring results independent of discretization artifacts. The impact of the discretization can be clearly seen in Fig.\ref{uNWsinc} which shows the quasi-particle weight and the vertex as a function of the number of clusters employed in the flow. Convergence is fast with reliable results available at $N_\omega\approx 10$. The figure also show the particle-particle ($\Pi_{mn}$) and particle-hole ($\Delta_{mn}$) expansions of the vertex at $N_\omega=22$ which shows the sparsity of the expansions. We note that a majority of the spectral weight ($\sum_{mn}|\Pi_{m,n}(0)|^2$) is contained in $\sim 2N_\omega$ entries. This allows for further truncations based on spectral weight which is ideal for constructing qualitative flows of high dimensional models that occupy a large parameter space. A similar plot for the Legendre expansion is given in Fig.\ref{uNWleg}.

Our next focus of study is the temperature dependence of the self-energy and the vertex of the SIAM. Generally for models in which a gap opens at a finite temperature, the fRG results diverge with strong signals in the vertex as to the nature of the gap. Though such divergent flows are rich in information they dramatically suppress competing phases, making the quantitative determination of a phase line difficult. We can avoid such divergences by scanning models at temperatures sufficiently higher than the critical scale, ensuring not only convergence but a stable vertex. With this in mind we look at the stability of the decoupled fRG flow in the SIAM for a wide variety of temperatures. The SIAM results for large coupling are well known\cite{hewson1997kondo}. As the temperature is increased, we expect a freezing of the charge fluctuations. This can be seen at the single particle level as a suppression of the quasiparticle weight which is clearly captured by the fRG in Fig.\ref{tdepQWsinc} for the cluster expansion and Fig.\ref{tdepQWleg} for the Legendre expansion. For comparison we have also plotted the zero temperature Bethe Ansatz results for the SIAM. We see a stable vertex at strong coupling both at high and low temperatures which is in part due to the dynamics of the SIAM with the conduction electrons screening the impurity for all couplings.
\begin{figure}
\centering
\includegraphics[scale=0.58]{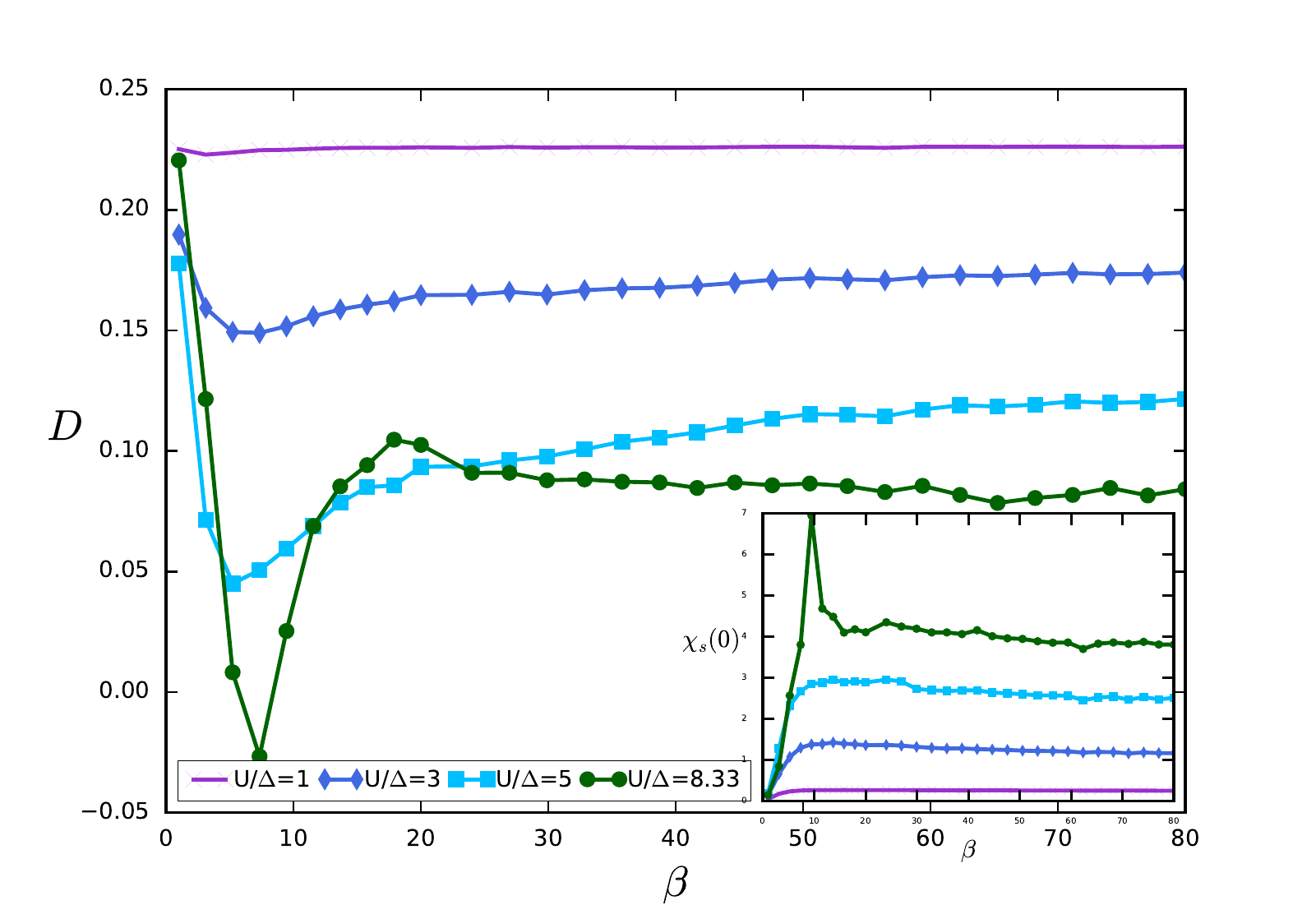}
\caption{The double occupancy, $D=\langle n_\uparrow n_\downarrow\rangle$, as a function of temperature for weak to strong coupling. Calculations where done via the cluster expansion with Litim regulator for $N_f=30$,$N_\omega=20$.}
\label{tdepDO}
\end{figure}

A more challenging aspect of the SIAM is the local moment regime, which exists at high temperatures when enough of the conduction electrons are yet to condense to screen the impurity. As we increase the coupling,  this regime occupies larger and larger temperature ranges with the coupling completely suppressing double occupancy in the SIAM. The double occupancy of the SIAM is given by
\begin{align}
\langle n_\uparrow n_\downarrow\rangle=&\sum_{\omega_1,\omega_2,\omega_{ph}}\mathcal{G}(i\omega_1)\mathcal{G}(-i\omega_{ph}+i\omega_1)f_m(\omega_{ph_x})\nonumber\\
&\Delta_{m,n}(\omega_{ph})f_n(\omega_{ph_y})\mathcal{G}(i\omega_2+\omega_{ph})\mathcal{G}(i\omega_2)
\label{dOccup}
\end{align}
where we have used the particle-hole expansion ($\Delta$) to simplify the expression and $\omega_{ph_{x/y}}$ depends only on $\omega_{1/2}$ and $\omega_{ph}$. Our results for the double occupancy are shown in Fig.\ref{tdepDO}. The decoupled fRG clearly captures the suppression of double occupancy as a function of coupling with the SIAM trapped in the local moment regime for longer temperatures at higher couplings. The inset of the figure shows the spin susceptibility of the SIAM for the same values of the coupling. In the local moment regime we see the expected linear growth in the spin susceptibility with the susceptibility leveling off as the spin is screened. Unfortunately the results for $U=8.33\Delta$ show a violation with the occupancy dipping below zero. Violations of Ward identities by the fRG are expected\cite{katanin2004quasiparticle} with this error persisting despite changes in $\Delta\tau$ and $N_\omega$. This indicates a need to account for the full hierarchy of the multi-loop fRG in order to tackle the full range of couplings and temperatures of the SIAM.

\section{Model Hamiltonians in Higher Spatial Dimensions}
 \label{1DModels}
Interacting electron systems in one and two dimensions are rich in phases with various instabilities leading to a variety of ordered phases. Functional renormalization methods have been instrumental as an unbiased tool for evaluating the competition among these orders. Decoupling of the momentum modes has been accomplished for static fRG with the SM-fRG and TU-fRG\cite{lichtenstein2017high}, with multi-loop flows that account for the frequency dependence of vertex implemented in conjunction with the TU-fRG for the Hubbard model\cite{hille2020quantitative,tagliavini2019multiloop}. As the TU-fRG has been rigorously tested,  we adopt it to decouple the momentum modes in conjunction with the frequency decoupling scheme described above. We implement the TU-fRG for two scenarios with the first being a direct expansion of the two-particle vertex for a finite size system. The expansion requires solving the decoupled fRG equations for an $N$-site system. The expansion is tractable, although computational resources which are split between the frequency and momentum modes limit the size of the systems we can treat effectively to $N\leq 256$. This can be problematic in evaluating Kosterlitz-Thouless (KT) type transitions and two dimensional systems where such a system size corresponds to a small lattice ($N\leq 16\times 16$), well below the thermodynamic limit. To overcome this limitation to a finite size lattice, we employ a patching scheme cutting the BZ into $N$ clusters and averaging over them. The computational cost of both methods is similar, requiring the solution of an $N$-site system for the fRG and an $N$-cluster system for the cluster-averaged variant. For two-dimensional systems, the cluster averaged fRG appears ideal for studying thermodynamic properties within our computational limits.

In many test cases, we found that the DCA-fRG approach leads to a converged self-energy and vertex at moderate coupling, even in cases where the direct fRG expansion may diverge or lead to unphysical vertices. We note that coarse-graining over the BZ may dilute the role of Van Hove points which are crucial in driving density and superconducting orders. One can get around this issue via a judicious choice of patch geometry\cite{staar2016interlaced} but such specificity is counter to a generalized solver. Instead we consider both approaches to check the consistency of our results. For our ultimate aim of treating multi-band Hamiltonians, DCA-fRG offers the ideal compromise in that the cluster averaging in the DCA feeds the full BZ into our flow and allows for a controlled study of how it affects local order parameters. A review of DCA methods can be found in Refs. \onlinecite{jarrell2008dynamical,maier2005quantum}.

Standard approaches for solving the cluster-averaged DCA Hamiltonian involve mapping the system to an effective cluster embedded in a bath and solving the auxiliary system in a fashion consistent with the lattice. The fRG allows us to eschew the self-consistent formulation, which can be computationally expensive for large clusters, by coarse-graining our vertices along the flow. The approximation at the heart of DCA methods involves the relaxation of momentum conservation at the vertices to cluster modes only. So in the patched BZ of the larger lattice ($N>N_c$) the $N$ clusters each with an average momentum $\vec{K}$. Lattice momentum is conserved in the interaction of these modes through a vertex and this conservation can be expressed via the Laue function. The Laue function is given by
\begin{equation}
\Delta(k_1,k_2,k_3,k_4)=\sum_r\exp(i\textbf{r}(k_1+k_2-k_3-k_4))
\end{equation}
where r labels the sites. For the full BZ the function above is a delta function ($\Delta=\delta(k_1+k_2-k_3-k_4)$) but we relax this constraint and enforce momentum conservation between the $N$ clusters only ($\Delta\approx \delta(K_1+K_2-K_3-K_4)$). Within the fRG we can institute a similar cluster averaging by softening the Laue function present in the irreducible vertices. As the interacting vertices are constructed along the flow, the process of softening the Laue function simply reduces the momentum modes of the vertex to the cluster modes. On the other hand, the new Laue function modifies the propagators, as we now have to sum over the momentum modes within a cluster along the flow. The averaged single scale propagator is given by
\begin{equation}
 \bar{\mathcal{S}}(\omega,\vec{K})=\frac{N_c}{N}\sum_k \mathcal{S}(\omega,\vec{K}+k,\Sigma)
\end{equation}
where $\Sigma$ corresponds to the full self-energy calculated from the coarse-grained vertex and $N_c$ is the number of clusters. This coarse-graining step is performed on the single scale and the full propagator at all scales in the fRG flow. The coarse-grained propagators are then used to calculate the (full) exchange propagators (Eq.\ref{exchangeP}) over the clusters. The final modification is a change of variables in the equation for the self-energy (Eq.\ref{selfEnergy}) so we can evaluate it over the entire lattice. We have
\begin{align}
\partial_\Lambda\Sigma_p=\sum_x\partial_\Lambda G_{x-p}\Gamma_{p,x-p,x-p,p}^{(4),\Lambda}
\end{align}
with $\Gamma^{(4)}$ projected to the nearest cluster modes. The pure frequency regulator significantly reduces the cost of cluster averaging, as  momentum dependence of the propagator is unaffected by the regulator, but averaging over systems with sharp momentum structures at each point of the flow can quickly become computationally demanding, especially in two dimensions.

\begin{figure}
\centering
\hspace*{-0.5cm}
\includegraphics[scale=0.43]{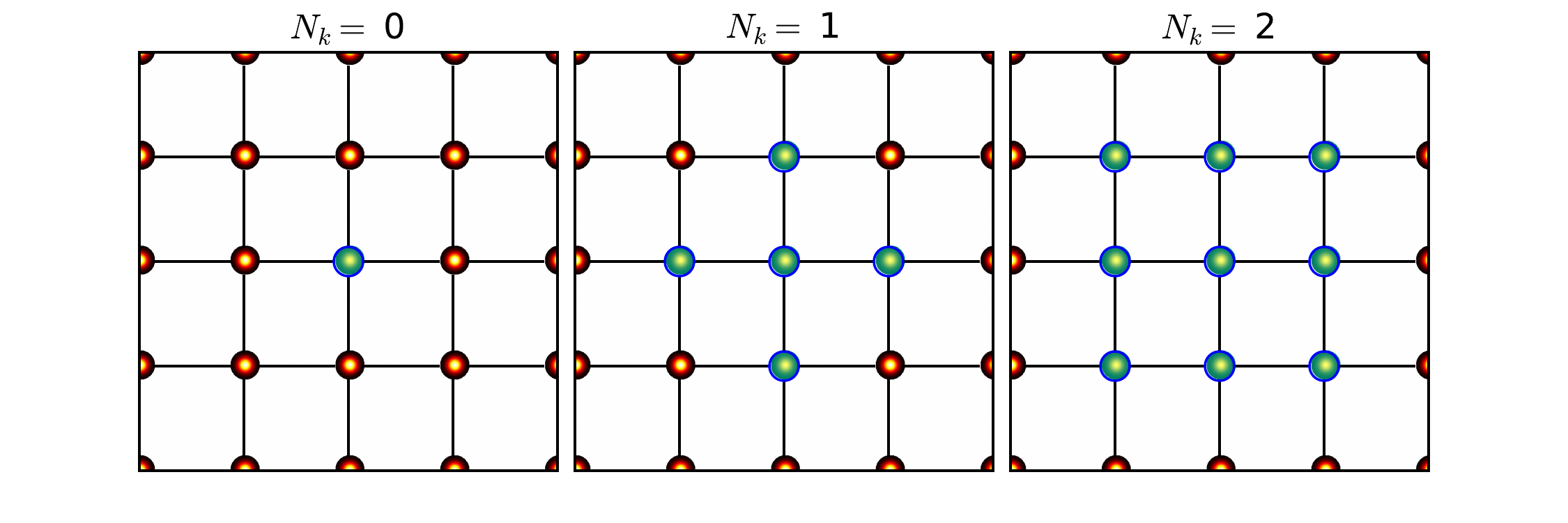}
\caption{The corresponding lattice sites (green) for different choices of $N_K$ for the two dimensional square lattice.}
\label{latNK}
\end{figure}

For either method, we need to solve the fRG equations for $N^3$ momenta modes of the vertex. As the momentum structure of the equations is identical to the structure of the frequency axis, we expand the two-particle vertex around the singular channels ($s_{pp}\rightarrow k_{pp}$). The finiteness of the BZ removes any need for separating asymptotic terms, and we expand the vertex via a basis of complex exponentials. We adopt the parametrization of the auxiliary momentum variables introduced in the TU-fRG\cite{eckhardt2018truncated}. The full particle-particle expansion of the vertex is given by
\begin{align}
\Gamma_{p_1p_2p_3p_4}&=\sum_{m,n,i,j}\Pi_{m\times i,n\times j}(\omega_{pp},k_{pp})\times\nonumber\\
&f_m(\omega_{pp_x})f_n(\omega_{pp_y})g_i(k_1)g_j(k_4)
\end{align}
where m,n correspond to the frequency basis ($N_\omega$) and i,j correspond to the momentum basis ($N_k$), with similar expansions in the other two channels. These representations scale linearly with the frequency and momentum modes ($\mathcal{O}(N_fNN_\omega^2N_k^2)$) so for an efficient choice of basis functions we can handle relatively large systems (N$\leq$256). The $N_k$ basis sets approximately represent the space over which the incoming or outgoing objects that interact through the two-particle vertex are formed. A selection of three $N_k$ values for the square lattice is shown in Fig.\ref{latNK}.

Implementing the full vertex expansion reduces the fRG equations at each scale to a series of matrix multiplications for each singular frequency($N_f$) and momentum modes($N$). For each singular mode we calculate $[F]_{N_t\times N_t}[L]_{N_t\times N_t}[G]_{N_t\times N_t}$, with $N_t$ equal to the product of the frequency and momentum basis sets ($N_t=N_\omega N_K$), $L$ being the exchange propagator and $F$,$G$ representing the three expansions of the vertex or in the two-loop case their scale derivative. For large frequency and momentum basis sets the computational cost of these calculations can become quite high, so for qualitative surveys of models we can utilize the sparsity of the vertex to retain just $N_r$ of the relevant entries in $F$,$G$. A more detailed look at truncations of the vertex expansion based on spectral weight for the extended Hubbard model is given below. To complete the flow we need to calculate the exchange propagators ($\mathcal{O}(4N_\omega N_f^2)\times\mathcal{O}(N_K^2N\ln N)$) and projections of the expansions between the channels ($\mathcal{O}(N_\omega^2N_fN_L)\times\mathcal{O}(N_K^2N\ln N)$). The flow iterates these calculations through the scales until all modes are integrated out. For the general case the flow cannot be full integrated, with instabilities driving the electrons towards various kinds of orderings. Depending on the coupling and temperature, these instabilities lead to divergences in the two-particle vertex and the susceptibility corresponding to the ordering in the system. Thus, we can construct the charge, spin and superconducting susceptibilities and in combination with high temperature sweeps capture the ground state phase diagram of any system of interest.

Access to the frequency content of the vertex allows us to compute the full correlator with the susceptibilities for the ordered phases corresponding to the long time limit ($\tau\rightarrow\infty$). The correlator for an arbitrary order ($\mathcal{O}_i(\tau)$) requires the evaluation of the quantity
\begin{align}
\chi_{\mathcal{O}}(\tau,\vec{R})=\sum_i\int_0^\tau dt \langle\mathcal{O}_{i+\vec{R}}^\dagger(\tau+t)\mathcal{O}_i(t)\rangle
\end{align}
which can be Fourier transformed to the frequency and momentum domain. For general charge and spin density orders with a density profile characterized by $f_{\mathcal{O}}$ the expression above simplifies to
\begin{align}
\chi_{c/s}(\Omega,\vec{q})&=\sum_{p_1,p_2,\sigma_1,\sigma_2}\langle s_{\sigma_1}f_{\mathcal{O}}(p_1)c_{p_1,\sigma_1}^\dagger c_{p_1+p_q,\sigma_1}\times\nonumber\\
&s_{\sigma_2}f_{\mathcal{O}}(p_2)c_{p_2,\sigma_2}^\dagger c_{p_2-p_q,\sigma_2} \rangle_c
\end{align}
where $p_q$ corresponds to the nesting vector ($p_q=(\Omega,\vec{q})$) and $s_{\sigma}$ is 1 for the charge wave and $\sigma$ for spin waves. As noted previously\cite{PhysRevB.85.035414}, the expansions of the vertex dramatically simplify the calculation of the susceptibilities, as the nesting vector of a susceptibility corresponds to a singular frequency in one of the channels. Expanding the vertex in the appropriate channel, the SU(2) symmetric charge and spin susceptibilities become
\begin{align}
\chi_c(\vec{q},\Omega)&=\textbf{S}(\vec{q},\Omega)(2\Delta(\vec{q},\Omega)-X(\vec{q},\Omega))\textbf{S}^T(\vec{q},\Omega)\nonumber\\
\chi_s(\vec{q},\Omega)&=-\textbf{S}(\vec{q},\Omega)X(\vec{q},\Omega)\textbf{S}^T(\vec{q},\Omega)
\end{align}
where $\textbf{S}$ ($[1\times N_\omega N_k]$) represents a sum over the incoming/outgoing particles for each frequency and momentum basis. For density susceptibilities, this term is given by
\begin{align}
S_{0,(m,i)}(\vec{q},\Omega)=\sum_{\omega,k}f_{\mathcal{O}}(\vec{k})\mathcal{G}(\omega,\vec{k})\mathcal{G}(\omega+\Omega,\vec{k}+\vec{q})\times\nonumber\\
f_m(\omega_{ph_x})g_i(k_1)
\end{align}
with a similar expression for the superconducting order parameters.

For a stable convergent vertex, the susceptibilities given above are finite and unambiguous in the determination of the ordering in the system. However with the fRG instabilities that break the symmetries of the system and generate a gap lead to divergences in the vertex. With an adaptive step solver we can always stop the flow before any of our 1PI vertices diverge, but the stopping scale ($\Lambda_s$) can take on a wide range of values for different parameters of the system. This complicates the determination of phase boundaries, as we would be comparing vertices at different scales. One approach around this problem is to conduct temperature sweeps, which for the right temperature ensures a uniform stopping scale. However, even in the single band case this is an expensive solution. Alternatively we can use the Bethe-Salpeter integral equation in the diverging channel to compute an initial vertex at $\Lambda=0$ from the vertex at the scale, $\Lambda_s$. Previous works have used a similar approach for the static fRG to construct an accurate mean field Hamiltonian and to estimate the gap. Utilizing our frequency and momentum basis sets the equation in the particle-particle channel is given by
\begin{align}
\boldsymbol{\Pi}^{\Lambda}(\omega_{pp},\vec{k}_{pp})=&\boldsymbol{\Pi}^{\Lambda=0}(\omega_{pp},\vec{k}_{pp}) -\boldsymbol{\Pi}^{\Lambda=0}(\omega_{pp},\vec{k}_{pp})\times\nonumber\\
& \boldsymbol{L}_F^{\Lambda,pp}(\omega_{pp},\vec{k}_{pp})\boldsymbol{\Pi}^\Lambda(\omega_{pp},\vec{k}_{pp})
\end{align}
which can be inverted to calculate the initial vertex
\begin{align}
\boldsymbol{\Pi}^{\Lambda=0}(\omega_{pp},\vec{k}_{pp})=\boldsymbol{\Pi}^\Lambda(\omega_{pp},\vec{k}_{pp})\left(1-\boldsymbol{L}_F^{\Lambda,pp}\boldsymbol{\Pi}^\Lambda\right)^{-1}
\end{align}
where $\boldsymbol{L}_F$ is the full exchange propagator. For fRG flows restricted to a single channel, the initial vertex derived from this expression is identical to the starting vertex, but as the decoupled flow treats all channels equally the calculated vertex will contain fluctuations from all channels. Similar vertices can be constructed for the three channels. This allows us to construct comparable susceptibilities from vertices diverging at a wide range of scales. We note that for some parameters in the extended Hubbard model the cluster expansion shows sensitivity to the scale at which the flow is terminated ($\Lambda_c$). With generality in mind we do not set the termination point and only stop the flow when the adaptive step size becomes unacceptably small ($|\Lambda_l-\Lambda_{l+\delta}|<<t$). Further refinements to solver are possible, although a simpler approach is to use a vertex at a scale slightly below $\Lambda_c$. With all these pieces in hand, we benchmark the performance and accuracy of our momentum discretization, by applying the DCA-fRG and fRG to the one-dimensional extended Hubbard model.

\subsection{The One-Dimensional Extended Hubbard Model}
\label{EHModel}
\begin{figure}
\centering
\includegraphics[scale=0.58]{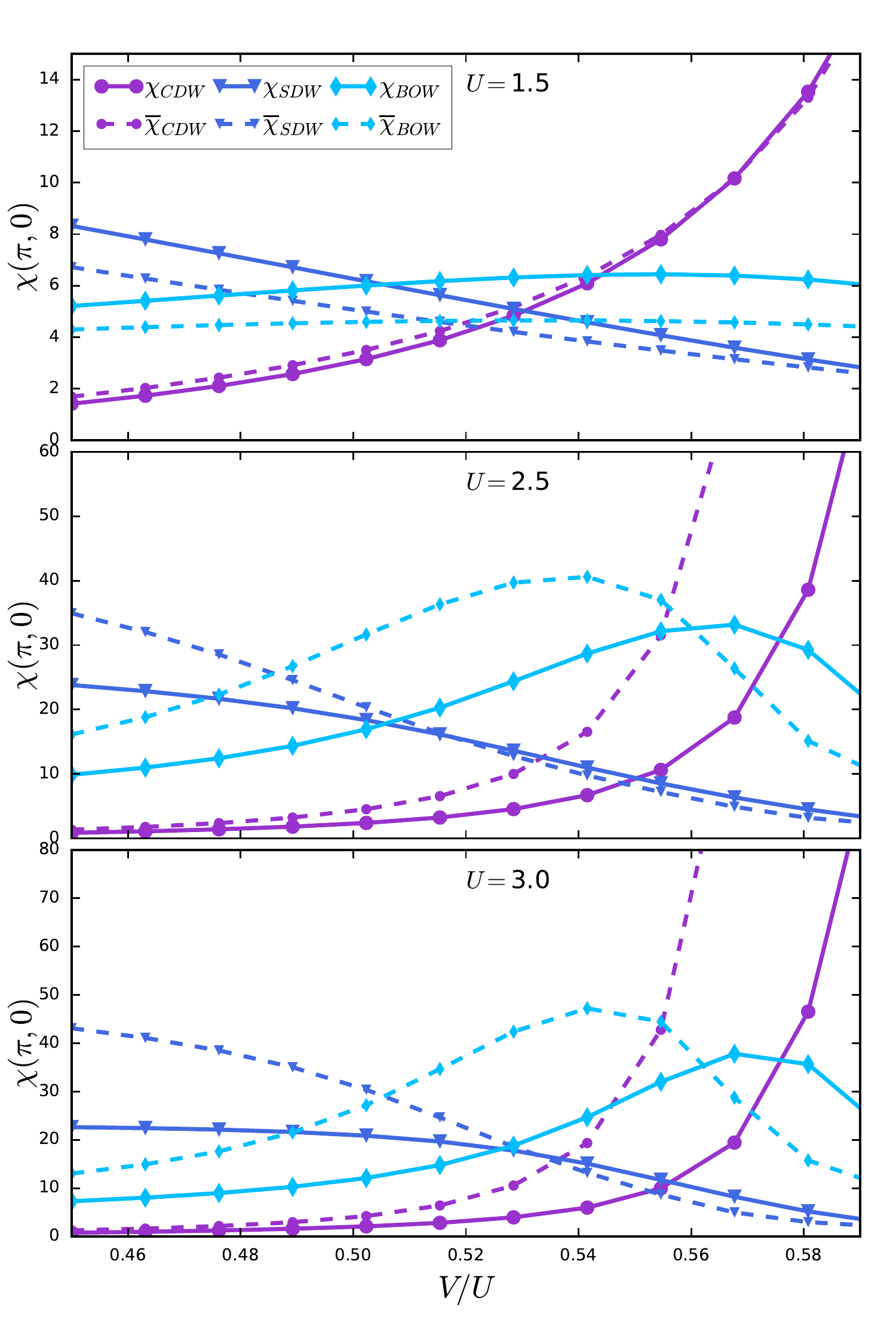}
\caption{The bond-charge ($\chi_c(\pi)$) and spin ($\chi_s(\pi)$) susceptibilities for the 1D EHM via the 2-loop decoupled fRG ($\chi_x$), in solid lines, and the decoupled DCA-fRG ($\overline{\chi}_x$), dashed lines, for various values of the Hubbard coupling. The calculations were performed at $\beta=50$ for $N_k=4$, $N_\omega=4$ for a 64 site (cluster) system.}
\label{1DSus}
\end{figure}
The Hubbard model and its generalizations have long been the prototypes for the study of correlation effects in Fermionic systems\cite{bourbonnais1991renormalization,nakamura2000tricritical}. These models show strong deviations from Fermi-liquid theory and can have a wide variety of interactions beyond the on-site Hubbard $U$,  including  nearest neighbor density-density interactions (Extended Hubbard), correlated hoping (bond-charge) and exchange interactions\cite{campbell1990modeling}. The level to which the Coulomb interaction is screened differs dramatically among different materials, which leads us to a large family of Hubbard models. In low dimensions these models are rich in phases with strong fluctuations driving a varied phase diagram. The first extension that is relevant to materials of interest is the nearest neighbor density interaction. In many materials the value of this interaction as derived from the full Coulomb potential is repulsive and weaker than the local, on-site interaction. Including this nearest-neighbor interaction leads the so-called extended Hubbard models, which in one-dimension is described by the Hamiltonian
\begin{align}
\mathcal{H}&=-\sum_{\langle i,j\rangle}t_{i,j}(c_{i,\sigma}^\dagger c_{j,\sigma} +h.c.) + \mu\sum_{i\sigma}n_{i,\sigma}\nonumber\\
&U\sum_in_{i\uparrow}n_{i\downarrow} + V\sum_{\langle i,j\rangle}n_in_j
\end{align}
where the term $V$ represents the nearest-neighbor density interaction. The phase diagram of the 1D EHM, particularly at half-filling, has been studied extensively via exact diagonalization\cite{nakamura1999mechanism}, Monte-carlo\cite{sengupta2002bond}, bosonization\cite{tsuchiizu2002phase} and the fRG\cite{tam2006functional}. The model has a rich phase diagram owing to the competition between the nearest-neighbor interaction $V$ and the on-site interaction $U$. For positive values of the couplings a charge density wave (CDW) forms for large values of $V  \gg U/2$, while competing spin fluctuations (SDW) dominate parts of the diagram where the local spin interaction is strong, $U \gg 2V$. In weak to strong coupling there exists a novel bond order wave (BOW) phase in a small area around the phase boundary ($U=2V$) between the CDW and SDW  phases. The BOW phase is characterized by the charge in the system condensing on the bonds and strong spin fluctuations. The BOW phase persists up to strong coupling where it terminates at a critical point end point\cite{sandvik2004ground}. The transition between the SDW and BOW phases is continuous up to the critical endpoint but the transition in the charge sector between the BOW and the CDW phase changes to a first-order type at the tricritical point\cite{sandvik2004ground,ejima2007phase}. This model presents the perfect test ground for comparing the effectiveness of finite-size fRG in comparison with the cluster averaged formulation. We restrict all calculations for the EHM to half filling ($\mu=0$).

We initialize our fRG flow for the EHM at a high energy scale ($\Lambda>>W$) far above the bandwidth of our system. We then redo the flow at a series of temperatures looking for a divergence in the spin and charge susceptibilities. In terms of numerics, we are able to reach low temperatures ($\beta=60$) by retaining a logarithmic subset of the Matsubara frequencies ($N_f=30-60$). The momentum basis for all calculations is chosen in a symmetric fashion, which leads to $2\times N_K+1$ basis functions at order $N_k$. $N_k$ is the distance from a local site with a one dimensional chain of Fig.\ref{latNK} as reference. The choice of $N_k$ limits the density profiles ($f_\mathcal{O}$) we can study with $N_k$=0 allowing only s-wave ordering. Without truncations we are able to reach the sixth neighboring site ($N_k=6$) to check the stability of phase diagram. The calculation of the self-energy is simplified by expanding the singular channels of the vertex expansions in an appropriate basis set which allows us to compute the momentum piece of the expression at the start of the flow. The frequency structure of the single scale propagator is peaked around a particular scale cannot be efficiently expanded by a Fourier basis. Thus, we work in a mixed basis calculating the momentum modes in real space while the frequency dependence is calculated along the Matsubara axis via the continued fraction expansion.
\begin{figure}
\centering
\includegraphics[scale=0.47]{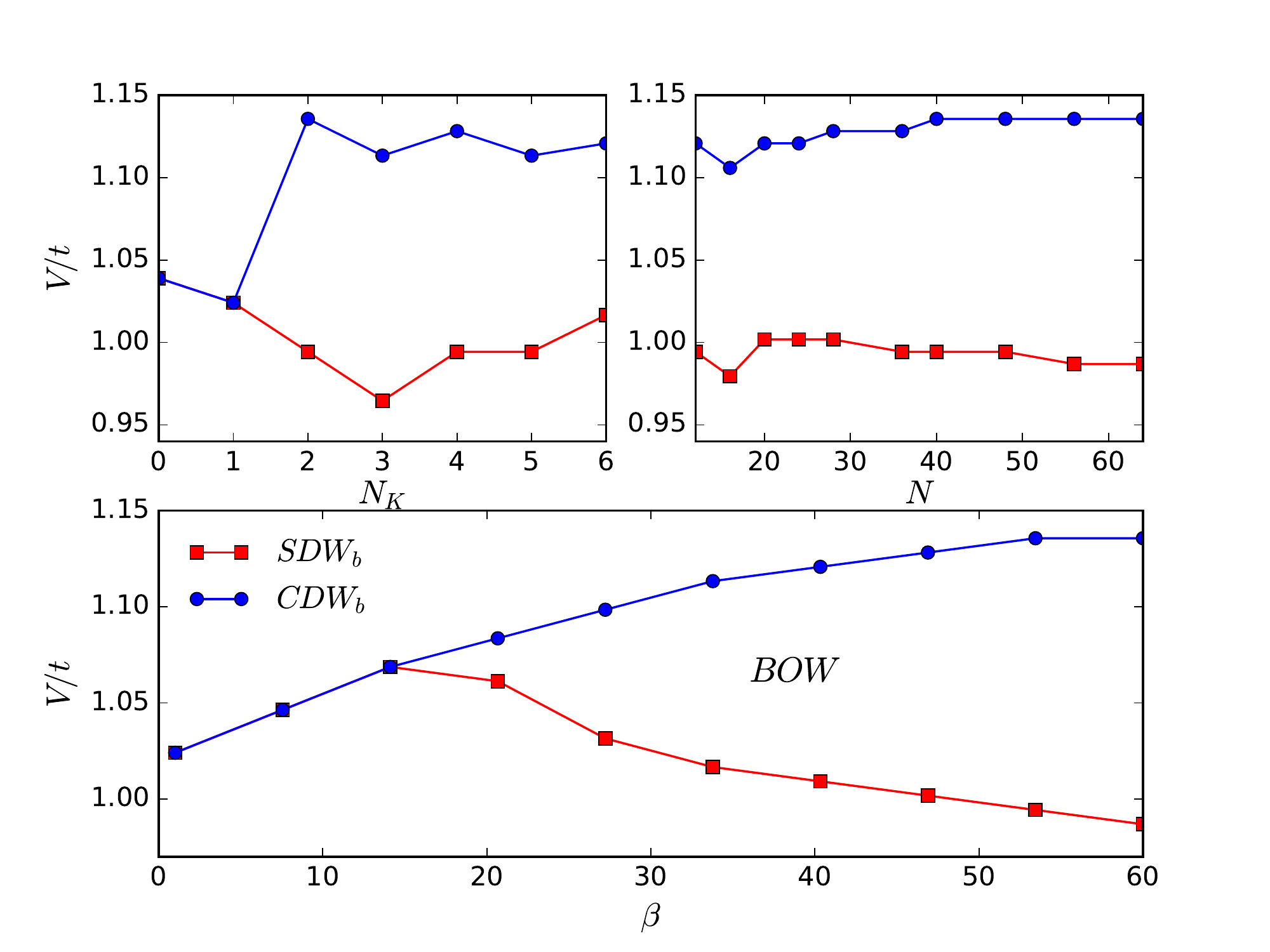}
\caption{The BOW phase at U/t=2.0 as a function of momentum basis (top-left:$N_\omega=4$ $N=32$, $\beta=50$), number of sites (top-right:$N_\omega=4$, $N_k=$,$\beta=50$) and at different temperatures (bottom: $N_\omega=5$, $N_k=4$, $N=32$).}
\label{EHMnn}
\end{figure}

For the repulsive EHM ($U,V>0$), following the equation above, we construct the susceptibility for the charge, spin and bond-ordered phases. We focus our study near the line $U=2V$, which is the mean-field prediction for a transition from a CDW to an SDW. As our expansion has various parameters, for our results to be reliable we need to see how the phase diagram changes as a function of the basis functions, the size of the system.  and the temperature at which the calculations are performed. Fig.\ref{EHMnn} shows the dependence of the results on these parameters. We should note here that the number of momentum basis functions is bounded by the number of sites. Thus to perform a comparison between different system sizes, we fixed our momentum resolution by limiting the number of basis functions ($2N_k+1=N/4$). Despite some variability, all three figures show the emergence and stability of the BOW.
\begin{figure}
\centering
\includegraphics[scale=0.6]{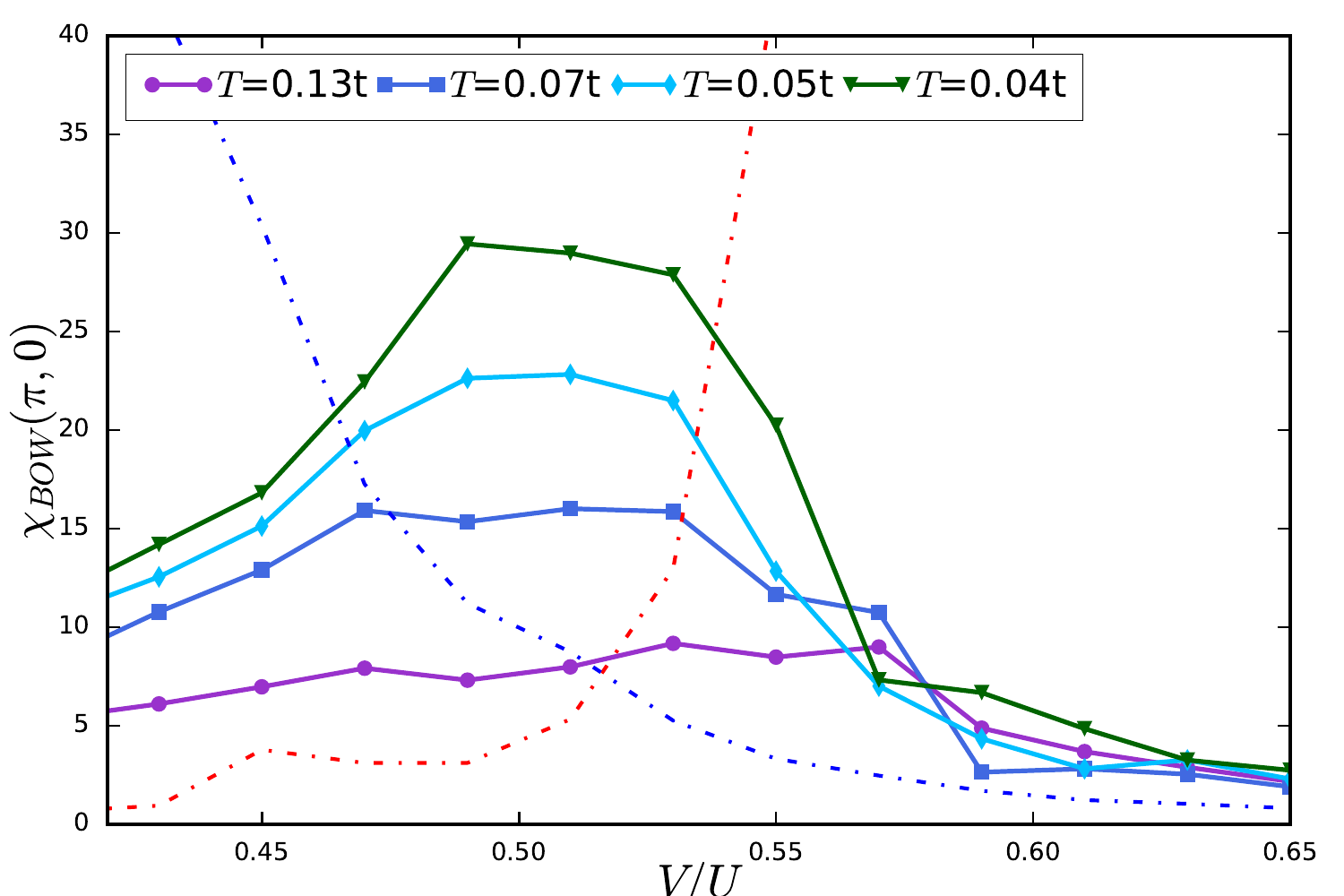}
\caption{The bond-order susceptibility calculated via DCA-fRG at $U/t=4.0$ for $N_\omega=4$, $N_K=5$ and $N=64$ at a range of temperatures. The basis set is truncated with $N_r=256$ of the most relevant functions retained.  The charge (spin) susceptibilities at $T=0.04t$ are plotted in red (blue) for reference.}
\label{tBOWsus}
\end{figure}

\begin{figure}
\centering
\includegraphics[scale=0.47]{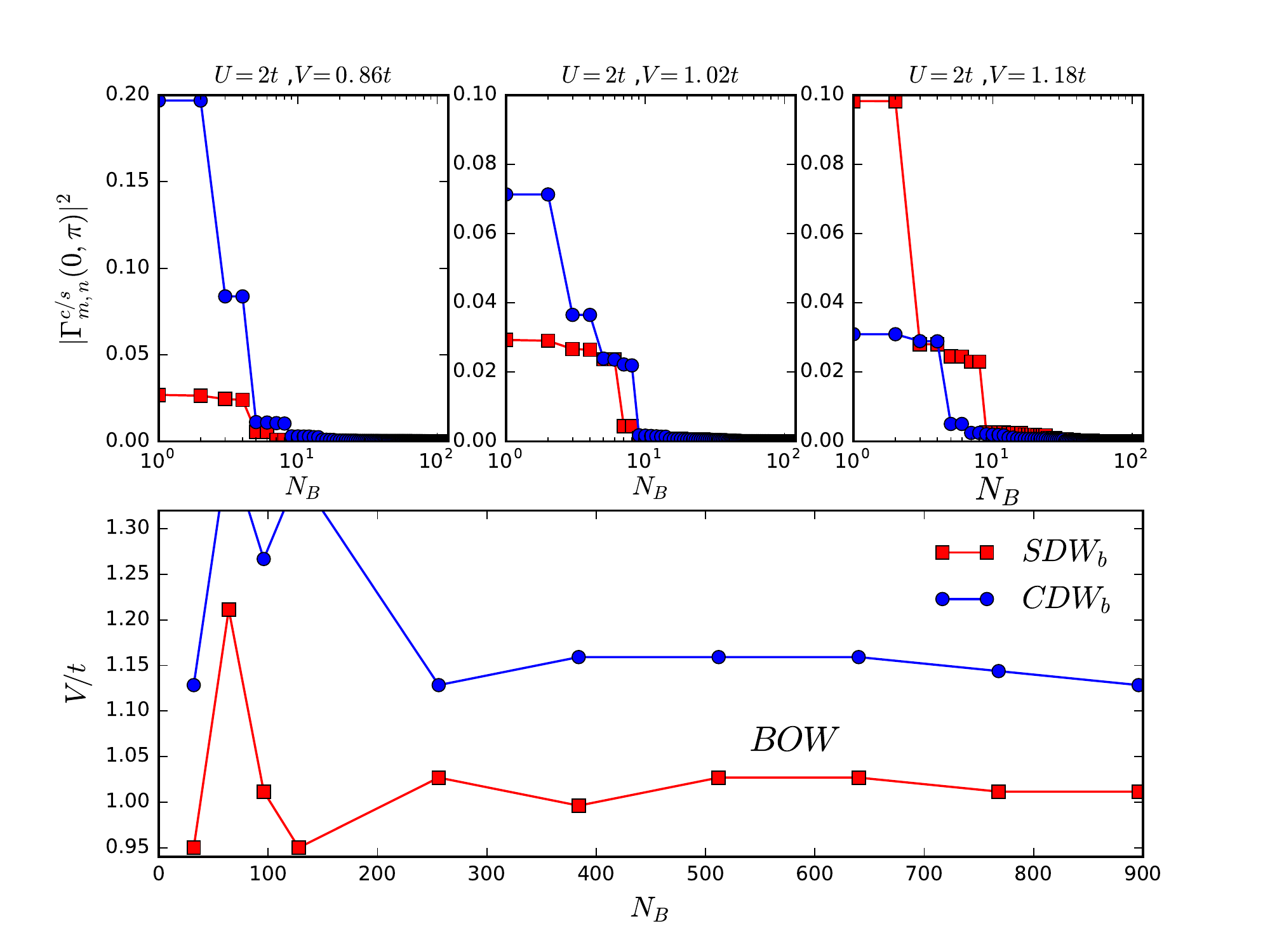}
\caption{The BOW phase at U/t=2.0 as a function of a variable basis set, $N_r$ calculated for $N_\omega=8$, $N_k=5$, $N=64$, at $\beta=50$.The relative spectral weight of the charge and spin vertices in the SDW, BOW and CDW phases is shown.}
\label{EHMnbDep}
\end{figure}

Our access to the frequency and momentum content of the fRG comes at a cost with a typical calculations at $N_\omega=4$, $N_k=4$ ($25\times 81\times N_fN$) requiring $\sim 10^7$ elements for a $64$ site lattice with $\sim 40$ singular frequency modes for each expansion. As we noted above and shown in Fig.\ref{uNWsinc} truncating the vertex based on spectral weight can offer further computational gains. With this in mind we truncate the basis set used for the EHM calculation and run the full flow. The truncation is applied along the flow which limits contributions from the projections of the other channels and those generated in channel to the $N_r$ most relevant entries. In Fig.\ref{tBOWsus} we show results from the from the flow at a series of temperatures with the basis truncated to 256 functions (of a possible 3025,$N_\omega=4$, $N_k=5$) does not affect the emergence of the $BOW$ as a function of temperature at $U=4t$. The impact of truncation for a large basis set ($N_\omega=8,N_k=5$, $9801$ basis sets) for the EHM is shown in Fig.\ref{EHMnbDep}. The relative spectral weights of the charge vertex ($2\Delta_{m,n}(0,\pi)-X_{m,n}(0,\pi)$) and the spin vertex ($-X_{m,n}(0,\pi)$) are also shown.

\begin{figure}
\centering
\includegraphics[scale=0.55]{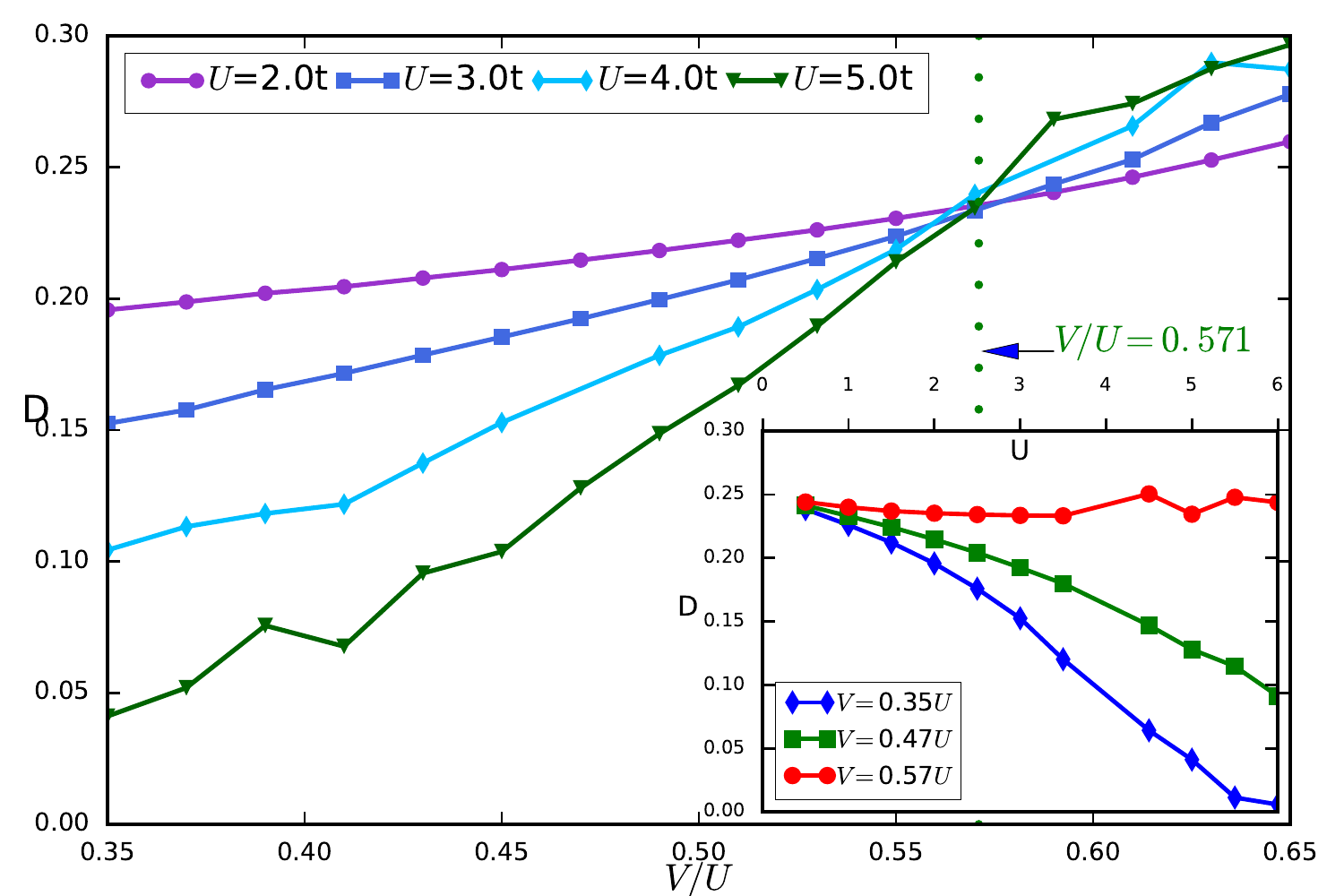}
\caption{The double occupancy ($D=\langle n_\uparrow n_\downarrow$) of the EHM calculated for a range of couplings $U$,$V$ at $T=0.02t$ as a function of $V/U$ (main) and as a function of the Hubbard coupling (inset). Calculations where carried out via the DCA-fRG for $N_\omega=4$, $N_K=4$ and $N=64$ with $N_r=512$ basis sets retained.}
\label{uvDoccup}
\end{figure}

\begin{figure}
\centering
\includegraphics[scale=0.47]{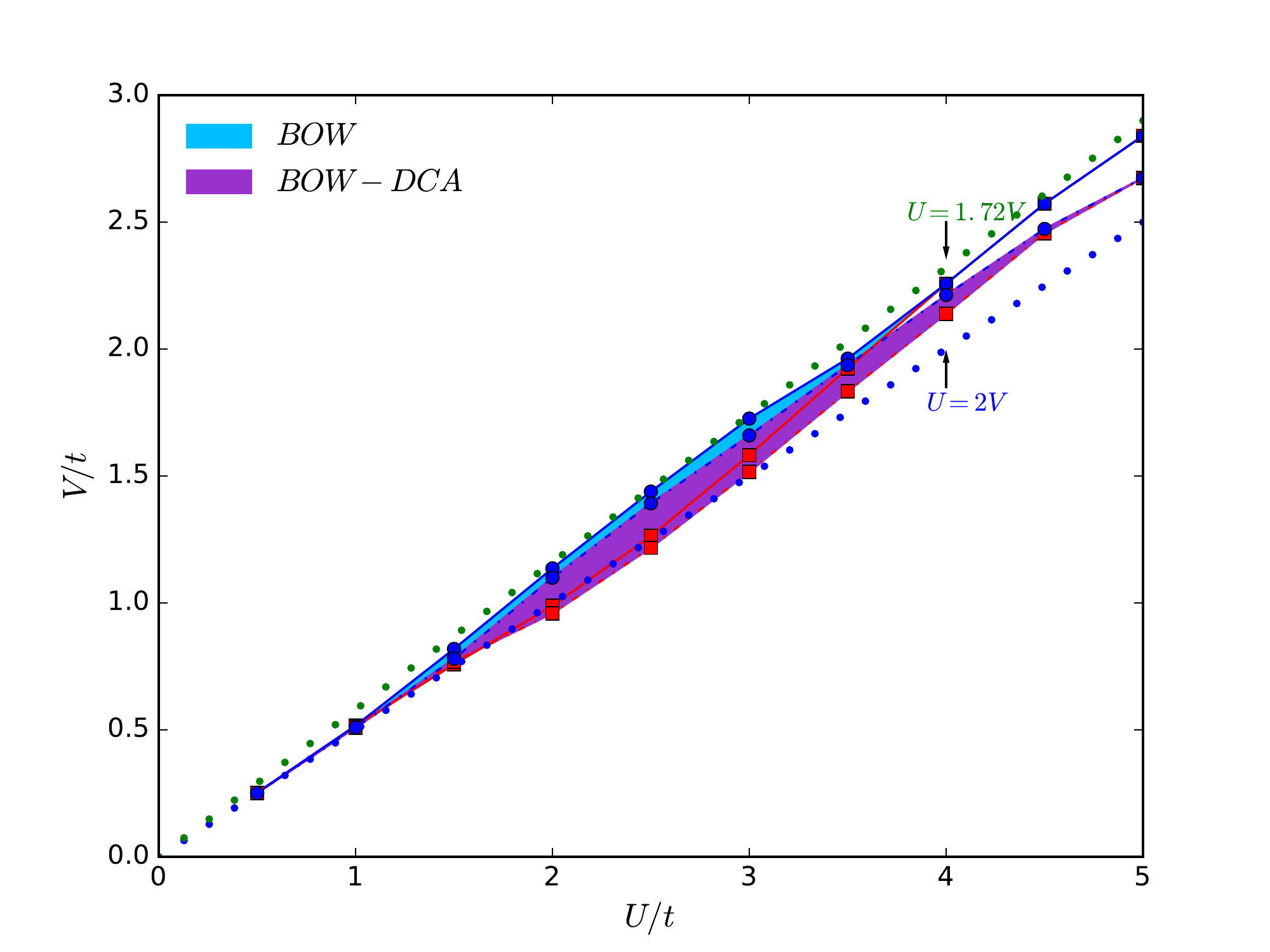}
\caption{The phase diagram of the EHM via 2-Loop fRG and DCA-fRG at $\beta=50$ for $N_\omega=4$, $N_K=4$ and $N=64$. The BOW phase predicted by the methods is shown. The mean-field transition line form CDW to SDW ($U=2V$), and the transition line for the double occupancy ($U=1.72V$) are plotted for reference.}
\label{EHMphase}
\end{figure}

The susceptibilities for the charge and spin waves are long time objects ($\omega=0$) accessible within the static fRG and do not utilize the frequency content of the general unequal time correlator. A quantity of interest for the EHM that requires the frequency data is the double occupancy. The double occupancy is suppressed in SDW phase and slowly restored as charge fluctuations due to an increasing $V$ drive us towards the CDW phase. The strong spin fluctuations in the BOW phase should restrict a saturated double occupancy to the CDW phase. The expression for the double occupancy is given in Eq.\ref{dOccup} with the results for a 64-site system plotted in Fig.\ref{uvDoccup}. We note that our results in the charge density wave sector do violate the Ward identity for charge conservation, with the double occupancy peaking above $0.25$. The divergence of the vertex in much of the CDW regime, along with the basis truncation scheme implemented for the flow, are the likely causes. The inset of Fig.\ref{uvDoccup} shows the double occupancy as a function of the onsite coupling, $U$. We see that far from the transition the double occupancy is completely suppressed at $U=6t$ but as we reach the transition it flattens showing essentially no change for a wide range of $U$ values. Consequently, this transition line is the crossing point for the double occupancies as a function of the nearest neighbor coupling ($V/U$). This line also marks the termination of the BOW phase with no BOW observed above this line (Fig.\ref{EHMphase}). Additional plots that show the frequency dependence of the correlators and the self energy as well as their dependence of the size of the momentum basis set ($N_K$) is given in Appendix.\ref{freqOfVert}.

The phase diagram of the EHM, as constructed by the fRG and the DCA-fRG, is shown in Fig.\ref{EHMphase}. Both methods capture the opening and closing of a finite phase with BOW correlations. Our results clearly show (Fig.\ref{1DSus}) the susceptibility for the bond ordered-wave peaking during the transition from a charge density wave to a spin density wave. As the DCA is approximately in the thermodynamic limit, the susceptibilities of our gapped phases (BOW, CDW) are enhanced and lead to a sharper divergence in the CDW phase. All our calculations were performed at temperatures ($\beta\leq 60$), which are low enough to have a convergent susceptibility but can introduce error in the determination of the phase boundary for the EHM. Previous studies have noted that at strong-coupling, very low temperatures ($\beta>100$) are needed to converge the phase transitions for the ground state phase diagram.

The critical end point we calculate from the 64-site system via the fRG happens at $U/t=4$ and $V/t=2.4$ which disagrees with previous DMRG results ($(U_m,V_m)\approx(9.25t,4.76t)$)\cite{ejima2007phase}. This disagreement with DMRG results seems to appear primarily along the SDW-BOW boundary and is actually expected since the KT transition at the boundary leads to the opening of an exponentially small spin gap which requires very large system sizes and low temperatures to properly resolve. On the other hand, the BOW obtained via the DCA at $\beta=50$ persists until $U/t=4.6$ and $V/t=2.58$, further extending the fRG results. The BOW phase peaks at $U\sim 4t$ and decreases until vanishing at the critical endpoint. The spin gap in this region is small and combined with the strength of the coupling makes a determination of the phases via the fRG very difficult.  Our results at lower coupling though can be further refined by lowering the temperature of our calculations. This refinement is already seen for $U/t=2$ in Fig.\ref{EHMnn} and should provide a complete picture at weak coupling. The significant computational gains and demonstrated stability of the basis truncation scheme suggest that we adopt it for all subsequent calculations capping the elements of the basis set at $N_r=512$. We also fix the filling along the flow to avoid any violation of charge conservation.

\subsection{The Two-dimensional Hubbard Model with $t$, $t'$ and $V$} \label{2DModels}
The two-dimensional Hubbard model has attracted overwhelming interest as a prototype Hamiltonian for describing the mechanism of high-temperature superconductivity in the Cuprates. The Cuprate family of compounds are rich in phases with strong competition between antiferromagnetic, charge and superconducting correlations. These materials show a strong link between spin fluctuations and d-wave superconducting correlations with spin correlations appearing to be the primary drivers of d-wave superconductivity. Quantum cluster methods (DCA\cite{maier2005systematic,macridin2006phase},CDMFT\cite{fratino2017effects,bragancca2018correlation}) and the fRG\cite{halboth2000renormalization,honerkamp2001magnetic,katanin2009two,eberlein2014superconductivity} have been successful in clarifying many of phases in the model. These studies have uncovered a $d_{x^2-y^2}$-type superconducting order that underlies the SDW at half filling. The coexistence of AF and d-SC correlations has been well established in various model Hamiltonians though within the fRG divergent flows coupled with a large parameter space make an accurate identification of the phase diagram difficult. An ingredient that is missing from computational studies of the Hubbard model are charge density fluctuations which are known to exist in the hole doped regime regime in competition with SDW and d-SC. In light of the vast number of superconducting compounds which have a wide range of critical temperatures $T_c$ , and varying levels of charge fluctuations the nearest neighbor interaction (V) may be crucial in modeling these systems\cite{reymbaut2016antagonistic}. Unlike fluctuations in the spin sector the charge order is in strong competition with d-SC with studies of Cuprates under high magnetic fields showing charge density waves emerging as magnetic field disrupts the superconducting correlations\cite{wu2011magnetic,loret2019intimate}. The final piece normally employed in modeling the Cuprates at the single band level is the nearest neighbor hopping ($t^\prime$) which can be tuned to capture the round Fermi surface associated with high $T_c$ Cuprates. These will be the parameters we consider in the models below. For each model we study the response of the system to doping. Previous studies of the Hubbard model have shown that doping quickly destroys the AF leaving a small but finite region of d-wave superconductivity. A detailed review of the phases of the Cuprate superconductors can be found in Ref. \onlinecite{proust2018remarkable}.

Besides the competing ground state phases both the Cuprates and the Hubbard model show a pseudogap phase that sets in at high temperatures as we dope the system. The pseudogap sits atop the low temperature AF and d-SC phases and appears as doping, temperature destroy the antiferromagnet stabilized on the nested fermi surface. Consequently the pseudogap appears to be sensitive to the fermi-surface topology with recent experiments establishing the need for a hole-type fermi surface for the psuedogap\cite{doiron2017pseudogap}. The decoupled-fRG described above allows us access to the self energy at finite temperatures thus is the ideal tool to investigate the impact of the fermi-surface topology on the correlations in the Hubbard model at moderate coupling. The restriction to moderate coupling is problematic as the pseudogap appears to be a strong coupling phenomena with reported values of the Hubbard coupling for the Cuprates all in the strong coupling regime ($U=8-11t$). We note that the fRG at moderate coupling does show the suppression of quasi-particle weight along the fermi-surface which can perhaps be considered precursor to the strong variation seen in the strong coupling regime. The restriction to moderate coupling is especially necessary at the two-particle level as the spin and charge susceptibilities show a strong response at moderate $U$, $V$ values leading to strong divergences in the vertex. A possible cause for the quick saturation of our susceptibilities is the limited basis set which restricts us to $N_k=3$ ($13\times 13$) with the set for $N_k=2$ shown in Fig.\ref{latNK}. The increased momentum resolution is necessary to account for the nearest neighbor interaction which has an initial momentum structure and the values interest ($V\approx U/4$) which push us further into the moderate coupling regime.
\begin{figure*}
  \includegraphics[scale=0.45]{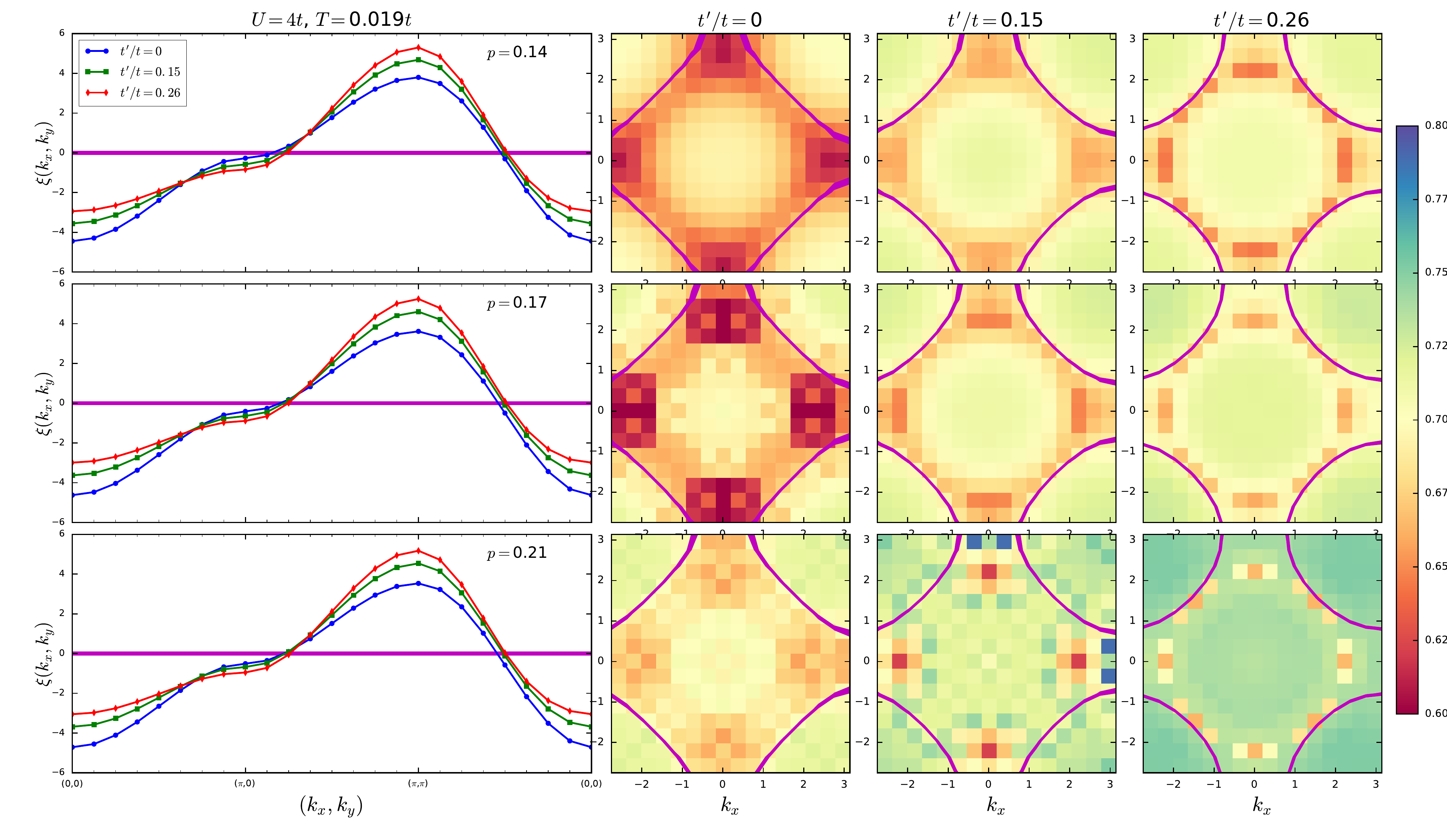}
  \caption{The band(left) and quasi-particle weight(right,$\mathcal{Z}_k=1-\partial_{i\omega}\Sigma_k(i\omega)$) of the Hubbard model with coupling $U=4.0t$, $V=0$ and $t^{\prime}/t$=0,0.15,0.26 on a $16\times16$ lattice at $p=0.14,0.17,0.21$. The frequency and momentum resolutions were $N_\omega=4$, $N_k=2$ with $N_r=512$ basis sets retained. The temperature of the system was set at $T=0.019t$.}
  \label{qWU2V0}
\end{figure*}

Considerations of fermi-surface topology, charge fluctuations and doping significantly expands our model, making a full scan unfeasible. We instead focus on two compounds ($\rm{La_2CuO_4,HgCuO_4}$), which show a large difference in $T_c$ ($40K$,$90K$). An ab-initio low energy effective model for these two systems has been derived recently at the one, two and three band levels\cite{hirayama2018ab}. In this work, we focus on the one band picture of these two systems ($t^\prime=0,0.15t,0.26t$).

\begin{table}[h]
\caption{\label{cValues} Ab-initio values for two Cuprate compounds\cite{hirayama2018ab}.}
\begin{ruledtabular}
\begin{tabular}{l c r r r}
\textrm{-$CuO_4$}&
\textrm{t}&
\textrm{$|t^{\prime}/t|$}&
\textrm{$U/t$}&
\textrm{$V/t$}\\
\colrule
$Hg$ & -0.461ev & 0.26 & 9.48 & 2.106\\
$La_2$ & -0.482ev & 0.15 & 10.4 & 2.6
\end{tabular}
\end{ruledtabular}
\end{table}

The values of the couplings for the two compounds are interesting, as the nearest-neighbor interactions in both compounds appears to be right at the value of the CDW-SDW boundary which should lead to a strong competition with the local spin coupling. The critical end point for the one dimensional EHM is at $(U=9.25t,V=4.76t)$ with a peak at $U/t\sim 4$ which places the values of the coupling for the Hg-compound and the La-compound near a similar regime. We note that for the two dimensional EHM the CDW-SDW transition is at $U=4V$ and adjusting for the larger bandwidth of the two dimensional system the values for the two compounds appear poised near criticality. Apart from the nearest neighbor density term, the one-band models for the two compounds have different nearest neighbor hopping terms ($t^{\prime}=0.15$, $t^{\prime}=0.26$). Previous studies have looked at the role of the nearest neighbor hopping($t^\prime$) in enhancing superconducting correlations we consider the added impact of the nearest neighbor interaction ($V$). We begin this section by looking at the Hubbard model at the single particle level to explore the impact of strong correlations, doping on the Fermi surface and the quasi particle weight. We then consider our results at the two particle level and the impact of $V$.

The Hamiltonian for the 2D Extended Hubbard model is given by
\begin{align}
\mathcal{H} &= -\sum_{\langle i,j\rangle}t_{i,j}(c_{i,\sigma}^\dagger c_{j,\sigma} + h.c.) + \mu\sum_{i\sigma}n_{i,\sigma} +\nonumber\\
 &U\sum_i n_{i\uparrow}n_{i\downarrow}+V\sum_{\langle i,j\rangle}n_in_j
\end{align}
where $t_{i,j}$ are the hopping parameters of the model. With the nearest neighbor hopping term ($t^{\prime}$) the dispersion of the model is given by $\xi(\vec{k})=-2t\cos(k_x)-2t\cos(k_y)+4t^{\prime}\cos(k_x)\cos(k_y)$.  Models appropriate to the Cuprates\cite{3bandCuprates} require particle-hole asymmetry with the value of the next-nearest hopping term ($t^{\prime}$) controlling the degree of asymmetry. This is inline with experiments many of which show the emergence of superconductivity at small hole doping. In keeping with our approach to the 1D EHM, we initialize the fRG at a large scale ($\Lambda\geq 4W$) and track the flow at a series of temperatures. We set our momentum basis for the decoupled fRG at $N_k=2$ (Fig.\ref{latNK}). In two dimensions this corresponds to nine basis functions ($N=(2N_{K,x}+1)(2N_{K,y}+1)=9\times 9$).  The computational complexity of the decoupled fRG equations is identical in 1D and 2D, but as the 2D Hubbard model is on lattice with discrete symmetries further reductions can be achieved. For the square lattice we retain a reduced triangular wedge in the first Brillouin zone. The symmetries of the lattice (inversion, $\pi/4$ rotations, reflections) can be used directly to extend single particle vertices to the full BZ. Extending the expansions of the two particle vertex to the full BZ is somewhat involved as the vertex representation is mixed between real space and momentum space. Extensions for the single band case is given in Appendix. \ref{symmV}. Lattice based reductions are crucial for large system sizes ($N>16$) and especially in the case of multi-band models which naively require tracking of $n_b^4$ vertices. We set the basis set for the frequency decoupling at $N_\omega=4$ and retain a variable set of basis functions($N_r=128-512$) based on spectral weight.

\begin{figure}
\hspace*{-0.5cm}
\includegraphics[scale=0.6]{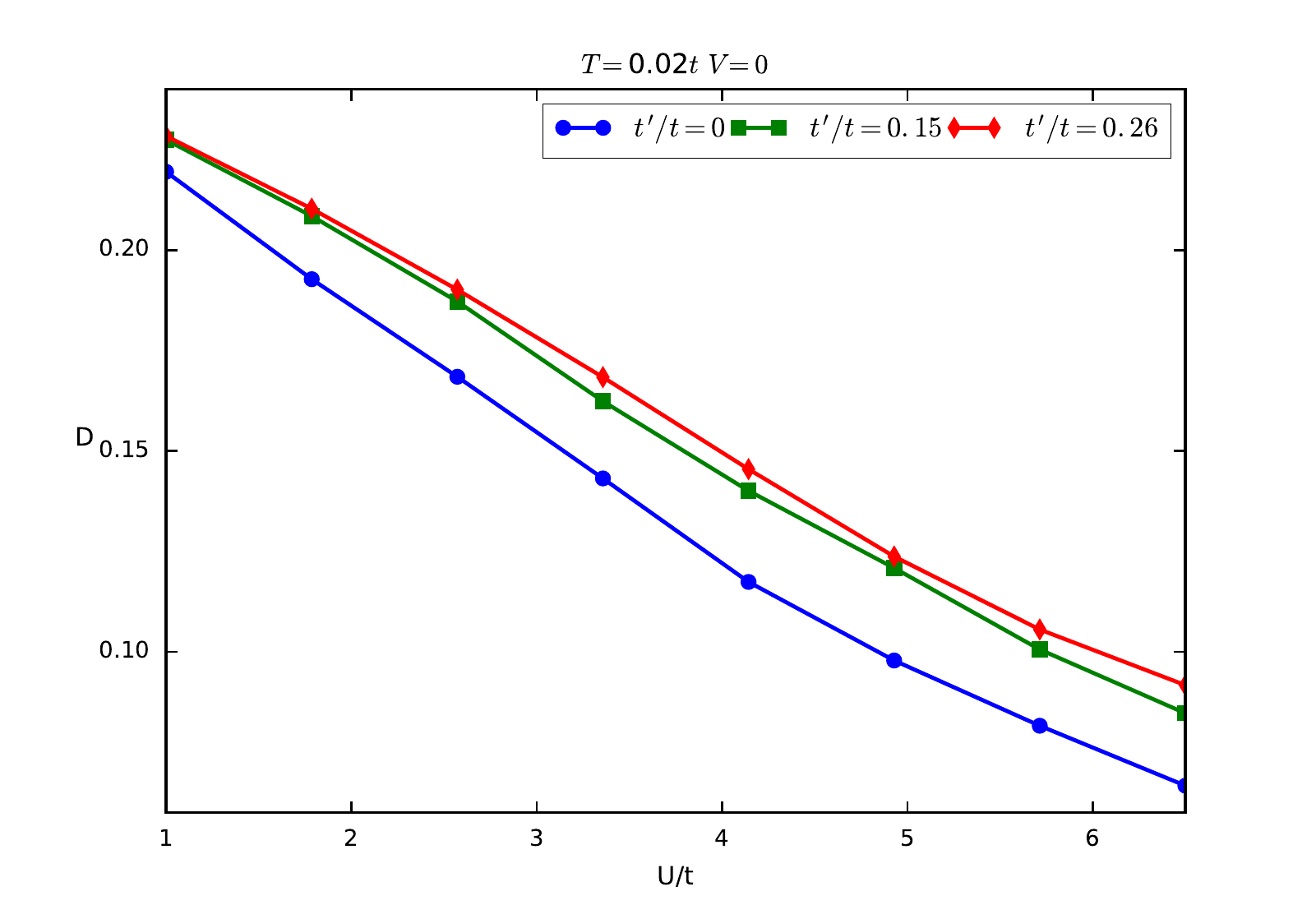}
\caption{The double occupancy of the Hubbard model for three values of $t^\prime =0,0.15t,0.26t$ calculated via DCA-fRG at $T=0.017t$ with $N_\omega=4$, $N_k=2$ and $N_r=512$}
\label{dFillU0}
\end{figure}
\begin{figure*}
\hspace*{-1.5cm}\includegraphics[scale=0.63]{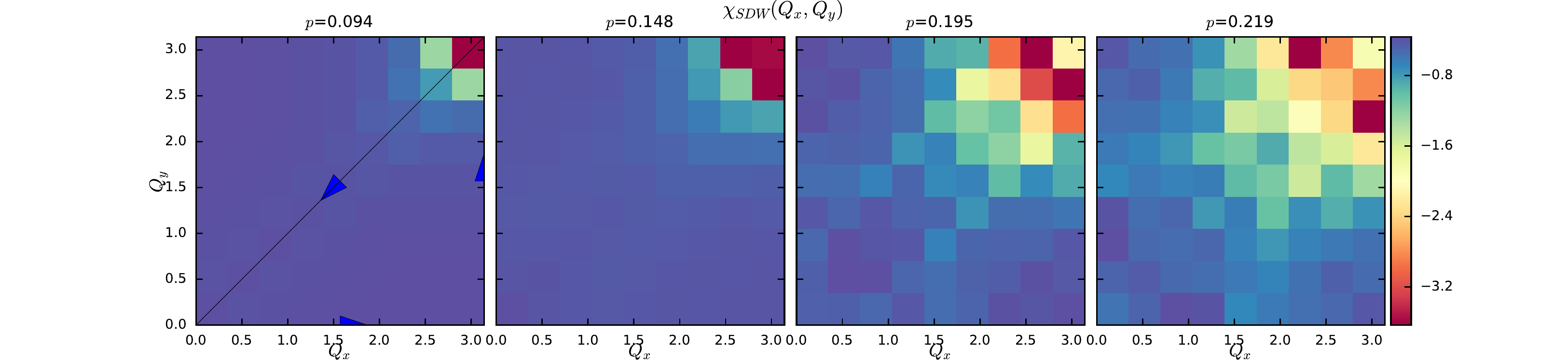}
\caption{Formation of incommensurate spin fluctuations in the first quadrant of the BZ as a function of hole-doping for the Hubbard model with $U=4t$, $t^{\prime}/t=0$ at $T=0.017t$ via the 2-loop DCA-fRG $N_\omega=4$,$N_k=2$, $N_r=512$.}
\label{incommSpin}
\end{figure*}
\begin{figure*}
\includegraphics[scale=0.57]{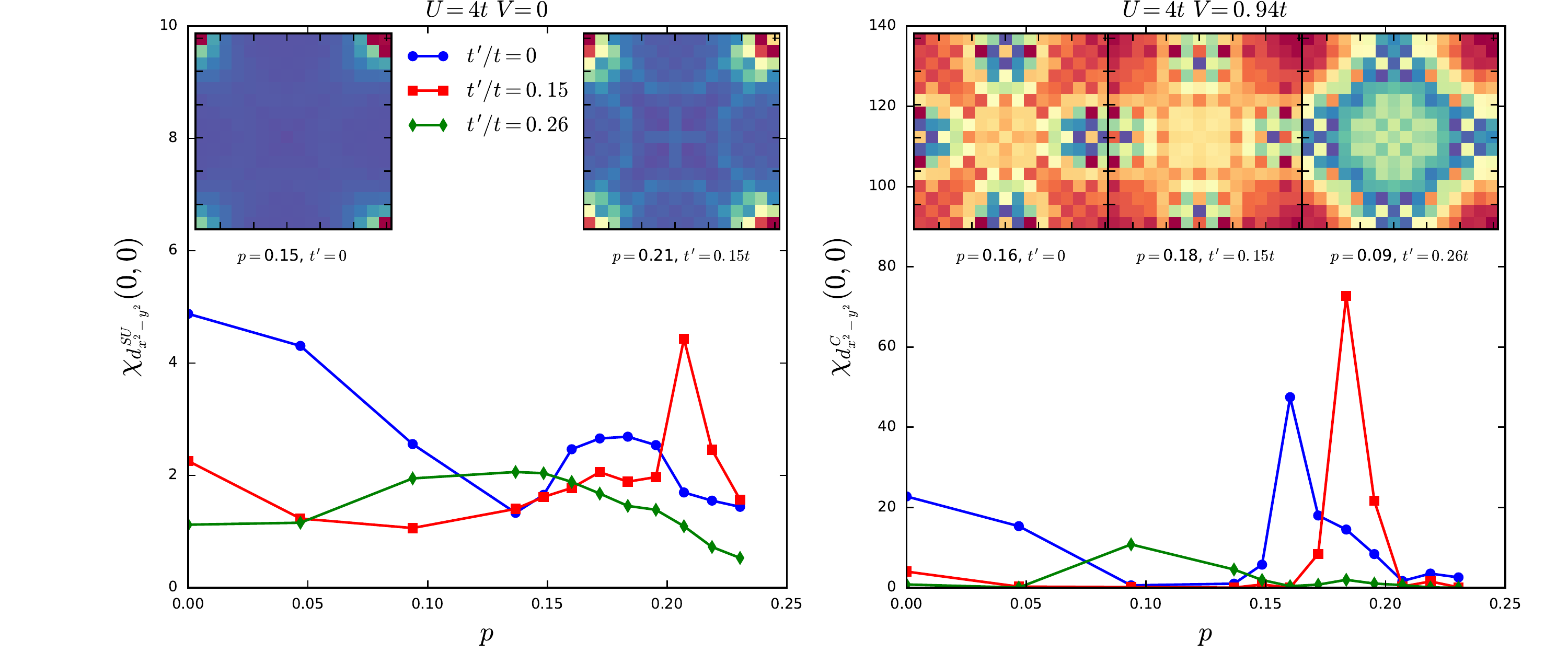}
\caption{The susceptibilities for $d_{x^2-y^2}$-type charge and superconducting orders of the two dimensional Extened Hubbard model $V=0$(left), $V=0.94t$(right) as a function of doping for three values of $t^\prime$ ($U/t=4$, $N_\omega=4$, $N_k=2$,$N=16\times 16$,T=0.017). The spin susceptibilities of the $V=0$ model at the peak of superconducting correlations are shown (left Inset). The quasiparticle weight of the $V=0.94t$ model is shown at point of peak charge correlations for the three values of $t^{\prime}$.}
\label{sus2DU}
\end{figure*}

The critical temperature of the Hg-Cuprates ($T_c=0.0168t_{Hg}$) and the La-Cuprates ($T_c=0.00795t_{Hg}$) in conjunction with the coupling ($U\approx 10t_{Hg}$) put model parameters squarely in the regime of divergent flows. Thus, we reduce the Hubbard coupling to a moderate value ($U=4.0t$) and focus our studies around an adjusted temperature range ($T\geq 0.015t$). A similar adjustment sets the nearest neighbor coupling ($V=0.94t$). At the single particle level our primary result for the EHM are the quasiparticle weight and the approximate fermi surface.
The bands of the model along with the quasiparticle weight are shown in Fig.\ref{qWU2V0} for $V=0$ at the the three values of $t^\prime$. Both plots show an approximate Fermi-surface extrapolated from the $16\times 16$ lattice at $T=0.019t$. The quasiparticle weight shows broad variations along the fermi-surface with clear nodal structure in the case of $t^{\prime}=0,0.15t$. The nodal structure at the chosen values of hole doping corresponds with peaks in the superconducting correlations shown in Fig.\ref{sus2DU} and the incommensurate spin correlations that emerge as the system is doped (Fig.\ref{incommSpin}). Past the optimal point ($p=0.17$ for $t=0$, $p=0.21$ for $t^\prime=0.15$) increasing the doping serves to destroy the nodal structure signaling a transition to a simple Fermi liquid.

\begin{figure}
\includegraphics[scale=0.56]{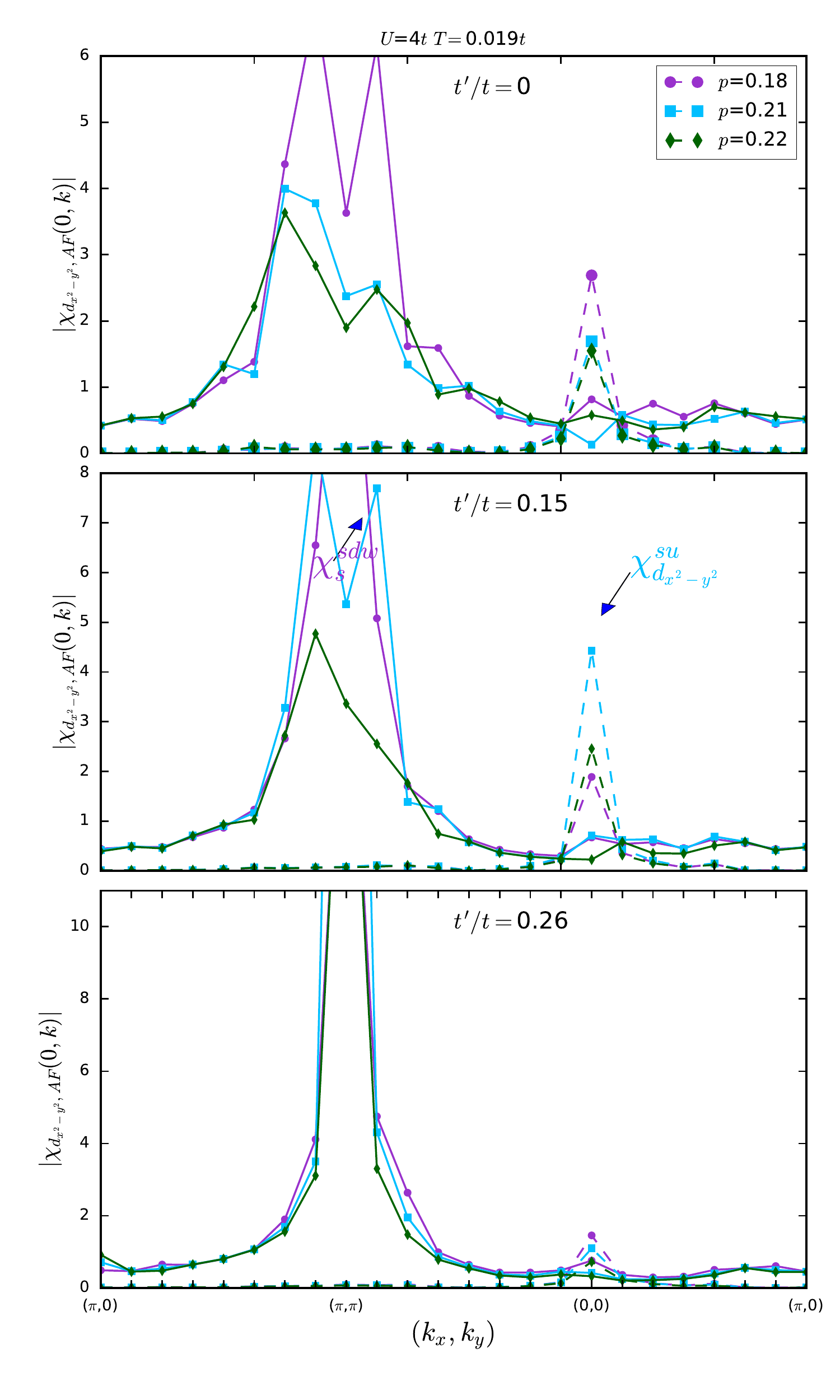}
\caption{The spin and $d_{x^2-y^2}$-superconducting susceptibilities as a function of the nesting vector for $U=4t$, $V=0$, $t^{\prime}/t=0,0.15,0.26$ at $p=0.16,0.21,0.23$.}
\label{qWU2D2}
\end{figure}

\begin{figure}
\hspace*{-0.5cm}
\includegraphics[scale=0.6]{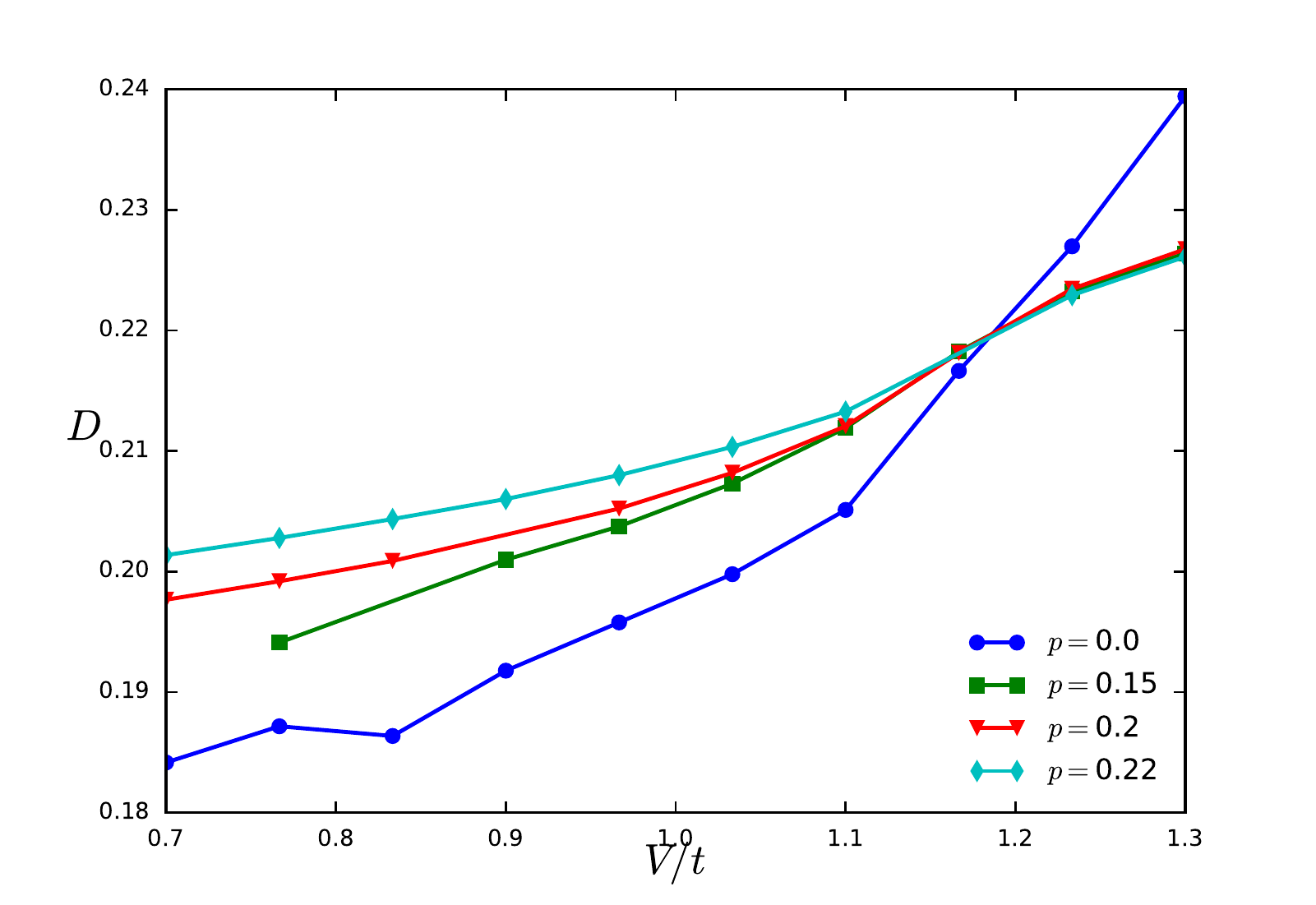}
\caption{The double occupancy of the Extended Hubbard model as a function of $V$ at $U=4t$, $t^\prime=0$ and $T=0.02t$ for various values of hole doping $p=0,0.147,0.198,0.22t$.}
\label{dFillV}
\end{figure}
\begin{figure}
\includegraphics[scale=0.56]{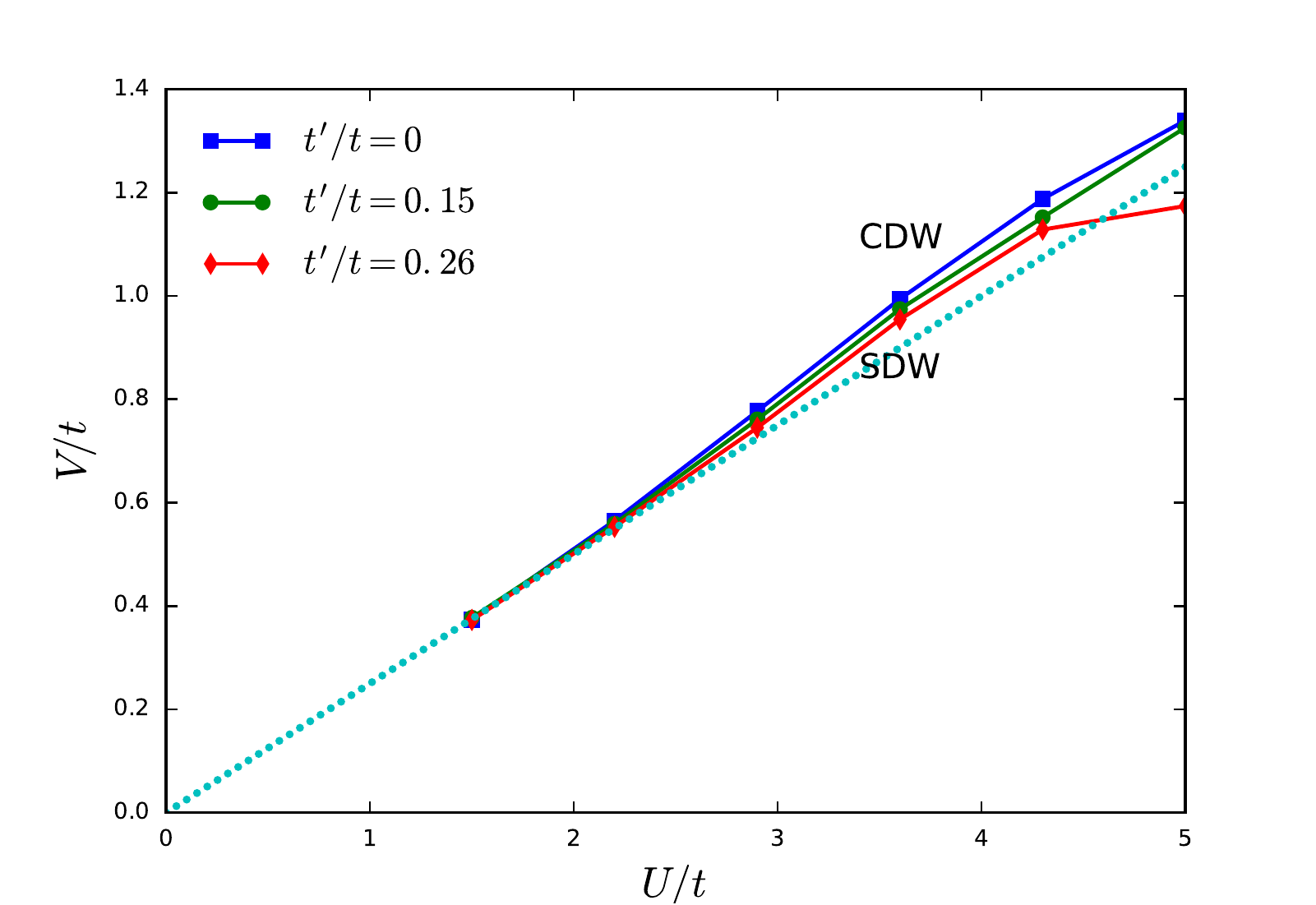}
\caption{Phase diagram of the two dimensional EHM via the DCA-fRG with $N_\omega=4$, $N_k=3$ and $N_r=256,512$. The mean field transition $U=4V$ is also plotted for reference.}
\label{UV2Dphase}
\end{figure}
At the two particle level we began by calculating the double occupancy which has been used to benchmark a variety of methods used to study the two dimensional Hubbard model\cite{leblanc2015solutions}. Our results for the three nearest neighbor hopping terms are shown in Fig.\ref{dFillU0}. As expected spin fluctuations due to the coupling ($U$) suppress the double occupancy with changes to band structure due to $t^{\prime}$ showing little change beyond an initial jump. From the susceptibilities we find a strong response for the AF, spin, d-superconducting correlators at $V=0$. At finite values of $V=0.94t$ we find the superconducting correlations replaced by $d_{x^2-y^2}$ charge order. As we show in Fig.\ref{sus2DU} the charge correlations dominate even as we dope the system. The spin and s-superconducting correlations appear to be sub-dominant phases with d-superconductivity fully suppressed. Antiferromagnetic correlations dominate the phase diagram up to moderate doping (p$\leq 0.1$). Along with these fluctuations we see the beginnings of $d_{x^2-y^2}$-type superconductivity for the case of $V=0$ and a $d_{x^2-y^2}$-type charge density wave for a moderate $V$ consistent with the reduced values of the La/Hg-Cuprates ($U=4t$, $V=0.94t$). The spin and superconducting fluctuations for our two models at various dopings are shown in Fig.\ref{qWU2D2}. For the particle-hole symmetric case we see strong $d_{x^2-y^2}$-type fluctuations at p=0 with superconducting correlations emerging again ($V-0$) with a smaller peak at $p=0.17$. For moderate coupling ($U=4t$) the small superconducting peak seems on par with the remnants of the spin fluctuations. This is expected as the spin fluctuations are the primary mechanism driving the superconducting order. Furthermore, the antiferromagnetic fluctuations are unstable to hole doping with incommensurate spin correlations appearing at $p\approx 0.125$ and moving to smaller nesting vectors as the doping is increased. The appearance and growth of the incommensuate spin correlations are shown for the particle hole symmetric case in Fig.\ref{incommSpin} while Fig.\ref{qWU2D2} shows the correlations for different nearest neighbor hopping along the high symmetry lines. Previous fRG studies \cite{yamase2016coexistence} have found incommensurate spin density waves appearing alongside superconducting correlations up to $p\approx 0.26$.

With out the nearest neighbor interaction ($V=0$) the AF, d-SC correlations peak at half filling for $t^{\prime}=0$ and drop as we dope away from half-filling. The d-SC correlations are strongest in the doped system with $t^{\prime}=0.15t$ where they appear on par with the antiferromagnetic correlations. On the other hand fluctuations in the charge channel appear negligible showing the strongest response for charge density wave order ($Q=0,t^\prime=0$) with values much weaker than spin and superconducting correlations at all temperatures and dopings. This is consistent with our results in 1D, as deep in the SDW phase charge density fluctuations are exponentially suppressed at $Q>0$. When we turn on $V$, charge correlations in the d-channel are significantly enhanced with a strong V ($V\gg U$) driving the expected charge density wave at $Q=(\pi,\pi)$ for all three choices of $t^\prime$. As we approach the transition we see expected changes in the double occupancy as a function of $V$. The double occupancy saturates across the transition at half filling (Fig.\ref{dFillV}) with the transition occurring further than the mean field line. Doping the system destabilizes the $CDW$ with the double occupancy showing weaker dependence as we increase $p$. The saturation of the double occupancy at $V=1.3t$ for $U=4t$ suggests at the minimum as shift from the mean field transition line. The $U-V$ phase diagram of the EHM for the three values of nearest neighbor hopping ($t^\prime$) is shown in Fig.\ref{UV2Dphase}. We find the transition occurring away from the mean field line for all three values of $t^\prime$. Bond ordered phases are enhanced along the transition but computational limits prevent us from accessing a larger momentum basis set $N_k>3$ which our 1D-EHM results indicate is necessary for observing such phases. Constructing the full phase diagram is difficult as we are plagued with divergences at large $V$ that prevent a deep foray into this sector.
\begin{figure}
\includegraphics[scale=0.58]{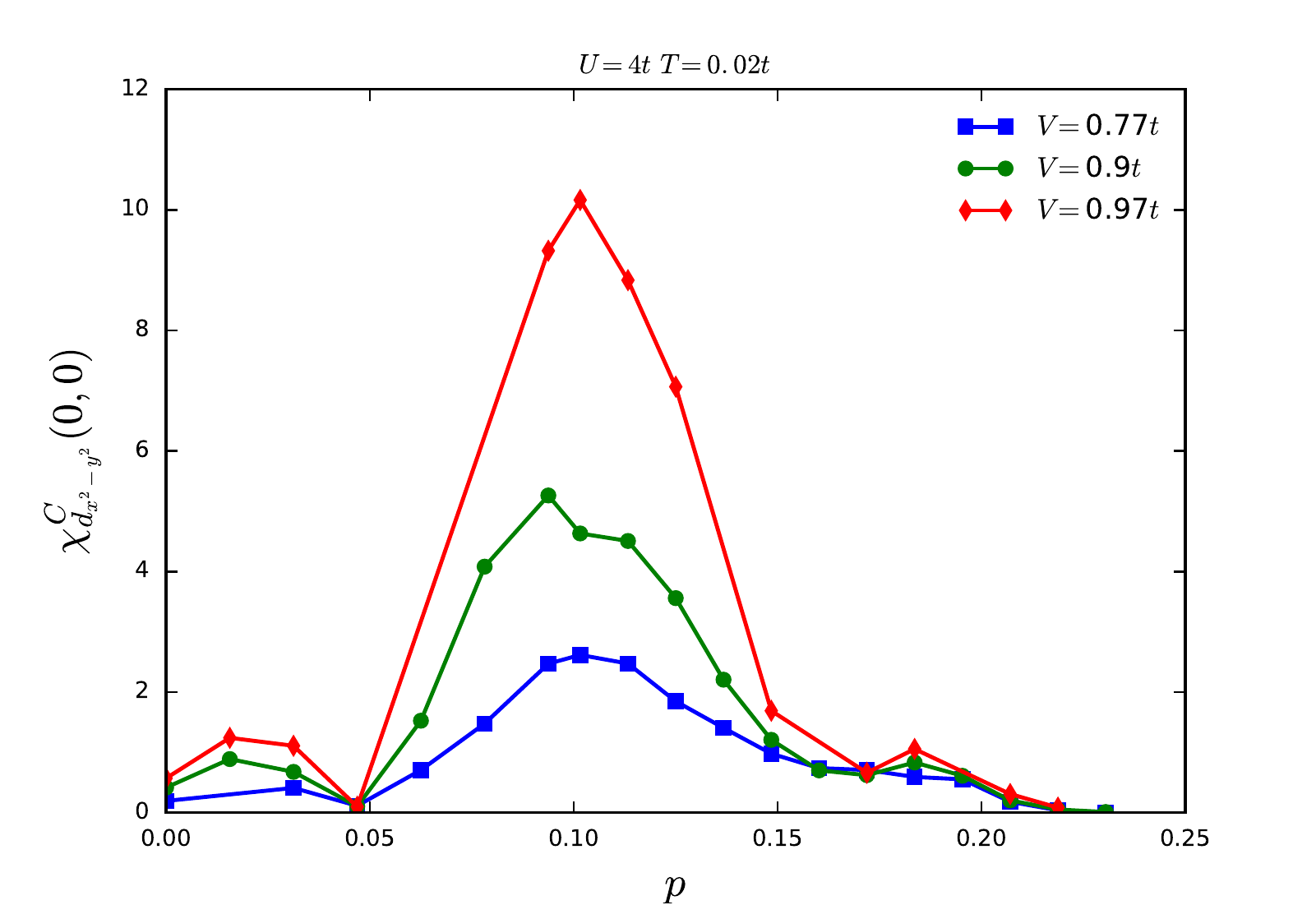}
\caption{The $d_{x^2-y^2}$-charge correlations of the EHM for $t^\prime=0.26$ at values of $V$ up to the transition ($V=t$).}
\label{chargeCorr}
\end{figure}

Doping the model at finite $V=0.94t$ in the SDW phase leads to a strong $d_{x^2-y^2}$-charge order at $Q=0$ that easily overtakes the spin and superconducting fluctuations. This charge order appears roughly in the same hole doping region as superconducting correlations at $V=0$ and is the dominant order for $t^\prime=0,0.15t$. The quasiparticle weight associated with the charge response is shown in Fig.\ref{sus2DU}. The case of $t^\prime=0.26$ is curious in that much like the response in the $V=0$ case the correlations are much weaker than the other values of $t^\prime$. The response does grow are we increase $V$ which given that $V$ drives charge correlations is unsurprising. The peak however shifts away from those of $t^\prime=0,0.15t$. The correlations as a function of hole doping are shown in Fig.\ref{chargeCorr}. Going back to the Cuprate compounds we chose as a basis for the values of $t^\prime$ we see a contradiction. Both superconducting and charge correlations are stronger for $t^\prime=0.15$ which corresponds to the La-system with much lower reported $T_c$. This discrepancy is well documented with a  resolution requiring an expanded model with two orbitals ($d_{x^2-y^2},d_{z^2}$)\cite{sakakibara2010two}. This expanded model will be the focus of future work.

\section{Conclusions}
\label{Conclusion}
In this work, we have studied a decoupled scale-dependent functional renormalization group (fRG) for zero, one and two dimensional fermionic Hamiltonians. The full channel decomposition extends previous fRG formulations to the frequency domain and significantly reduces the computational weight of the equations.  Advancements in methodology at the one and two-loop level were studied by solving the Single Impurity Anderson model and benchmarking against Bethe Ansatz results. The scale dependent basis set along with a similar decoupling in the momentum domain has allowed us to construct flows for the study of extended Hubbard interactions in one and two dimensional systems. Our treatment of frequency and momentum modes in these systems was carefully cross checked against results in the literature for the one dimensional Extended Hubbard model and the two dimensional Hubbard model. Access to the frequency content of the vertex, allows us to compute the full correlator from which one can extract the double occupancy, the charge and spin structure factors and the frequency dependence of susceptibilities as a function of temperature.

The decoupled frequency dependent fRG forms the core of our approach and was rigorously examined at the level of the SIAM. At the one loop level the results obtained from the decoupling are in agreement with previous fRG results but come at a much reduced computational price. The two-loop calculation improves upon our results in two ways. The fRG equations show a much weaker dependence on our choice of regulator and we obtain much better agreement with the Bethe Ansatz results up to moderate coupling ($U/\Delta=6$). The error in the converged flow show the expected dependence on temperature and scale with coupling, with weak coupled flows showing fast convergence and small errors. The basis sets used to decouple the flow showed rapid convergence as the number basis functions used to represent the vertex was increased. Given their relatively similar quantitative performance the (truncated) cluster expansion is the preferred choice as it allows for the utilization of larger basis sets truncated in each channel to the modes that show the strongest dependence to the parameters of study. Issues of noise can lead to instability in the truncated cluster expansion but the sparsity of the basis set makes it the ideal choice for higher dimensional models where computational cost is the primary concern. Further computational gains can be achieved via a judicious choice of regulator and utilization of the symmetries of the vertex.

After benchmarking the method, we used the Litim regulator and the optimum representation of the two-particle vertex at the two-loop level to study the one-dimensional Extended Hubbard model. Our primary focus was the issue of convergence and the size of the basis set needed to capture the phases in the EHM. Our results are convergent at low temperatures ($\beta=50$) and capture a finite bond ordered wave phase around $U=2V$. Convergence is fast as a function of temperature, number of sites and the number of momentum basis functions. In particular, we found the fRG equations to be convergent at large T, even at strong coupling ($U\sim 4t, V\sim 2t$), which allowed us to more fully explore the phases in the moderate coupling regime of the 1D EHM. Restrictions to system size ($N<256$) especially at low temperatures were overcome by adopting the clustering methods of the DCA. The DCA clustering approach can be especially important in two dimensional systems where higher computational cost along with larger memory requirements limit us to $16\times 16$ lattices for single band Hamiltonians. Studies of the impact of clustering in two dimensional systems is currently underway as coarse-graining over the Brillouin zone may reduce or even average over the role of van-hove points in enhancing density and superconducting orders.

Applications of the two-loop fRG in two dimensions focused again on the EHM with additional considerations for the filling and the next-nearest neighbor hopping, $t'$. Our choices particularly those of coupling and filling where intended to be representative of models for $\mathrm{La}$ and $\mathrm{Hg}$-cuprates. We calculated the self-energy and the vertex at a variety of couplings, temperatures, fillings with the next-nearest neighbour hopping adjusted to match the fermi-surface of the cuprate systems. From the flow for the imaginary piece of the self-energy, we were able to calculate the quasiparticle weight, which shows variance and node like structures along the Fermi surface. These structures persist even as we introduce charge fluctuations by turning on the nearest neighbour density interaction. At the two particle level the charge fluctuations due to the extended interaction destroy strong $d_{x^2-y^2}$-superconducting correlations seen in the Hubbard model. These charge correlations are dominant over superconductivity as the system is doped with $d_{x^2-y^2}$-charge order emerging as the leading instability above $p\geq 0.15$ at $t^\prime=0,0.15t$. Strong charge correlations are a presence in the Cuprates with \textit{ab-initio} Hamiltonians for these compounds reporting a large nearest neighbor density interaction. To further clarify this order we conducted sweeps in the $U-V$ space around the transition from antiferromagnetic ordering to charge density wave (the line $U=4V$). We found the transition line shifted from the mean field value with a marked dependence on the next-nearest neighbor hopping ($t^{\prime}$). Along the critical line we see strong enhancement of bond-ordered charge density waves ($d_{x^2-y^2}$,$d_{xy}$) in lieu of the expected charge density wave. Deviations also emerge in the double filling of the two dimensional EHM which unlike its one dimensional counterpart saturates at value much further than the transition to a charge density wave. These results suggest bond-ordered phases along this line. The ideal next steps towards a full resolution of the phase diagram are denser momentum basis sets along with larger system sizes. Work towards these goals along with an extension of the decoupling to arbitrary loop order is currently underway.
\section{Acknowledgements}
We wish to thank Carsten Honerkamp, Ka-Ming Tam and Shan-Wen Tsai for illuminating discussions during the course of this work. We would also like to thank Boston University's Research Computing Services for their technical support and computational resources.
\bibliography{Methods.bib}

\appendix
\section{Derivation of the fRG equations}
\label{dfRGequations}
The fRG allows for an unbiased study of instabilities in Hamiltonians of the form $\mathcal{H}=\mathcal{H}_0+\mathcal{H}_{I}$. One avenue to a complete description of interactions within theses systems can is via the one particle irreducible(1PI) vertex functions. A generating functional for the 1PI functions can be easily constructed within the action formalism. Given the Hamiltonian $\mathcal{H}$, we have the corresponding action
\begin{align}
\mathcal{S}[\bar{\psi},\psi]=\int_0^\beta d\tau \left(\sum_\alpha \bar{\psi}_\alpha (\tau)\partial_\tau\psi_\alpha (\tau) + \mathcal{H}_0 +\mathcal{H}_{I}(\bar{\psi},\psi)\right)
\end{align}
where $\bar{\psi},\psi$ are fermionic grassmann fields and the label $\alpha$ runs over all momentum, spins and bands present in the system. We map the action to the frequency and momentum domain where our quadratic action is diagonal with respect to frequency and momentum and it can be diagonalized in the band basis via a unitary transform. While the spatial dependence of the action varies with the model of interest the frequency dependence is universal thus we insert a regulating function into the frequency domain. As the free action is diagonal the regulator serves to suppress all excitations below the current scale $\Lambda$. At high scales ($\Lambda\rightarrow\infty$) the regulator freezes all excitations which in turn leads to zero contributions from the interacting Hamiltonian making the action trivial to solve. The regulated propagator is given by
\begin{align}
G_{\alpha}^{\Lambda}(i\omega,k)^{-1}=i\omega-\xi_{\alpha}(k)+R_\Lambda(\omega)
\end{align}
with $\xi_\alpha(k)$ corresponding to the eigenvalues of the noninteracting Hamiltonian ($\mathcal{H}_0$) and $R_\Lambda$ is our regulator of choice. From this starting point one can step down scale by scale until the regulator is fully removed from the action ($\Lambda\rightarrow 0$). Due to the boundary conditions at $\Lambda\rightarrow\infty$ it is convenient to track the evolution of the 1PI vertex functions rather than the connected moments of the action, $\mathcal{S}$. The generating functional for the 1PI functions of the scale dependent action is given by
\begin{align}
\Gamma^{\Lambda}[\phi,\bar{\phi}]&=\sum_{\alpha\beta}\bar{\eta}_\alpha\phi_\beta +\bar{\phi}_\alpha\eta_\beta\nonumber\\
&-\ln\int D[\bar{\psi},\psi]\exp\left(-\mathcal{S}^{\Lambda}[\bar{\psi},\psi] +\sum_{\alpha\beta}\bar{\eta}\psi+\bar{\psi}\eta\right)
\end{align}
which is the Legendre transform of generating function of the connected moments. The fields $\phi$ is defined via the source $\eta$ as $\phi=\delta\mathcal{G}/\delta\eta$ where $\mathcal{G}$ is the generating functional for the connected moments. We can expand $\Gamma^\Lambda$ up to the order of interest in the fields $\phi$. We have
\begin{align}
\Gamma^{\Lambda}[\phi,\bar{\phi}]=\sum_{n=0}^{\infty}\frac{1}{(n!)^2}\sum_{\alpha_{i}}\Gamma^{\Lambda,2n}_{\alpha_{i_1},\alpha_{i_2},...,\alpha_{i_{2n}}}\bar{\phi}_{\alpha_{i_1}}...\phi_{\alpha_{i_{2n}}}\
\end{align}
where $\Gamma^{\Lambda,2n}$ are the n-particle vertex functions. The flows for each vertex function can then be obtained by inserting the expansion above into the generating functional, taking a derivative with respect to the scale $\Lambda$ and matching the terms order by order. The quantities $\mathcal{H}_0$ initialize the propagator and the interacting Hamiltonian, $\mathcal{H}_I$, initializes higher order vertex functions, $\Gamma^{\Lambda,(2n)\geq 4}$. The flows for the first three vertex functions is given by
\begin{align}
\partial_\Lambda\Gamma^{(2),\Lambda}&=tr\left(\partial_\Lambda G^\Lambda\Gamma^{\Lambda,(4)}\right)\nonumber\\
\partial_\Lambda\Gamma^{(4),\Lambda}&=tr\left(\partial_\Lambda G^\Lambda\Gamma^{\Lambda,(6)} + \Gamma^{\Lambda,(4)}\partial_\Lambda \left(G^\Lambda G^\Lambda\right)\Gamma^{\Lambda,(4)}\right)\nonumber\\
\partial_\Lambda\Gamma^{(6),\Lambda}&=\partial_\Lambda tr\left(G^\Lambda\Gamma^{\Lambda,(4)}G^\Lambda\Gamma^{\Lambda,(4)}G^\Lambda\Gamma^{\Lambda,(4)}\right)
\end{align}
with the flows for vertices beyond $\Gamma^{(6)}$ discarded. A more thorough derivation of the fRG equations can be found in previous works\cite{metzner2012functional,kopietz2010introduction}. The one-loop and two-loop variants employed in this work are attempts at capturing the relevant pieces of $\Gamma^{(6)}$ for cases with no initial three particle vertex ($\Gamma^{(6),\Lambda_{start}}=0$). Specifically the one-loop approximation employs a term derived by Katanin\cite{katanin2004fulfillment} that corrects the scale derivative ($\partial_\Lambda\rightarrow \frac{d}{d\Lambda}$) which accounts for terms of the form
\begin{align}
\partial_\Lambda\Gamma_{k_1k_2k_3k_4}^{(4)}=& (...)-\sum_{k,q,s}\underline{\partial_\Lambda\mathcal{G}_k\Gamma_{kssk}^{(4)}}\mathcal{G}_q(\mathcal{G}_s)^2\Gamma_{k_1k_2ks}^{(4)}\Gamma_{sqk_3k_4}^{(4)}
\end{align}
with the correction to the scale derivate ($\partial_\Lambda\Sigma$) corresponding to the underlined piece. The two-loop scheme employed is due to Eberlein\cite{eberlin2Loop} which utilizes derivatives of the contributions in each channel to more fully capture three particle contributions. The corrections at two loop are of the form
\begin{align}
\partial_\Lambda\Gamma_{k_1k_2k_3k_4}^{(4)}=& (...)-\sum_{k,q,s}\underline{\Gamma_{k_1k_2kr}^{(4)}\partial_\Lambda\left(\mathcal{G}_k\mathcal{G}_r\right)\Gamma_{krqs}^{(4)}}\mathcal{G}_q\mathcal{G}_s\Gamma_{sqk_3k_4}^{(4)}
\end{align}
which corresponds to the scale derivative of the particle-particle piece ($\dot{\Phi}_{pp}$) connected to the two particle vertex, $\Gamma^{(4)}$. A more recent multi-loop framework\cite{kugler2018multiloop} accounts for contributions from the full hierarchy. The multi-loop fRG fully sums all parquet diagrams with calculations at $m-loop$ requiring approximately $m\times(1-loop)$ run times. Implementation of decoupling for multi-loop flows is the focus of current work.
\section{Auxiliary Channels in the fRG flow}
\label{auxVarChoice}
Within the one particle irreducible formalism the contributions of the fRG flow at the two particle level are classified into the particle-particle, particle-hole and particle-hole exchange channels. Specifically these two particle objects can be thought of as precursors to pair and density condensates in the pp and ph/phe channels. Contributions in each channel can be defined by a singular variable that characterizes interaction between two body objects in the channel and two auxiliary variables that label the composition of the objects. Far away from any critical point, one expects contributions primarily from objects with a small separation ($\Delta\tau<<\beta,\Delta L\approx\mathcal{O}(a)$) between the constituents that form these virtual fluctuations. Away from the critical point convergence is rapid as thermal fluctuations suppress long-range interactions. As we approach and transition through the critical point the few terms we retain from the series may not capture the full picture but the ordering of the system will drive only one of the three channels making contributions from the other channels irrelevant.

The interaction vertex is given by
\begin{align}
\Gamma_{i_1,i_2,i_3,i_4}^\alpha(\tau_1,\tau_2,\tau_3,\tau_4)\bar{\psi}_{i_1}(\tau_1)\bar{\psi}_{i_2}(\tau_2)\psi_{i_3}(\tau_3)\psi_{i_4}(\tau_4)
\end{align}
where $\Gamma$ is really a function of three independent time and spatial differences only. With this in mind we can expand the two particle vertex in three ways as a function of the separation of the constituents.
\begin{align}
&\Gamma_{\alpha,(r_1-r_2),(r_3-r_4)}^{pp}(\omega_{pp},k_{pp})=\mathcal{N}_{\omega,k}\sum_{\omega_i,k_i}e^{i\omega_x(\tau_1-\tau_2)}\times\nonumber\\
&\qquad e^{-i\omega_y(\tau_3-\tau_4)}e^{ik_x(i_1-i_2)}e^{-ik_y(i_3-i_4)}\Gamma_\alpha(s_1,s_2,s_3,s_4)\nonumber\\
&\Gamma_{\alpha,(r_3-r_2),(r_4-r_1)}^{ph}(\omega_{ph},k_{ph})=\mathcal{N}_{\omega,k}\sum_{\omega_i,k_i}e^{-i\omega_x(\tau_3-\tau_2)}\times\nonumber\\
&\qquad e^{-i\omega_y(\tau_4-\tau_1)}e^{-ik_x(i_3-i_2)}e^{-ik_y(i_4-i_1)}\Gamma_\alpha(s_1,s_2,s_3,s_4)\nonumber\\
&\Gamma_{\alpha,(r_3-r_1),(r_4-r_2)}^{phe}(\omega_{phe},k_{phe})=\mathcal{N}_{\omega,k}\sum_{\omega_i,k_i}e^{-i\omega_x(\tau_3-\tau_1)}\times\nonumber\\
&\qquad e^{-i\omega_y(\tau_4-\tau_2)}e^{-ik_x(i_3-i_1)}e^{-ik_y(i_4-i_2)}\Gamma_\alpha(s_1,s_2,s_3,s_4)
\end{align}
where $r_i$ labels the lattice index and imaginary time $(i,\tau)$, $s_i$ labels the frequency and momentum modes ($\omega,k$), $\mathcal{N}$ is the normalization constant and $\alpha$ labels the band index. These expansions can then be projected to the basis of choice to give the $\Pi$, $\Delta$ and $X$ vertex.
\begin{figure}
\centering
\includegraphics[scale=0.49]{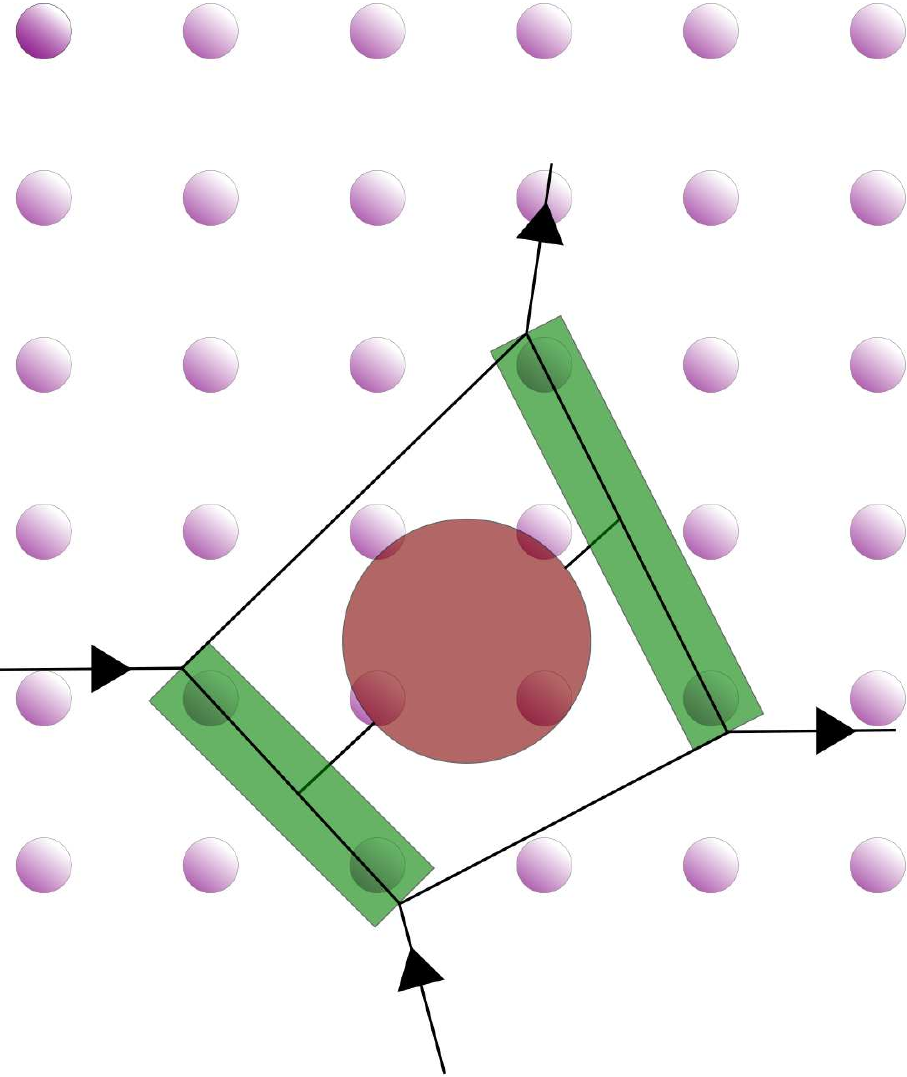}
\caption{The interacting vertex $\Gamma$ with the auxiliary channels in green and the singular channel in red. Such a decomposition is possible in all three channels leading to the three expansions, $\Pi_{m,n}(s_{pp})$, $\Delta_{m,n}(s_{ph})$, $X_{m,n}(s_{phe})$. Convergence of the expansion in the auxiliary channel is given in Fig. \ref{uNWsinc} for the frequency basis set and in Fig. \ref{EHMnn} for the spatial expansion.}
\label{vertexExpansion}
\end{figure}

\section{Expansion of the Vertex}
\label{exVert}
\begin{figure}
\centering
\includegraphics[scale=0.57]{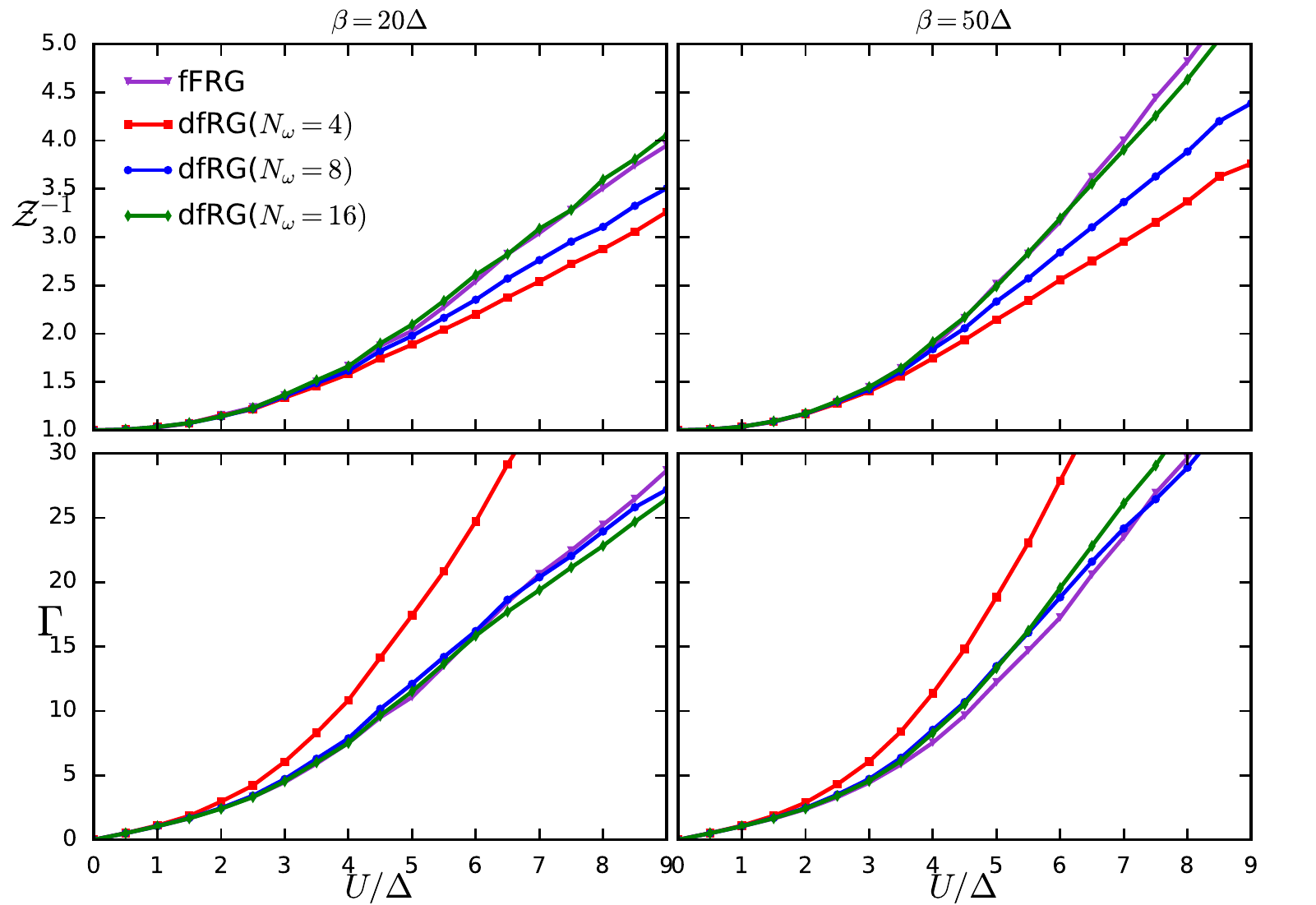}
\caption{Quasi-particle weight (Top) and vertex (Bottom) of the SIAM as a function of the local Hubbard interaction via the 1-Loop full fRG (ffRG) and the decoupled variant (at $N_\omega=4,8,16$). The flow was regulated with the Litim regulator with $N_f=60$ matsubara frequencies retained.}
\label{1LoopCompZG}
\end{figure}

\begin{figure}
\centering
\includegraphics[scale=0.7]{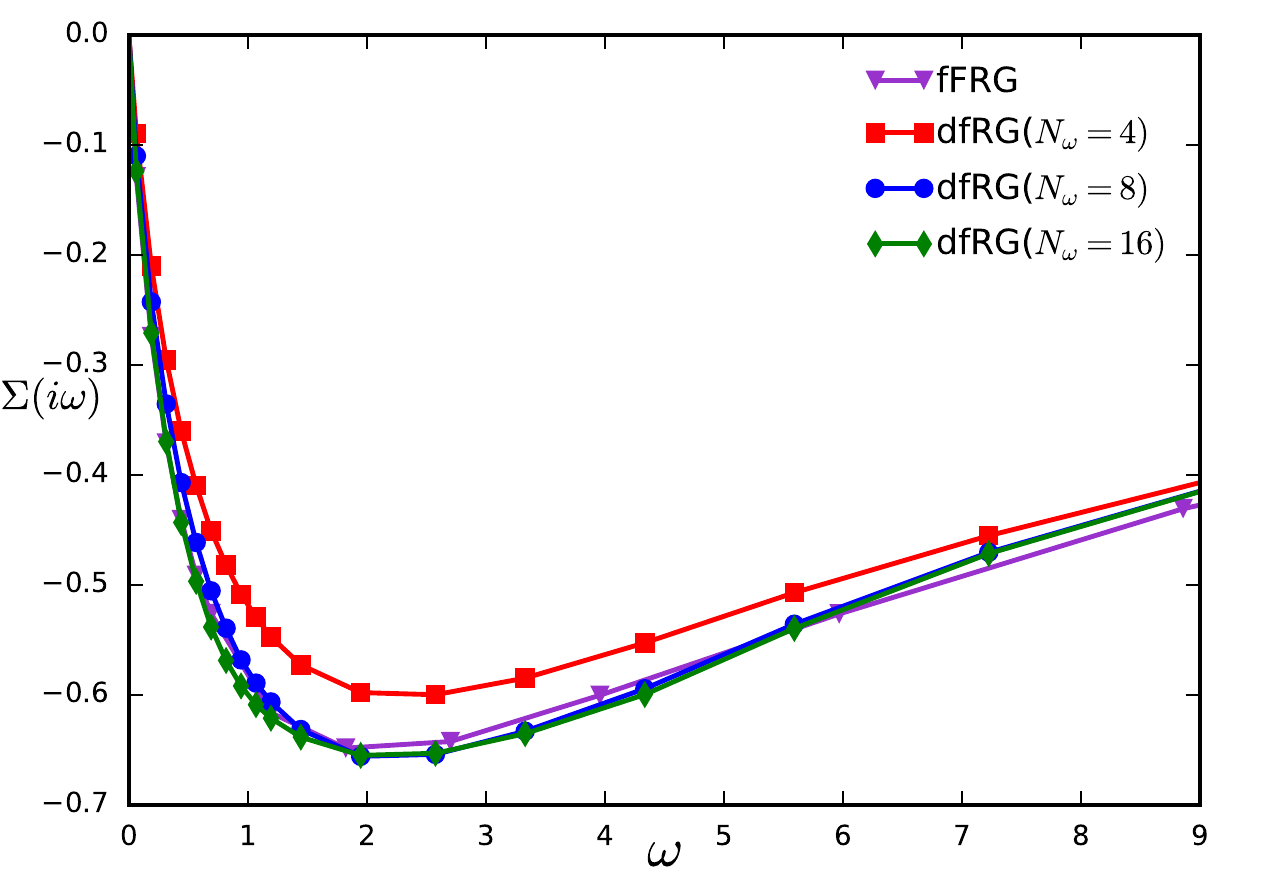}
\caption{The one-loop self energy of the SIAM calculated via the full and decoupled fRG. The litim regulator was used to construct the flow with $N_f=40$ frequencies retained.}
\label{1LoopFullSE}
\end{figure}

\begin{figure}
\centering
\includegraphics[scale=0.57]{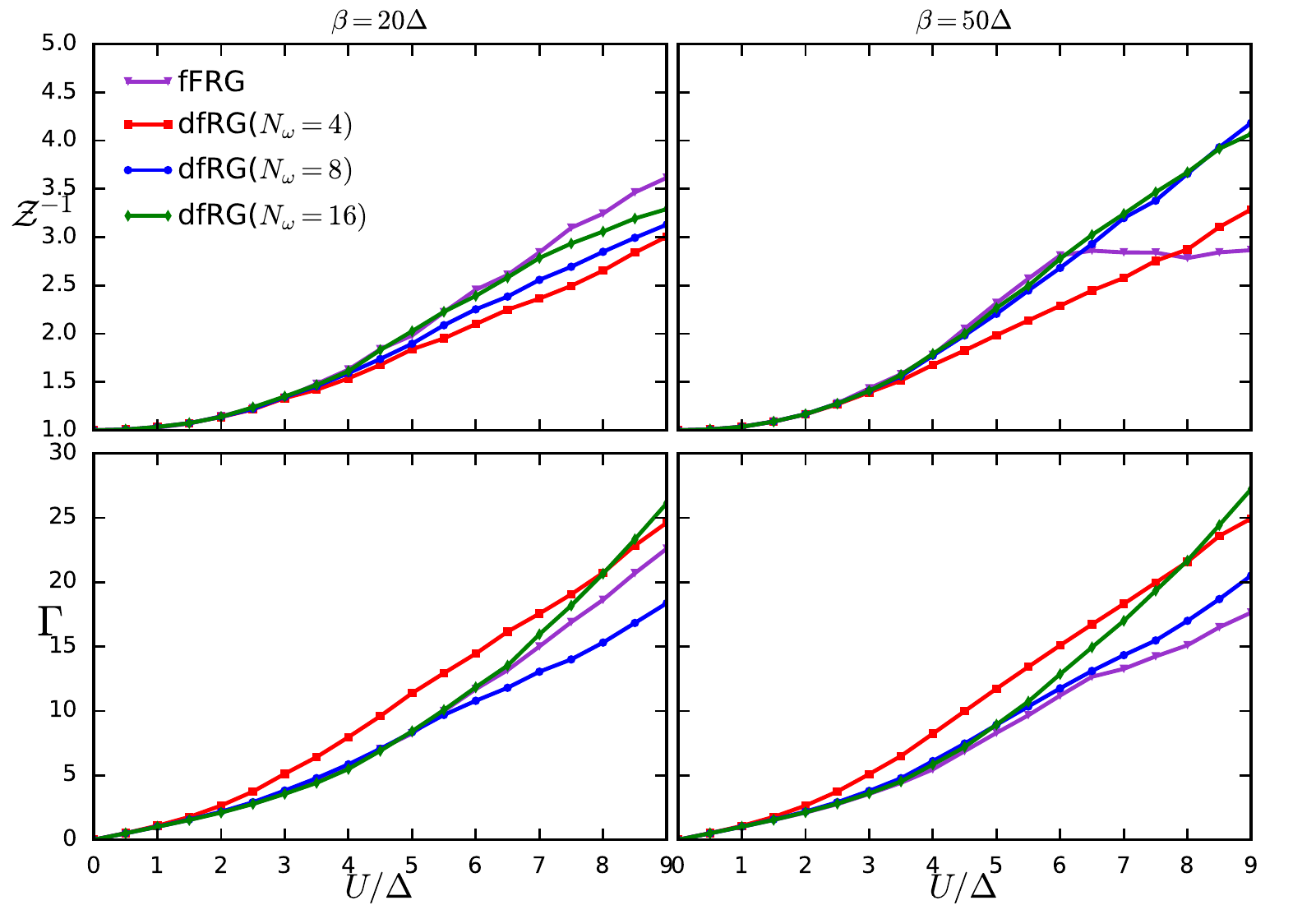}
\caption{Quasi-particle weight (Top) and vertex (Bottom) of the SIAM as a function of the local Hubbard interaction via the 2-Loop full fRG (ffRG) and the decoupled variant (at $N_\omega=4,8,16$). The flow was regulated with the Litim regulator with $N_f=60$ matsubara frequencies retained.}
\label{2LoopCompZG}
\end{figure}

\begin{figure}
\centering
\includegraphics[scale=0.7]{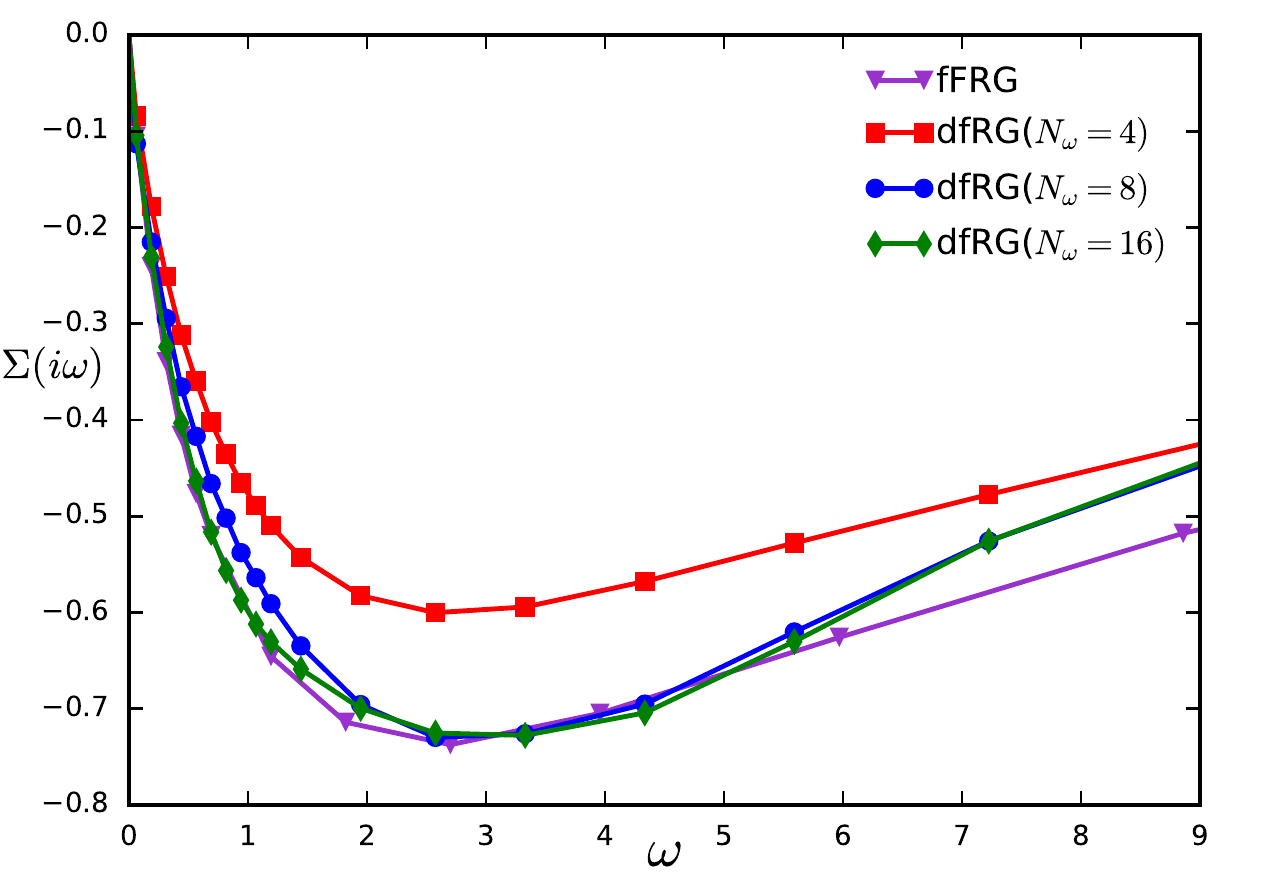}
\caption{The two-loop self energy of the SIAM calculated via the full and decoupled fRG. The litim regulator was used to construct the flow with $N_f=40$ frequencies retained.}
\label{2LoopFullSE}
\end{figure}

The convergences of the three expansions of the vertex ($\Pi_{m,n},\Delta_{m,n},X_{m,n}$) is difficult to demonstrate for the general model. With each expansion only good for describing the vertex around the corresponding singular frequency as we increase the basis set ($N_\omega\rightarrow\infty$) the computational cost of a run quickly becomes relevant. An approach implemented in this work for higher dimensional models involved varying the basis set ($N_\omega$, $N_K$). Sensitivity to such variation is apparent near phase transitions and should help clarify the extent of a phase. Such studies however do not answer the general question of the size of
basis set needed to reproduce the undecoupled fRG flow. A step towards this direction can be taken in  low dimensional models like the SIAM where we can directly solve the fRG equations and compare our results with those obtained via the decoupled fRG. The SIAM is the ideal candidate as the computational cost of runs is low and the model converges for all values of the coupling. Solving the full fRG equations for the SIAM involves Eq. \ref{fRGequations} with the vertices in the three channels given by
\begin{align}
\Gamma_{\omega_1\omega_2\omega_3\omega_4}^{(4)} = U +\Phi_{\omega_1\omega_2\omega_3\omega_4}^{pp} + \Phi_{\omega_1\omega_2\omega_3\omega_4}^{ph} +\Phi_{\omega_1\omega_2\omega_3\omega_4}^{phe}
\end{align}
where $U$ is Hubbard coupling. We retain $N_f=40$ frequencies for each singular frequency and solve the flow of the
full $[N_f\times N_f\times N_f]$ set of couplings. The self energy obtained from this flow along with self energies from the decoupled flow are shown in Fig.\ref{1LoopFullSE} for the one-loop fRG equations and at two-loop in Fig.\ref{2LoopFullSE}. A similar comparison between the two fRG schemes for the quasi particle weight and the vertex is shown in Fig.\ref{1LoopCompZG} for one-loop and in Fig.\ref{2LoopCompZG} for the two-loop flow.
\begin{figure}
\centering
\includegraphics[scale=0.59]{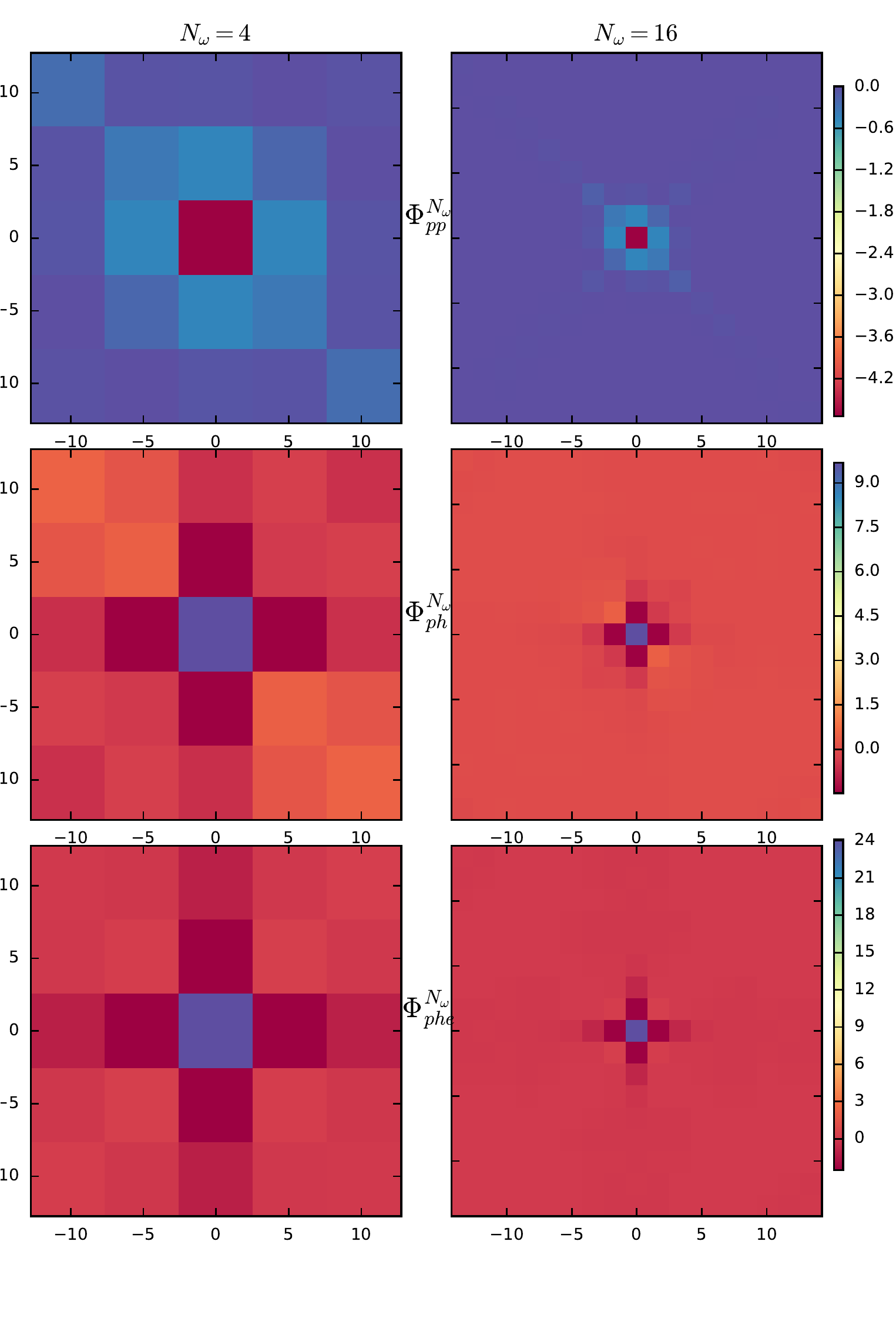}
\caption{Cluster averaged vertices in the three channels obtained via the full 2-Loop fRG with coupling $U=5\Delta$ at zero singular frequency ($\omega_s=0$) for $N_\omega=4,16$. The litim regulator was used with the temperature set to $\beta=50$.}
\label{vertExpandW0}
\end{figure}
\begin{figure}
\centering
\includegraphics[scale=0.59]{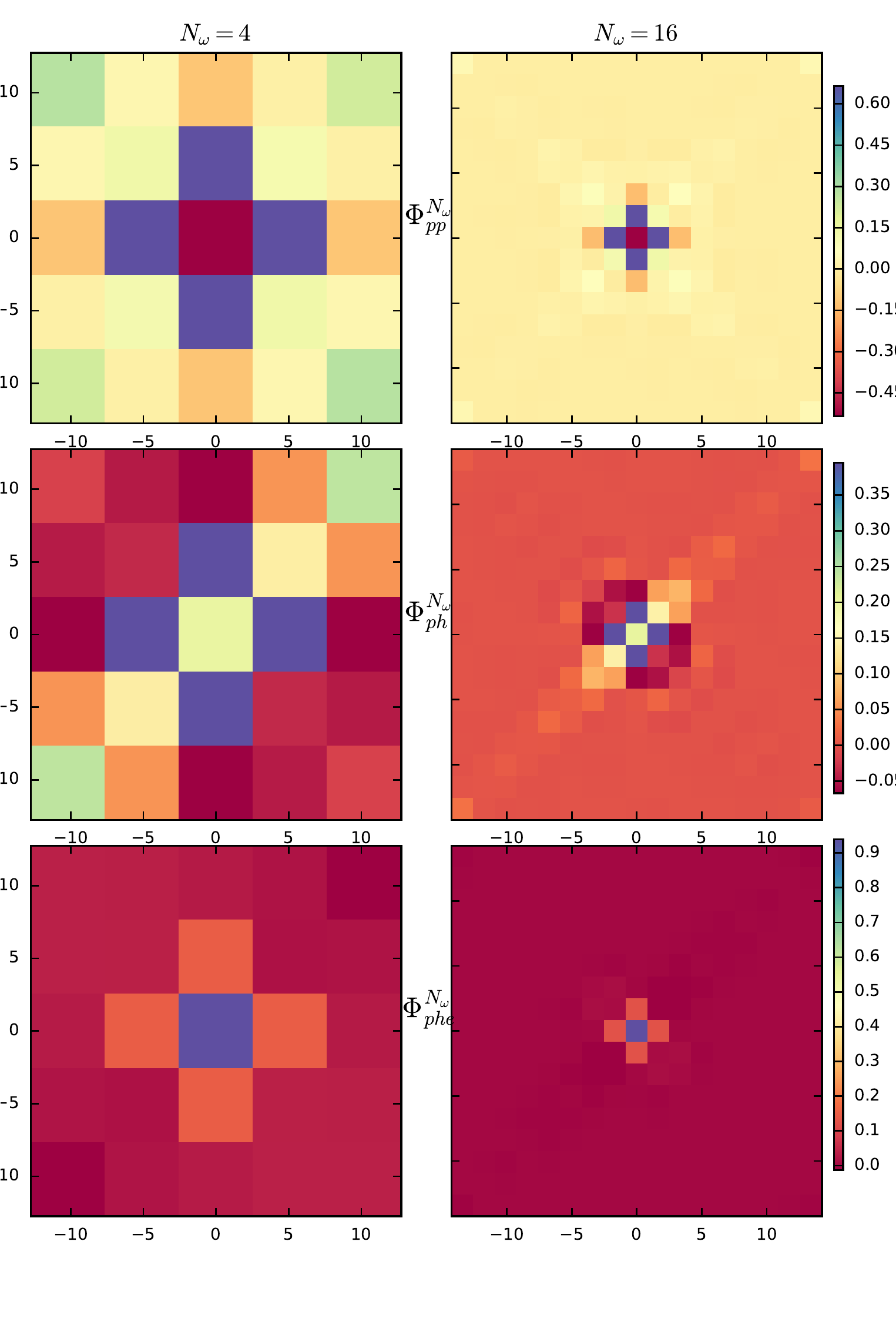}
\caption{Cluster averaged vertices in the three channels obtained via the full 2-Loop fRG with coupling $U=5\Delta$ at zero singular frequency ($\omega_s=\Delta$) for $N_\omega=4,16$. The litim regulator was used with the temperature set to $\beta=50$.}
\label{vertExpandW1}
\end{figure}
We can go beyond simple comparisons of the results from the two schemes by directly averaging over the final vertex ($\Phi^{pp},\Phi^{ph},\Phi^{phe}$) obtained from the full fRG equations. Within the decoupled fRG the full vertex ($\Gamma$) in the particular channel (s) is constructed by projection
\begin{align}
\Gamma_{m,n}^{s}(\omega_s)=\Gamma_{m,n}^{(0)}+\Phi_{m,n}^{s}(\omega_s)+P(\Phi_{m,n}^{r\neq s}) + P(\Phi_{m,n}^{r\neq s})
\end{align}
leading to the vertex in the channel via a product ($\boldsymbol{\Phi}^s(\omega_s)=\boldsymbol{\Gamma}^{s}(\omega_s)\boldsymbol{L}_\Lambda^s(\omega_s)\boldsymbol{\Gamma}^{s}(\omega_s)$). Direct averaging over the vertices from the flow shows us how well a basis set with $N_\omega$ functions captures the vertex. The averaged vertex is given by
\begin{align}
\Phi_{m,n}^{pp}(\omega_{pp}) =\frac{1}{\beta^2}\sum_{\omega_x\omega_y}\Phi_{\omega_{pp},\omega_{ph},\omega_{phe}}^{pp}f_m(\omega_x)f_n(\omega_y)
\label{averageVertex}
\end{align}
with similar expressions for the particle-hole and exchange channels. The function for the cluster averaging is just a shifted sinc function given by
\begin{align}
f_m(\omega)=e^{i\omega\tau_m} sinc(\omega\Delta\tau_m)
\end{align}
where $\Delta\tau_m$ corresponds to the width of interval. The averaged vertices for the SIAM with coupling $U=5\Delta$ are shown in Fig.\ref{vertExpandW0} for $\omega_s=0$ and in Fig.\ref{vertExpandW1} at $\omega_s=\Delta$. The figures clearly show that the contributions from each channel drops quickly to zero as $\tau_{x/y}>>0$. Increasing $N_\omega$ captures additional details which can be done more cleanly by rotating the discretization grid as shown in Fig.\ref{Patches}.

Despite the efficient expansion of the converged vertex ($\Gamma^{(4),\Lambda=0}$) intermediate vertices are not readily expandable in the same basis set. Exapnsion of the unconverged intermediate vertices over the frequency domain is complicated due to the scale dependence that is baked in all RG methods. In keeping with this structure we need to scale the basis set to expand the vertex. The intermediate vertex at scale $\Lambda$ can only describe the dynamics of our system at higher scales and thus can be associated with the vertex of a system at a higher temperature ($\beta_\Lambda$). For a system at some temperature $\beta$ the propagator ($\mathcal{G}$) is defined overt the domain $[0,\beta]\times\Omega_{BZ}$ where $\Omega_{BZ}$ is represents the Brillouin zone. One can decompose the function over $\Omega_{BZ}$ via fourier modes or a suitable chosen set of orthogonal polynomials. Similarly the imaginary time interval can be expanded via the Legendre polynomials. The Legendre polynomials have been shown\cite{boehnke2011orthogonal} to offer a compact representation of the Matsubara axis. For the projection of the frequency dependence in the singular channels we need scale the domain in accordance with cutoff ($\Lambda$) and average over the imaginary time axis for the cluster expansion or map the unbounded frequency domain for an expansion interms of the Legendre polynomials\cite{shen2009some}. The rescaling of the vertex generates a term $F_\Lambda$ which emerges from the basis functions given as
\begin{align}
\partial_\Lambda \Phi_{\omega_1\omega_2\omega_3\omega_4}^\Lambda=\partial_\Lambda\sum_{m,n}\Phi_{m,n}^\Lambda(\omega_{pp})f_m(\omega_x/f_\Lambda)f_n(\omega_y/f_\Lambda)
\end{align}
with the derivative generating three terms ($\partial_\Lambda\Phi f_mf_n + \Phi\partial_\Lambda(f_mf_n)$). Using the orthogonality of our basis set we simplify to obtain $F_\Lambda$
\begin{align}
\partial_\Lambda\Phi_{a,b}^\Lambda + \sum_{\omega_x\omega_y}\Phi_{m,n}^\Lambda(\omega_pp)f_a(\omega_x^\Lambda)f_b(\omega_y^\Lambda)\partial_\Lambda(f_m(\omega_x^\Lambda)f_n(\omega_y^\Lambda))
\end{align}
the second term which vanishes as $\omega/f_\Lambda\rightarrow \omega$. We note that for the Legendre expansion this term has to be mapped to finite domain in accordance with the vertex.


\begin{figure}
\centering
\includegraphics[scale=0.59]{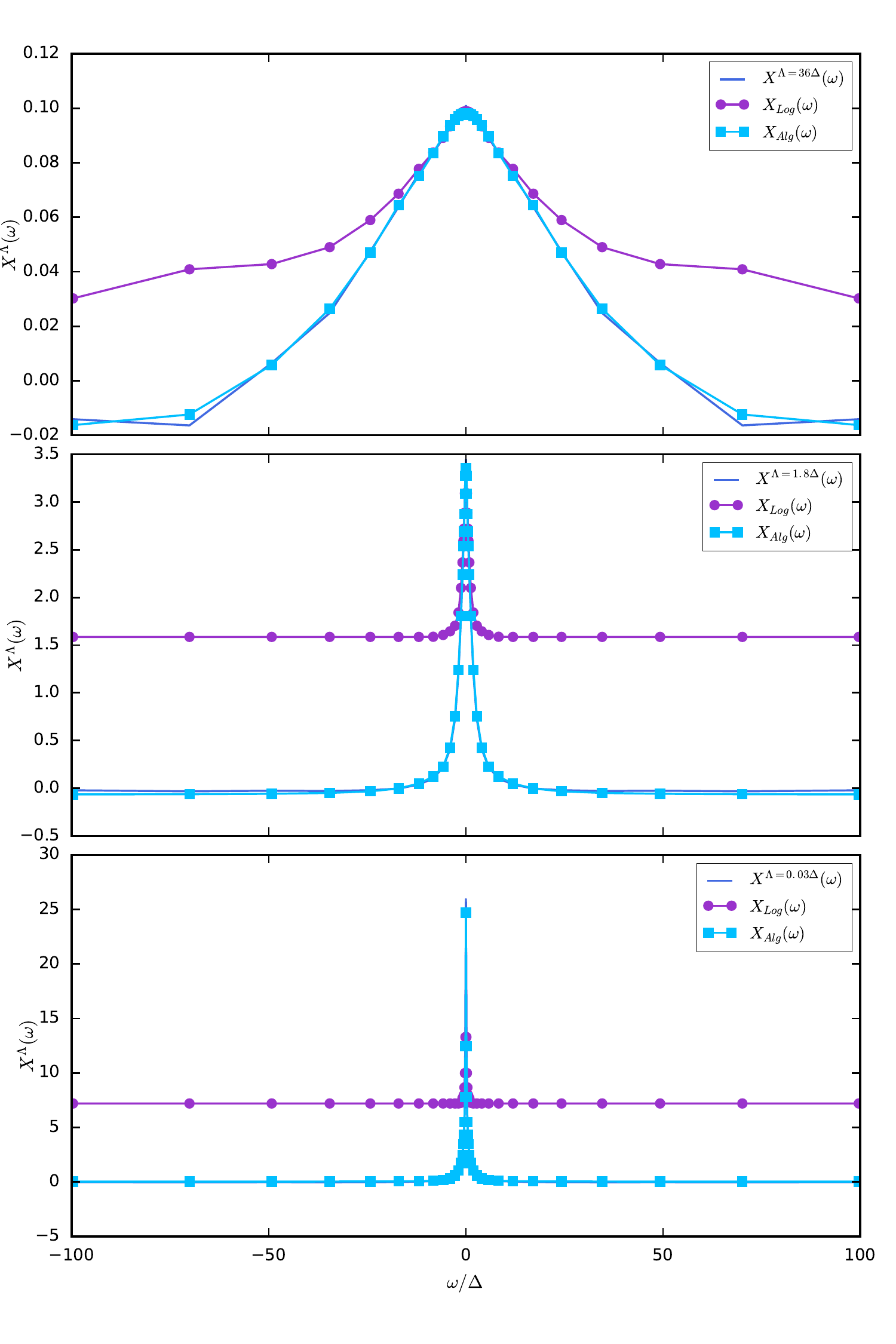}
\caption{The single channel phe vertex and the reconstructed phe vertex for the SIAM as a function of frequency at three scales in the flow ($N_\omega=10$, $U/\Delta=3$, $\beta=50$).($\Lambda=36\Delta$, $\Lambda=1.8\Delta$, $\Lambda=0.03\Delta$).}
\label{uPPdiffA}
\end{figure}

\begin{figure}
\centering
\includegraphics[scale=0.59]{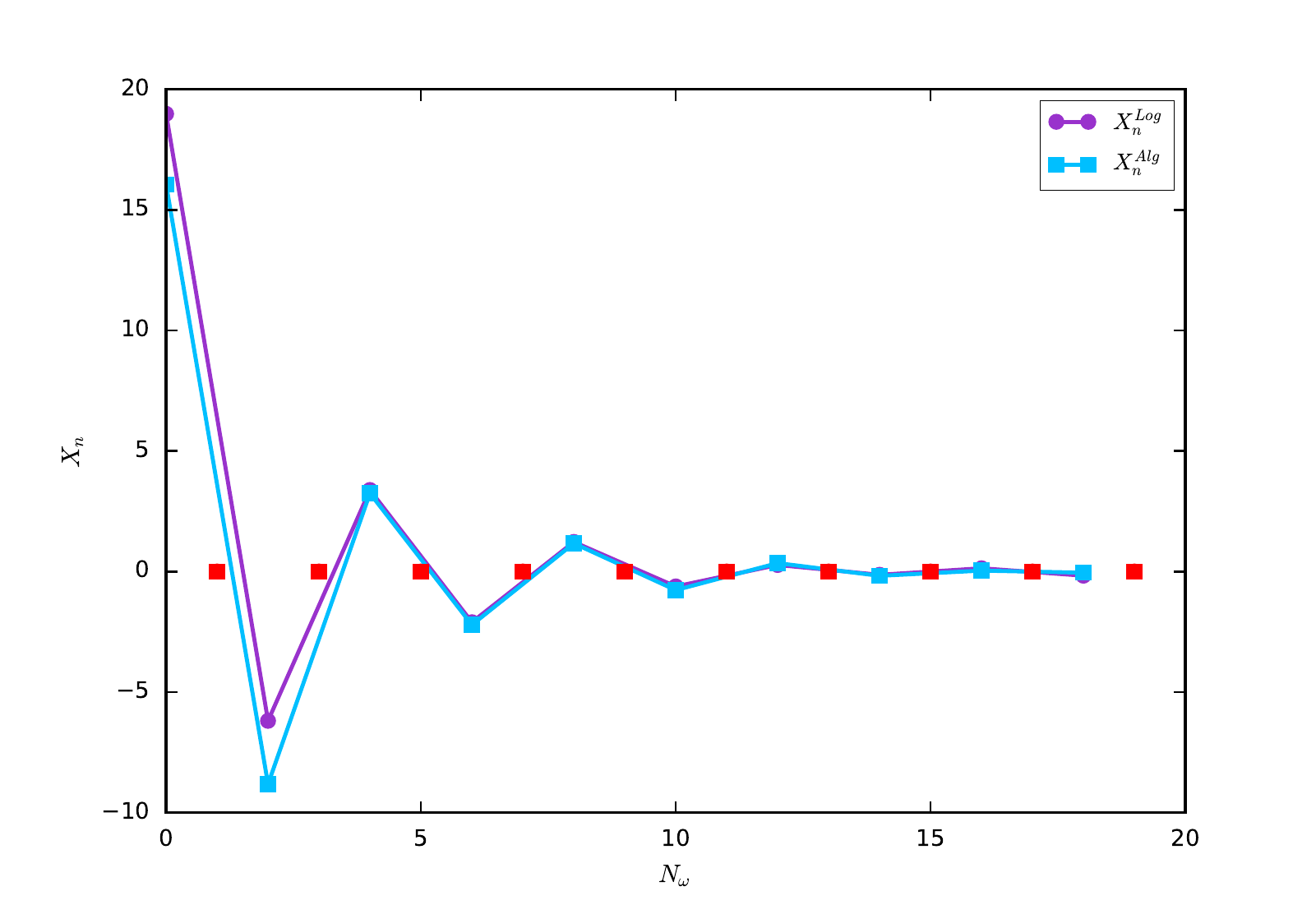}
\caption{Convergence of the algebraic and logarthmic Legendre coefficients of the single channel phe vertex at $\Lambda=0.0$ for the SIAM.}
\label{legendC}
\end{figure}

The map for the Legendre expansion of the frequency domain can be of the algebraic, logarithmic and exponential variant. We consider the first two as they capture much of the domain relevant to the systems we consider. The mappings and their inverse are given by
\begin{align}
z&=\frac{\omega}{\sqrt{\omega^2+f(\Lambda)^2}},\qquad &\omega=\frac{f(\Lambda)z}{\sqrt{1-z^2}}\\
z&=\tanh(\omega f(\Lambda)^{-1}), \qquad &\omega=f(\Lambda)\tanh^{-1}(z)
\end{align}
where $f(\Lambda)$ scales our domain along the flow. There is some freedom in the choice of $f$ as it depends weakly upon the system and our choice of regulator. In our numerical work we remove this dependence and construct it numerically via the flow of the pp/ph bubbles. To clarify our choice of an algebraic map for the Legendre expansions we apply both mappings to the flow of the single channel particle-particle vertex. The flow equation for the vertex is given by
\begin{equation}
\partial_\Lambda X^\Lambda(s_{phe})=-X^\Lambda(s_{phe})\partial_\Lambda\left(\mathcal{G}_s^\Lambda\mathcal{G}_{s+s_{phe}}^\Lambda\right)X^\Lambda(s_{phe})
\end{equation}
where $\mathcal{G}$ is the propagator. As the mapping is over the frequency domain let us specialize to the SIAM and choose a simple form for f ($f(\Lambda)=\sqrt{\Lambda^2+\Lambda_0(\beta)^2}$). The value of $\Lambda_0$ can be set to the hybridization ($\Delta$) or chosen in relation to the starting scale of the vertex. The vertex calculated from the flow at various scales is shown in Fig.\ref{uPPdiffA}. The reconstruction of the vertex via the two mappings is also shown. We can see that the algebraic map captures the vertex well especially at high frequencies. This is a direct consequence of the algebraic decay of the vertex. The Legendre coefficients for expansions of the converged vertex are shown in Fig.\ref{legendC}. We end on the final note that the difficulty of generalizing the basis sets required to reproduce the fRG flow to the higher dimensional systems considered lies not only in the high computational demands but also the instabilities present in these models.

\section{Asymptotics of the Vertex}
\label{asymP}
The asymptotics of the two-particle vertex is driven by the static interaction. The initial interaction vertices addressed in this work are instantaneous and have support at all frequencies. Thus, they cannot be expanded in any rational polynomial basis. The high frequency asymptotics of the vertex has been studied in previous works which has led to an efficient parameterization scheme\cite{wentzell2016high}. It has been shown that a proper accounting of the high frequency structure can offer up to three orders of magnitude improvement in terms of numerical accuracy\cite{wentzell2020high}. We found a separation of the static vertex from the dynamical vertex to offer numerous computational advantages. This separation occurs naturally as the initial vertex does not flow. In the single channel case we have
\begin{align}
\partial_\Lambda\Pi^\Lambda(s_{pp})=&(U+\Pi^\Lambda(s_{pp}))\partial_\Lambda\left(\sum_s\mathcal{G}_s^\Lambda\mathcal{G}_{s_{pp}-s}^\Lambda\right)\times\nonumber\\
&(U+\Pi^\Lambda(s_{pp}))
\end{align}
where the flow of dynamic vertex, $\Pi^\Lambda(\omega)$ can be expanded over the appropriate set of basis functions and projected to the other channels.

When dealing with the full fRG equations, the separation is a bit more involved, as the interaction vertex now depends on three variables. Focusing on the flow in the pp channel we have
\begin{align}
\partial_\Lambda\Phi_{s_1,s_2,s_3,s_4}^{pp,\Lambda}=&U^2\sum_s\partial_\Lambda\left(\mathcal{G}_s^\Lambda\mathcal{G}_{s_{pp}-s}^\Lambda\right)+\nonumber\\
&U\sum_s\Pi_{s_1,s_2,s_{pp}-s,s}\partial_\Lambda\left(\mathcal{G}_s^\Lambda\mathcal{G}_{s_{pp}-s}^\Lambda\right)+\nonumber\\
&U\sum_s\Pi_{s,s_{pp}-s,s_3,s_4}\partial_\Lambda\left(\mathcal{G}_s^\Lambda\mathcal{G}_{s_{pp}-s}^\Lambda\right)+\nonumber\\
&\Pi_{s_1,s_2,s_{pp}-s,s}\partial_\Lambda\left(\mathcal{G}_s^\Lambda\mathcal{G}_{s_{pp}-s}^\Lambda\right)\Pi_{s,s_{pp}-s,s_3,s_4}
\end{align}
which, in addition to the static piece described by the pp singular frequency ($s_{pp}$) and particle-particle interaction term, contains two pieces each of which depend on the incoming and outgoing particles. Our parameterization is identical to previous studies\cite{wentzell2020high} and that adopted for the decoupling above. We can construct the full vertex in the particle-particle channel from the flow of these four terms. The vertex in the pp channel is given by
\begin{align}
\Gamma_{s_1,s_2,s_3,s_4}^\Lambda&=U + \Pi^\Lambda(s_{pp}) +\Pi_{s_{pp_X}}^\Lambda(s_{pp}) +\nonumber\\
&\Pi_{s_{pp_y}}^\Lambda(s_{pp}) + \Pi^\Lambda_{s_{pp_x},s_{pp_y}}(s_{pp})
\label{intTerm}
\end{align}
with identical expansions in the $ph$ and $phe$ channels. The three additional terms describe the contributions from renormalization of an incoming pair objects parameterized by $s_{pp_x}$, an outgoing object $s_{pp_y}$, with the last term describing the interaction between the two two-particle objects. We track the flow of each piece separately. Finally, we should note the difference between the flow in the channel $\Phi^{pp,\Lambda}$ and the vertex in the channel, $\Pi^{\Lambda}$. In the single channel case these two functions are the same but when we consider the flow in all channels, the vertex $\Pi^{\Lambda}$ in addition to $\Phi$ contains projections from the $pp$ and $phe$ channels.
\section{Cross Projections between the Channels}
\label{projectionD}
The contribution from some channel X to another channel Y is always contained within the interaction term which is expressed in terms of the vertex in Eq.\ref{intTerm}. The cross channel projection can then be calculated by expressing the vertex from the channel we are projecting in terms of the appropriate frequencies. This expression is given for the $ph$ channel in Eq.\ref{projectV}. In general we insert and expansion of the vertex into our equation for the expansion, Eq.\ref{threeExpansions}. If we don't separate out the asymptotics the projections are cleaner but may prove costly from a numerical perspective (Evaluation of the flow equations for the self energy can be slow as a proper description of the self energy at high frequency will require an evaluation of the the vertex which is a slow operation). The projection is given by
\begin{align}
\Pi_{m,n}(\omega_{pp}^\Lambda)&=\int\int dz_xdz_y U(\omega_{pp}^\Lambda,\omega_{ph}^\Lambda,\omega_{phe}^\Lambda)f_m(z_x)f_n(z_y)\nonumber\\
=&\sum_{a,b,l}\int\int dz_xdz_y\Delta_{a,b}(l)f_l(z_{ph})f_a(z_{ph_x})f_b(z_{ph_y})\nonumber\\
&f_m(z_x)f_n(z_y)
\end{align}
 where $\omega^\Lambda$ are the scaled frequencies. We have an identical expression for the projection of the other vertices. For the case where the asymptotics are separated the projection of the frequency dependence of the vertex from the other channels to the particle-particle channel,
\begin{align}
&\Pi_{\tau_1,\tau_2}(\omega_{pp})=\delta(\tau_s)\Delta(2\tau_1) + e^{-i\omega_{pp}\tau_s}\Delta_{\tau_s}^{I}(\tau_d)+\nonumber\\
&e^{i\omega_{pp}\tau_s}\Delta_{\tau_s}^O(\tau_d) +\int d\tau e^{i\omega_{pp}(\tau_s-2\tau)}\Delta_{\tau,\tau_s-\tau}(\tau_d)\\
&\Pi_{\tau_1,\tau_2}(\omega_{pp})=\delta(\tau_d)X(-2\tau_1) + e^{i\omega_{pp}\tau_d}X_{-\tau_d}^{I}(-\tau_s)+\nonumber\\
&e^{-i\omega_{pp}\tau_d}X_{-\tau_d}^O(-\tau_s)+\int d\tau e^{i\omega_{pp}(-\tau_d-2\tau)}X_{\tau,-\tau_d-\tau}(-\tau_s)
\end{align}
particle-hole channel,
\begin{align}
&\Delta_{\tau_1,\tau_2}(\omega_{ph})=\delta(\tau_s)\Pi(2\tau_1) +e^{-i\omega_{ph}\tau_s}\Pi_{\tau_s}^{I}(\tau_d)\nonumber\\
&e^{i\omega_{ph}\tau_s}\Pi_{\tau_s}^{O}(\tau_d)+\int d\tau e^{i\omega_{ph}(\tau_s-2\tau)}\Pi_{\tau,\tau_s-\tau}(\tau_d)
\end{align}
\begin{align}
&\Delta_{\tau_1,\tau_2}(\omega_{ph})=\delta(-\tau_d)X(-2\tau_1) + e^{i\omega_{ph}\tau_d}X_{\tau_d}^{I}(-\tau_s)+\nonumber\\
&e^{-i\omega_{ph}\tau_d}X_{-\tau_d}^{O}(-\tau_s)\int d\tau e^{i\omega_{phe}(2\tau-\tau_d)}X_{\tau,\tau-\tau_d}(-\tau_s)
\end{align}
and the particle-hole exchange channel
\begin{align}
&X_{\tau_1,\tau_2}(\omega_{phe})=\delta(\tau_s)\Pi(2\tau_1)+ e^{i\omega_{phe}\tau_s}\Pi_{\tau_s}^{I}(\tau_d)+\nonumber\\
&e^{-i\omega_{phe}\tau_s}\Pi_{-\tau_s}^{O}(\tau_d)+\int d\tau e^{i\omega_{phe}(2\tau-(\tau_s))}\Pi_{\tau,\tau-\tau_s}(\tau_d)
\end{align}
\begin{align}
&X_{\tau_1,\tau_2}(\omega_{phe})=\delta(-\tau_d)\Delta(-2\tau_1)+e^{i\omega_{phe}\tau_d}\Delta_{\tau_d}^{I}(-\tau_s)+\nonumber\\
&e^{-\omega_{phe}\tau_d}\Delta_{-\tau_d}^{O}(-\tau_s)+\int d\tau e^{i\omega_{phe}(2\tau-\tau_d)}\Delta_{\tau,\tau-\tau_d}(-\tau_s)
\end{align}
where I and O label the incoming and outgoing contributions defined above and the two times are defined as $\tau_s=\tau_1+\tau_2$ and $\tau_d=\tau_1-\tau_2$. One can then construct the appropriate Legendre coefficients for the projection. The singular contributions from each channel are expanded via $N_m$ Legendre polynomials. The expansion is full at every step of the RG in the sense that we choose $N_m$ such that we reproduce, within tolerance, the singular contributions at all frequencies. The non-singular frequencies above are expanded up to some order N that is within computational reach.

Cross channel projections for systems with momentum modes involve the additional step of momentum projection. The order of projections can be chosen depending on the size of the momentum and frequency basis sets.  The projection of momentum modes into the particle-particle channel is given by
\begin{align}
\Pi_{a,b}(k_{pp})&=e^{-ik_{pp}b}\sum_l\Phi_{a-l,l-b}^{ph}(l)e^{ik_{pp}l}\nonumber\\
&=\sum_l\Phi_{a+l,-l+b}^{phe}(l)e^{-ik_{pp}l}
\end{align}
with $\Phi_{m,n}^r(l)$ corresponding to the inverse Fourier transform of the momentum dependent piece of the channel r ($\mathcal{F}^{-1}(\Phi_{m,n}^{r}(k_r))$). The sums above can be calculated via an FFT for each a and b. This puts the overall effort of expansion and projection at $2N_k^2N\ln(N)$ for a system with N sites and $N_k$ modes in the auxiliary channels. The expansion for the particle-hole channel is given by
\begin{align}
\Delta_{a,b}(k_{ph})&=e^{ik_{ph}a}\sum_m\Phi_{m,m-b-a}^{pp}(-b)e^{-ik_{ph}m}\nonumber\\
&=e^{-ik_{ph}(b-a)}\sum_m\Phi_{m,m+b-a}^{phe}(-b)e^{-ik_{ph}m}
\end{align}
and for the particle-hole exchange channel as
\begin{align}
X_{a,b}(k_{phe})&=\sum_l\Phi_{a-l,l+b}^{pp}(l)e^{-ik_{phe}l}\nonumber\\
&=e^{ik_{phe}(b-a)}\sum_m\Phi_{m,m+b-a}^{ph}(b)e^{ik_{phe}m}
\end{align}
For cases where the asymptotic separation has been implemented these projections still hold but have to be implemented for all four pieces in the three channels.
\begin{figure*}
\centering
\includegraphics[scale=0.47]{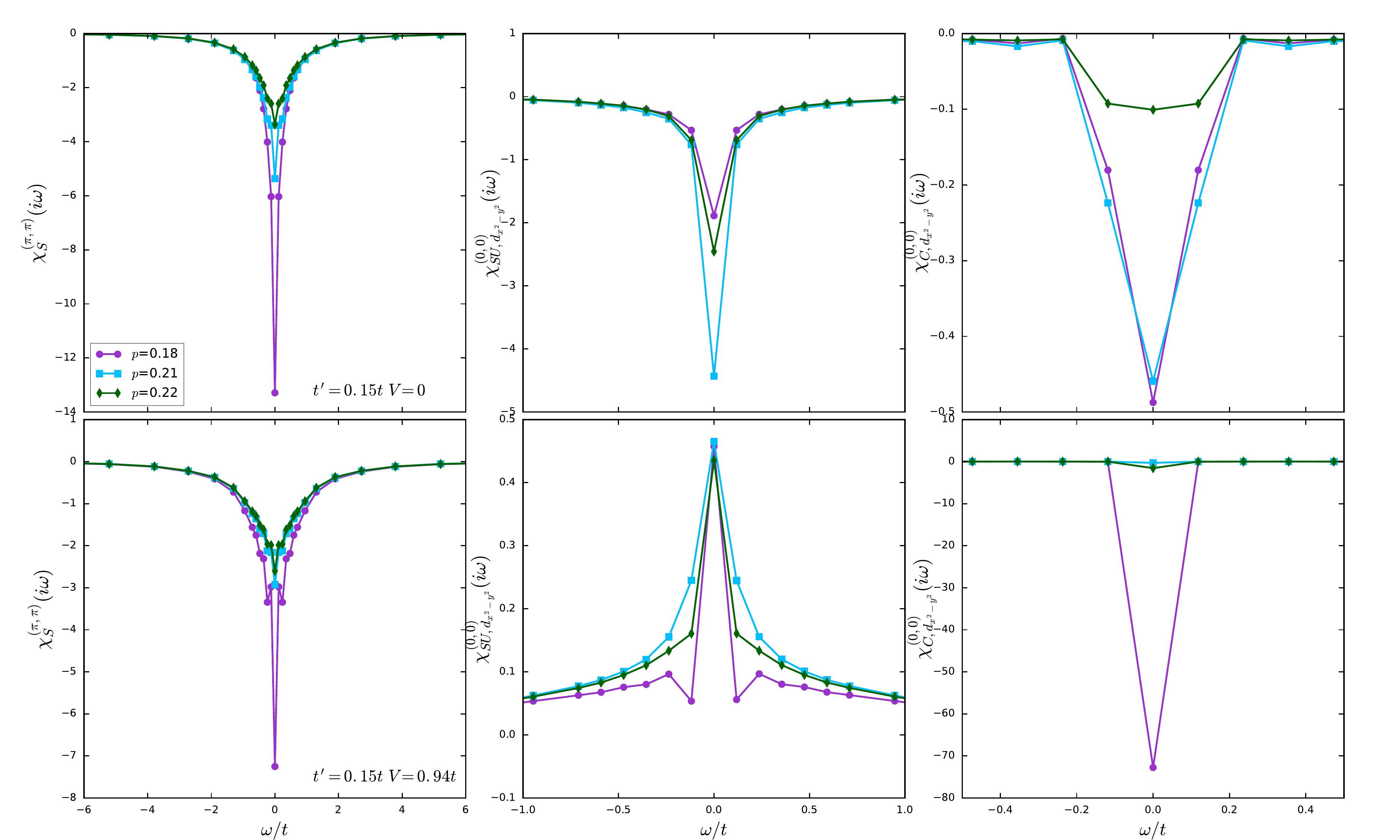}
\caption{The frequency dependence of the antiferromagnetic, $d_{x^2-y^2}$-superconducting and charge susceptibilities of the 2D EHM at $T=0.017t$ calculated via the 2-loop fRG ($N_\omega=4$, $N_K=2$, $N=16$) for different values of hole doping.}
\label{freqSusSSUC}
\end{figure*}
\section{Symmetries of the Vertex}
\label{symmV}
Further computational gains for the symmetric fRG are possible by employing the various symmetries of the vertices and the Hamiltonians of interest. In the discussion above we have exploited the translational invariance of the systems to parameterize the vertices via the singular frequencies. Additionally most of the systems we study are time-reversal invariant and are described by a hermitian Hamiltonian leading to an Osterwalder-Schrader positive action\cite{maier2014functional}. These two symmetries represent a constraint on the vertex given by
\begin{align}
\Gamma^{(2)}(\hat{p_4},\hat{p_3},\hat{p_2},\hat{p_1})=\Gamma(p_1,p_2,p_3,p_4)^{*}
\end{align}
where $\hat{p_1}=(-\omega_1,k_1)$. Applying this to our three expansions leads to
\begin{align}
\Pi_{m\times a,n\times b}(\omega_s,k_s) &= \Pi_{n\times b, m\times a}^{*}(-\omega_s,k_s)\nonumber\\
\Delta_{m\times a,n\times b}(\omega_s,k_s)&=e^{ik_s(a-b)}\Delta_{m\times-a,n\times-b}^{*}(\omega_s,-k_s)\nonumber\\
X_{m\times a,n\times b}(\omega_s,k_s) &= X_{n\times b,m\times a}^{*}(-\omega_s,k_s)
\end{align}
with the particle-hole case requiring the additional symmetry under exchange of particles ($\Gamma_{p_1p_2p_3p_4}=\Gamma_{p_2p_1p_4p_3}$) to construct values of the vertex at negative frequencies. The relation between the expansions under particle exchange is given by
\begin{align}
\Pi_{m\times a,n\times b}(\omega_s,k_s)&=e^{ik_s(a-b)}\Pi_{-m\times -a,-n\times -b}(\omega_s,k_s)\nonumber\\
\Delta_{m\times a,n\times b}(\omega_s,k_s) &=e^{ik_s(a-b)}\Delta_{n\times -b,m\times -a}(-\omega_s,-k_s)\nonumber\\
X_{m\times a,n\times b}(\omega_s,k_s) &=e^{ik_s(b-a)}X_{n\times-b,m\times-a}(-\omega_s,-k_s)
\end{align}
where $\omega_s$ and $k_s$ represent the singular frequency and momentum in the channel. For discrete lattice we go can even further and utilize the lattice to reduce the momenta points tracked throughout the flow. We consider the general case of space inversion symmetry which applies even to the one dimensional models considered in this work. We have
\begin{align}
\Gamma^{(2)}(R(p_1),R(p_2),R(p_3),R(p_4))=\Gamma^{(2)}(p_1,p_2,p_3,p_4)
\end{align}
where $R(p)=(\omega,-k)$. For our expansions this leads to
\begin{align}
\Pi_{m\times i,n\times j}(p_{pp}) &= \Pi_{m\times -i, n\times -j}(R(p_{pp}))
\end{align}
with an identical expression for the ph and phe channels. Invoking these symmetries can cut computational expenditures by a fourth with even further gains achieved by employing the full symmetries of our lattice. Full application of symmetries is not computationally necessary at the single band level but as we tackle more complex systems utilization of these reduction will quickly become a necessity.

\section{Frequency content of the Vertex}
\label{freqOfVert}
The primary focus of this work was the construction of phase diagrams which requires as input static vertices ($\Gamma^{(2)}(\omega=0)$). The full frequency content of the vertices has only been used to construct the double occupancy (Fig.\ref{tdepDO},\ref{uvDoccup},\ref{dFillV}) and the quasiparticle weight of the models (Fig.\ref{tdepQWleg},\ref{qWU2V0}). In this section we consider the evolution of the frequency dependence of the susceptibilities as a function of doping and nearest neighbor interaction for the two dimensional Extended Hubbard model. As per the discussion above both the doping and the nearest neighbor interaction serve to destroy antiferromagnetic interactions which are dominant in the parameter regime of interest ($V<0.25U$, $p<0.15$). At high enough temperatures superconducting correlations are suppressed for the entirety of the parameter space. As the temperature is lowered spin fluctuations and the underlying superconducting or charge fluctuations show a strong response depending on the value of the nearest neighbor coupling.

In Fig.\ref{freqSusSSUC} we present the frequency dependence of the spin, charge and superconducting susceptibilities of the two dimensional EHM at $U=4t$. The figures show the role of doping in destroying antiferromagnetic fluctuations and enhancing superconducting and charge orders. Of particular note in the frequency dependence of the susceptibilities is the frequency window over which the orders are dominant. The antiferromagnetic order shows a strong response through a wide window ($\chi_s\sim t$) which indicates dominance at high temperatures over much of the parameter space. For the case with no nearest neighbor density-density interacting superconducting fluctuations emerge in a much narrower window ($\chi_{d_{x^2-y^2}}\sim 0.1t$). Given the strong incommensurate spin fluctuations that occupy a similar parameter regime we can conclude that they are the primary drivers of the order. Finite values of $V$ suppress antiferromagnetic fluctuations and destroy the underlying superconducting order. Instead of the superconducting order we see a strong charge order with $d_{x^2-y^2}$ density profile sitting at $\omega=0$. A larger basis sets, lower temperatures and  a systematic study as function of system size should help paint a fuller picture of the phase diagram of the EHM.
\end{document}